\documentclass[a4paper,twocolumn,11pt]{quantumarticle}
\pdfoutput=1

\usepackage{amssymb}
\usepackage{mathtools}
\usepackage{graphicx}
\usepackage{bbm}
\usepackage{color}
\usepackage{comment}
\usepackage{gensymb}
\usepackage{cite}
\usepackage[numbers,sort&compress]{natbib}

\begin{document}

\title{Hamiltonian Model for Fault Tolerant Singlet-Like Excitation: First Principles Approach}

\author{Donghyun Jin}
\affiliation{Georgia Institute of Technology, Atlanta, Georgia, United States}
%\email[]{djin40@gatech.edu}

\author{Grihith Manchanda}
\affiliation{Georgia Institute of Technology, Atlanta, Georgia, United States}
%\email[]{gmanchanda3@gatech.edu}

\author{Dragomir Davidovi\'c}
\affiliation{Georgia Institute of Technology, Atlanta, Georgia, United States}
\orcid{0000-0003-1486-4910}
\email[]{dragomir.davidovic@physics.gatech.edu}

\begin{abstract}
Deriving quantum error correction and quantum control from the Schr\"odinger equation for a unified qubit-environment Hamiltonian will give insights into how microscopic degrees of freedom affect the capability to control and correct quantum information beyond that of phenomenological theory. Here, we investigate the asymptotic reduced state of two qubits coupled to each other solely via a common heat bath of linear harmonic oscillators
and search for evidence of fault-tolerant excited qubit states. We vary the Hamiltonian parameters, including the qubit-qubit and qubit-bath detuning, the bath spectral density, and whether or not we use the Markov approximation in the calculation of our dynamics. In proximity to special values of these  parameters, we identify these states as asymptotic reduced states that are arbitrarily pure, excited, unique, and have high singlet fidelity. We emphasize the central role of the Lamb-shift as an agent responsible for fault tolerant excitations. To learn how these parameters relate to performance, we discuss numerical studies on fidelity and error recovery time.
\end{abstract}
\maketitle
\section{\label{sec:intro}Introduction}

Long-lasting coherence of quantum states is critical for quantum computation and quantum information processing.~\cite{Nielsen}
Alas, environmental coupling in realistic quantum systems leads to rapid decoherence of quantum states.
Surmounting this decoherence and other quantum information errors like damping remains the main challenge for the experimental realization of quantum computers. Quantum error correction (QEC)~\cite{Shor,Steane} opens a way to overcome decoherence and damping. It works by encoding quantum states into quantum codes, performing projective measurements of error syndromes, and recovering the true state via fast gates. Other approaches to fight decoherence include quantum feedback control for a continuously measured system~\cite{Wiseman,Vitali,wiseman_milburn_2009} and combination approaches.~\cite{Ahn}

A measurement apparatus can always be regarded as part of a larger isolated quantum system
subject to unitary dynamics. At this point, we know that overcoming decoherence can be derived from that unitary framework without invoking any measurements. Indeed, coherent feedback control is an example of that. It includes a Langevin equation approach to feedback mediated by continuous measurements, which can be implemented without measurements.~\cite{Wiseman} Unitary interactions alone between quantum systems can be utilized to implement feedback control.~\cite{Lloyd,Nelson} Another example where explicit measurements are not necessary comes from the principle of implicit measurement in quantum computing and QEC, which states that at the end of computation, any qubits which have not been measured may be assumed to have been measured.~\cite{Nielsen} The result is that QEC can be performed without explicit measurement.

\begin{figure}
\centering
\includegraphics[width=0.4\textwidth]{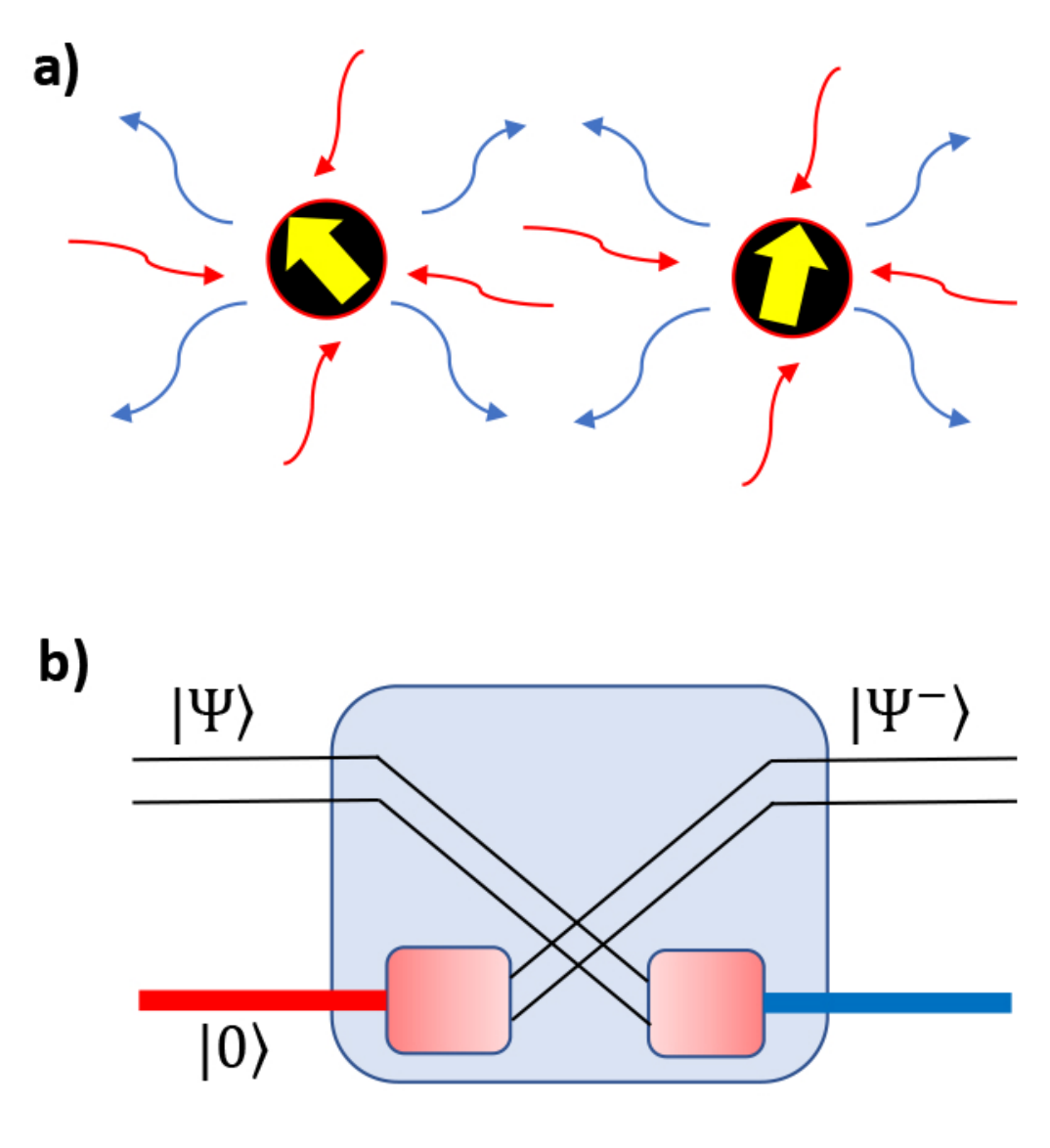}
  \caption{{\bf a)} Two qubits immersed in a common heath bath of linear harmonic oscillators at zero temperature. {\bf b)} Quantum circuit model for a heat bath as a stabilizer of singlet. On the left, the inputs
  are an arbitrary two-qubit state $\vert\Psi\rangle$ and the heat bath vacuum state $\vert 0\rangle$ (red, lower-left). On the right, in the qubit system, the output is the Bell or singlet state $\vert\Psi^-\rangle=\vert S\rangle$, while in the bath the output is  implicitly measured (blue, lower-right). The large blue box is a unitary gate.\label{fig:blackbox}}
\end{figure}

Much of the theory of QEC and quantum control is not based on first-principle derivations. Rather, it is based upon phenomenological error models that begin from a "higher level, effective description, most notably
that of Markovian dynamics."~\cite{Alicki_2006}
As an illustration, let us consider two qubits immersed in a heat bath of linear harmonic oscillators interacting solely via the bath, sketched in Fig.~\ref{fig:blackbox}~{\bf a)}.
We want to determine if the reduced state of the qubits in equilibrium can be all pure, excited, and fault-tolerant at the same time.
A black-box circuit model of the qubit-environment system that could lead to such a state could be like the one sketched in
Fig.~\ref{fig:blackbox}~{\bf b)}. Fault-tolerance would be a special interaction with the bath, where the qubit state is teleported,
transduced to the bath degrees of freedom, and swapped out by a reverse-transduced and reverse-ported ancillary singlet. By the principle of implicit measurement, quantum signal passed to the bath need not be measured in order for this to work.

However, without the microscopic Hamiltonian of a system-bath universe, the circuit model is only useful as a calculation tool,
rather than a way to gain insight into questions like:
What is the recovery time from errors? How does the recovery time affect state fidelity? What is the effect of the bath
spectral density on fidelity, especially at low frequencies? What are the roles of the bath counterterms? The Hamiltonian model presented in this paper opens a way to answering all these questions. Other approaches include quasi-phenomenological descriptions of fault tolerant QEC~\cite{Mohseni,Barbara,Aliferis,Aharonov} and continuous quantum feedback,~\cite{Ahn,Sarovar_2004,Sarovar_2005} and fully microscopic descriptions of dynamical decoupling.~\cite{Khodjasteh,mozgunov}

%In this paper, we present our search for numerical signatures of fault tolerant excited states (FTES) of qubits coupled to the %environment.
In most situations, the bath causes a damping in a system that it has contact with, which leads to finite lifetimes of the system excited states. Our central claim about the fault tolerant excited state (FTES) for a two-qubit system is summarized as follows: by adjusting {\it both} the detuning of the qubits {\it and} the asymmetry in their environmental coupling, an excited two-qubit state can be decoupled from the dissipative effect of the bath. In the case of equilibrium and zero temperature, the qubits are permanently in that pure and excited state. Any fluctuation of the state is recovered by the action of the bath, which means that the state is fault tolerant.

Additionally, the two lowest excited states of the two qubits are usually near two-fold degenerate, making it easy to generate a highly entangled state by a unitary transformation. This is provided here by the unitary environmental effect. The net result is that the bath mediated interaction places the qubits into a permanent, nearly maximally entangled, excited state.

We find the fault tolerant state not by carefully assembling the system-bath to get the desired result, but by sweeping the microscopic parameters to see if and when FTES emerges naturally on its own. The ab initio approach distinguishes our work from previous works.

The advancements made in this paper were made possible by recent developments of completely positive quantum master equations (MEs), namely the geometric-arithmetic master equation (GAME),~\cite{Davidovi__2020} similar to the universal Lindblad equation (ULE)~\cite{Nathan}. These two equations were developed independently and at the same time. While they have identical dissipative dynamics, they differ in the Lamb-shift. They both claim the same accuracy in the order for a system-environment coupling, equivalent to that of the Redfield master equation (RE). Indeed, a comparison of the solutions of ULE and GAME with the solutions of RE leads to nearly identical errors with respect to RE solutions,~\cite{Davidovi__2020} as well as errors comparable to the estimated accuracy of RE. The Lamb-shift in ULE is more self-consistent with the dissipator, but more difficult to calculate. The mathematical form of the Lamb-shift is simpler in GAME, making implementation faster. Rapid evaluation of the Lamb-shift will be paramount in extending FTES to many-qubit systems.

Throughout the paper, we use the following pairs of words interchangeably: heat bath and environment, and system and qubits. The outline of the paper is as follows. We begin by introducing the system-environment Hamiltonian in Sec.~\ref{secMES}.
The dynamics of the reduced qubit system is derived from the Hamitonian in Sec.~\ref{sec:MES}, in which we discuss the four
calculation methods we implemented. GAME and its logic is discussed more succinctly relative to Ref.~\cite{Davidovi__2020}. The Redfield and time-convolutionless master equation (TCL4) are also reviewed. The section is
concluded by discussing an exact method for solving reduced dynamics, time evolved matrix product operators (TEMPO),~\cite{Strathearn2018} which confirms and
extends some of the findings beyond the Markov approximation.
Section~\ref{sec:stabilizer} presents numerical simulations that study the heat bath's role as a stabilizer of qubit excited states and the critical role of the Lamb-shift. Scaling properties of FTES and the error recovery times are studied in Sec.~\ref{sec:scaling}. We conclude the paper by discussion of what can be done next in Sec.~\ref{discussion}.

\section{\label{secMES} Microscopic Model}

Two qubits are immersed in a common heat-bath and are described by the spin-boson
Hamiltonian
\begin{eqnarray}
 %to remove numbering (before each equation)
&& H_T = H_S+A\otimes B+\sum_\lambda\omega_\lambda b_\lambda^\dagger b_\lambda \\
&& B=\sum_\lambda g_\lambda (b_\lambda+b_\lambda^\dagger)\\
\label{eq:hc}
&&H_S =\frac{\Delta}{2}[\sigma_{z}^1+(1-\xi)\sigma_{z}^2]\label{eq:Hs} +H_c \\
&&A=\frac{1}{2}[(1+\eta)\sigma_{x}^1+
\sigma_{x}^2]-\frac{\zeta}{4}(\sigma_{y}^1\sigma_{z}^2-\sigma_{z}^1\sigma_{y}^2)\label{eq:A}.
\end{eqnarray}
Here, $b_\lambda$ and $b_\lambda^\dagger$ are boson annihilation and creation operators, respectively, acting in the Fock-space of the heat bath, $\sigma_{x,y,z}^{1,2}$ are the Pauli matrices with superscripts indicating the qubit number, and $\Delta$ is the drive frequency. Parameters $\xi$ and $\eta,\zeta$ are detunings in the qubit Hamiltonian and the couplings to the bath, respectively.

We study integrable heat baths of linear harmonic oscillators with spectral density
\begin{eqnarray}
\nonumber
J(\omega)&=&\pi\sum\limits_\lambda g_\lambda^2\delta(\omega-\omega_\lambda)\\
&=&\frac{\pi}{2}\alpha\omega_c\left(\frac{\omega}{\omega_c}\right)^s
\Theta(\omega)e^{-\frac{\omega}{\omega_c}},
\label{eq:spdensities}
\end{eqnarray}
where $\alpha$ is the dimensionless coupling constant, $\Theta(\omega)$ is the Heaviside step function
and $\omega_c$ is the bath cut-off frequency. Throughout the paper, we work at zero temperature ($T=0$K) and with the bath frequency cutoff $\omega_c=10\Delta$. Here, $s=1$ and $s=3$ for an Ohmic and super-Ohmic heat-bath, respectively.

$H_c$ in Eq.~\ref{eq:hc} is a contribution of the heat bath to the qubit Hamiltonian known as the counterterm,~\cite{CALDEIRA1983374}
\begin{equation}
H_c=\frac{1}{2}\alpha\omega_c\Gamma(s)A^2.
\label{eq:counterterm}
\end{equation}
The counterterm is perhaps most easily understood in the simple case in which both
the system and environment are linear harmonic oscillators
with coordinate-coordinate coupling. The counterterm in that case arises naturally after transformation of the bath Hamiltonian to the normal mode representation; see for example the Rubin model in Ref.~\cite{ulrich}
More generally, whether or not the counterterm should be included from the very beginning depends on
the system considered, as discussed in Ref~\cite{CALDEIRA1983374}.

Regardless of whether or not the counterterm is included, there is a second unitary contribution to the system dynamics from the bath, known as the Lamb-shift,~\cite{BreuerHeinz-Peter1961-2007TToO} that is distinct from the counterterm. The Lamb-shift has a linear dependence with the high frequency cut off $\omega_c$ of the bath.
Without the counterterm, the Lamb-shift would cause the reduced system dynamics to significantly depend on $\omega_c$ for $\omega_c\gg\Delta$, which is unphysical.
Adding the counterterm cancels this linear dependence,~\cite{Hartmann_2020} making the reduced system dynamics independent of
$\omega_c$. This cancellation is studied in Sec.~\ref{subsec:bir}. In contrast to the counterterm, the Lamb-shift can have higher order contributions in the power of $\alpha$, but those contributions do not depend on $\omega_c$ for $\omega_c\gg\Delta$. This issue is discussed in more detail in Appendix~\ref{appendix:tcl}.

\section{\label{sec:MES} Reduced Quantum Dynamics}

All equations that describe reduced qubit dynamics are derived from the Liouville equation for the density matrix of the total system consisting of the qubits and the heat bath,
\begin{equation}
\frac{d\rho_T}{dt}=-i[H_T,\rho_T].
\label{eq:rtotal}
\end{equation}
We determine the approximate reduced dynamics of two qubits using four techniques, in increasing order of accuracy with respect to exact dynamics, as described in the following four subsections.

\subsection{Redfield Master Equation~\label{sec:RED}}

After rotating to the interaction picture, e.g., $\varrho_T(t)=U_0(t)^\dagger\rho_T(t)U_0(t)$, where
$U_0=\exp[-i(H_S+\sum_\lambda\omega_\lambda b_\lambda^\dagger b_\lambda)t]$ is the free propagator,
Eq.~\ref{eq:rtotal} becomes
\begin{equation}
\frac{d\varrho_T}{dt}=-i[A(t)\otimes B(t),\varrho_T].
\end{equation}
Throughout the paper, symbols $\rho$ and $\varrho$ will represent the density matrix in the Schr\"odinger and interaction picture, respectively.

This equation is equivalent to integro-differential equation
\begin{eqnarray}
\nonumber
&&\frac{d\varrho_T}{dt}=-i[A(t)\otimes B(t),\varrho_T(0)]-\\
&&\int\limits_0^td\tau\big[A(t)\otimes B(t),[
A(\tau)\otimes B(\tau),\varrho_T(\tau)]\big],
\end{eqnarray}
where $\varrho_T(0)$ is the initial condition. We assume a factorized initial state, e.g., $\varrho_T(0)=\rho(0)\otimes\vert 0\rangle\langle 0\vert$,
where $\rho(0)$ is the initial qubit state and $\vert 0\rangle\langle 0\vert$ is the bath vacuum. Taking the partial trace over the bath, we arrive to the familiar expression for exact quantum dynamics for the reduced qubit state,
\begin{equation}
\frac{d\varrho}{dt}=
-\text{Tr}_B\int\limits_0^td\tau\big[A(t)\otimes B(t),[
A(\tau)\otimes B(\tau),\varrho_T(\tau)]\big].
\label{eq:integrodif}
\end{equation}

The Redfield master equation is obtained by applying the Born-Markov approximation, which approximates the reduced density matrix of the bath with the initial state $\vert 0\rangle\langle 0\vert$. In addition
the integral neglects the memory of the qubit reduced matrix. In other words, it replaces  $\varrho(\tau)$ with $\varrho(t)$. After the approximation, Eq.~\ref{eq:integrodif}
becomes
\begin{eqnarray}
\nonumber
\frac{d\varrho}{dt}&=&A(t)\varrho(t)\int\limits_0^td\tau C(\tau-t)A(\tau)\\
\nonumber
&-&A(t)\int\limits_0^td\tau C(t-\tau)A(\tau)\varrho(t)\\
\nonumber
&-&\varrho(t)\int\limits_0^td\tau C(\tau-t)A(\tau)A(t)\\
&+&\int\limits_0^td\tau C(t-\tau)A(\tau)\varrho(t)A(t).
\label{eq:ints}
\end{eqnarray}
Here $C(t-\tau)=\langle 0\vert B(t)B(\tau)\vert0\rangle$ is the bath correlation function (BCF), which satisfies  $C(t-\tau)=C(\tau-t)^\star$.
For the Ohmic and super-Ohmic baths, we obtain the BCF as the inverse Fourier transform of the SD given by Eq.~\ref{eq:spdensities},
\begin{equation}
C(t)=\int\limits_{-\infty}^\infty d\omega\frac{J(\omega)}{\pi}e^{-i\omega t}=\frac{\alpha\omega_c^2\Gamma(s+1)}{2(1+i\omega_ct)^{s+1}}.
\end{equation}

Equation~\ref{eq:ints} is next represented in the eigenbasis of $H_S$. After rotating back to the Schr\"odinger picture, e.g., $\rho(t)=e^{-iH_St}\varrho(t)e^{iH_St}$, it becomes
\begin{eqnarray}
\nonumber
\frac{d\rho}{dt}&=&-i[H_S,\rho]+A\rho \Lambda(t)^\dagger-\rho \Lambda(t)^\dagger A\\
&-&A\Lambda(t)\rho+\Lambda(t)\rho A.
\label{eq:RE2}
\end{eqnarray}
This is the time dependent Redfield equation. Here,
\begin{equation}
\Lambda(t)=\int_0^td\tau C(t-\tau)A(\tau-t)=A\circ \Gamma(t)^T,
\end{equation}
where the superscript $T$ indicates transposition
and  $\circ$ is the Schur, or Hadamard, product of matrices $A$ and $\Gamma^T$.
Thus, we get a matrix with elements $A_{ij}\Gamma(\omega_{ji},t)$. $E_i$ are the eigenenergies
of $H_S$ and $\omega_{ij}=E_i-E_j$ are the Bohr frequencies. The function
\begin{equation}
\Gamma(\omega,t)=\int\limits_0^td\tau C(\tau)e^{i\omega\tau}
\end{equation}
will be referred to here as the timed spectral density.

For the Ohmic and super-Ohmic bath, timed spectral densities
can be determined analytically, respectively as
\begin{eqnarray}
\nonumber
\Gamma(\omega,t)=-i\frac{\alpha\omega_c}{2}\Bigg\{
1-\frac{e^{i\omega t}}{1+i\omega_ct}-\frac{\omega}{\omega_c}e^{-\frac{\omega}{\omega_c}}\\
\bigg[
ei(\frac{\omega}{\omega_c})-ei(\frac{\omega}{\omega_c}+i\omega t)-i\pi\Theta(-\frac{\omega}{\omega_c})
\bigg]
\Bigg\}
\label{eq:sdtOhmic}
\end{eqnarray}
and
\begin{eqnarray}
\nonumber
&&\Gamma(\omega,t)=-i\frac{\alpha\omega_c}{12}\bigg\{
\frac{\omega^2}{\omega_c^2}\bigg(1-\frac{e^{i\omega t}}{1+i\omega_ct}\bigg)\\
\nonumber
&&+\frac{\omega}{\omega_c}
\bigg(1-\frac{e^{i\omega t}}{(1+i\omega_ct)^2}\bigg)+2\bigg(1-\frac{e^{i\omega t}}{(1+i\omega_ct)^3}\bigg)
\\
&&-\frac{\omega^3}{\omega_c^3}e^{-\frac{\omega}{\omega_c}}\bigg[
ei(\frac{\omega}{\omega_c})-ei(\frac{\omega}{\omega_c}+i\omega t)-i\pi\Theta(-\frac{\omega}{\omega_c})\bigg]\bigg\},
\label{eq:sdtSuperOhmic}
\end{eqnarray}
where $ei(x)=\int_{-\infty}^x\frac{e^t}{t}dt$.

\subsubsection{Markov Limit}

Master equation~\ref{eq:RE2} has time-dependent
coefficients. However, these coefficients become time independent in the limit $t\gg 1/\omega_c$.
Then, if the asymptotic state is unique,
the initial time dependence of the coefficients will have no effect on the asymptotic state. This can be obtained
from the master equation with asymptotic coefficients or the Markov limit of the master equation.
At time $t\gg 1/\omega_c$, the timed spectral density
can be replaced with
\begin{equation}
 \lim_{t\to\infty} \Gamma(\omega,t)=J(\omega)+iS(\omega).
 \label{eq:RDlimit}
\end{equation}
$J(\omega)$ is given by Eq.~\ref{eq:spdensities} and $S(\omega)$ is a real function we call principle density, which can be obtained
either directly from Eq.~\ref{eq:RDlimit} or by taking the Kramers-Kronig transform of $J(\omega)$.

For Ohmic bath, we find
\begin{equation}
S(\omega)=-\frac{\alpha\omega_c}{2}\left[ 1-\frac{\omega}{\omega_c}e^{-\frac{\omega}{\omega_c}}ei\left(\frac{\omega}{\omega_c}\right)\right],
\label{eq:SOhm}
\end{equation}
while for the super-Ohmic one,
\begin{equation}
S(\omega)=-\frac{\alpha\omega_c}{2}\Big[2+\frac{\omega}{\omega_c}+(\frac{\omega}{\omega_c})^2
-(\frac{\omega}{\omega_c})^3e^{-\frac{\omega}{\omega_c}}ei(\frac{\omega}{\omega_c})\Big].
\label{eq:SSOhm}
\end{equation}

In the Markov limit, the Redfield equation becomes
\begin{equation}
\frac{d\rho}{dt}=-i[H_S,\rho]+A\rho \Lambda^\dagger-\rho \Lambda^\dagger A - A\Lambda\rho+\Lambda\rho A,
\label{eq:RE}
\end{equation}
where $\Lambda=\lim_{t\to\infty}\Lambda(t)$.
In a more symmetric form, the equation can be written as
\begin{eqnarray}
\nonumber
\frac{d\rho}{dt}&=&-i[H_S+H_L,\rho]\\
&+&A\rho \Lambda^\dagger+\Lambda\rho A-\frac{1}{2}\{A\Lambda+\Lambda^\dagger A,\rho\},
\label{eq:Redfield}
\end{eqnarray}
where
\begin{equation}
H_L=\frac{1}{2i}(A\Lambda-\Lambda^\dagger A)
\label{eq:LambShift}
\end{equation}
is the Lamb-shift, as we will see in the next section.

Matrix elements of Eq.~\ref{eq:Redfield} are
\begin{eqnarray}
\nonumber %to remove numbering (before each equation)
\frac{d\rho_{nm}}{dt} &=&  -i[H_S+H_L,\rho]_{nm}+\sum\limits_{ij}\Big[\mathcal{G}_{ni,mj}\\
 &&-\frac{1}{2}\sum\limits_k\big(\delta_{in}\mathcal{G}_{km,kj}+\mathcal{G}_{ki,kn}\delta_{jm}\big)\Big]\rho_{ij},
 \label{Eq:kernelized}
 \end{eqnarray}
where
\begin{equation}
\mathcal{G}_{ni,mj}=A_{ni} A_{mj}^\star\{\Gamma(\omega_{in})+[\Gamma(\omega_{jm})]^\star\}.
\label{eq:GRE}
\end{equation}

In the basis $I_{ij}=\vert i\rangle\langle j\vert$ on the Banach space of linear operators acting in the qubit Hilbert space, we rewrite Eq.~\ref{Eq:kernelized} as
\begin{eqnarray}
\nonumber
\frac{d\rho}{dt}&=&\mathcal{R}\rho=-i[H_S+H_L,\rho]\\
 &&+\frac{1}{2}\sum\limits_{ijnm}\mathcal{G}_{ni,mj}\Big(
[I_{ni}\rho, I_{mj}^\dagger] + [I_{ni},\rho I_{mj}^\dagger]\Big).
 \label{Eq:LindGens}
\end{eqnarray}
Here, $\mathcal{R}$ is the generator of the one-parameter semigroup on the Banach space.

\subsection{Geometric Arithmetic Master Equation}

The generator $\mathcal{R}$ in Eq.~\ref{Eq:LindGens} is easily compared to the standard form of the generator of
reduced quantum dynamics, given by Eq. 38 in Ref.~\cite{AlickiBook}. The latter is the most general generator of a quantum dynamical contraction semigroup derived by Gorini, Kossakowski, and Sudarshan,~\cite{Gorini} and independently by Lindblad (GKSL).~\cite{lindblad1976}

It immediately follows that $H_S+H_L$ in the first line of Eq.~\ref{Eq:LindGens} maps to the effective qubit Hamiltonian, where $H_L$ is the unitary contribution of the bath to reduced qubit dynamics, or the Lamb-shift.
As for the second line in  Eq.~\ref{Eq:LindGens}, it will have
the GKSL form if and only if $\mathcal{G}$ is positive semidefinite.  Unfortunately, a simple exercise in linear algebra shows that for the Redfield  equation, $\mathcal{G}$ will be positive semidefinite if and only if the spectral density is flat. Thus, it is necessary to further approximate $\mathcal{G}$ with a positive semidefinite matrix, to obtain the quantum dynamical semigroup.

The GKSL equation was first rigorously derived from first principles by Davies~\cite{davies1974} in what is now known as the
rotating wave approximation (RWA).
Recently, there have been many strides in deriving GKSL equations from first principles, with a wider range
of applicability than the RWA.~\cite{chen2020hoqst} Those include coarse-graining,~\cite{Majenz,Schaller,mozgunov}, adiabatic master equations,~\cite{Albash_2012} partial RWA,~\cite{Vogt,Tscherbul},
and the quasi-phenomenological approach.~\cite{perlind} In the year of 2020, two papers were published claiming first principles derivation of the GKSL equation with accuracy equivalent to that of the Redfield ME. They are the Universal Lindblad Equation (ULE)~\cite{Nathan} and the  Geometric-Arithmetic Master Equation (GAME)~\cite{Davidovi__2020}, and they were independently developed.
While they have identical dissipators, they differ in the Lamb-shift.

In the derivation of GAME,~\cite{Davidovi__2020} we made a very crude positive semidefinite approximation in Eq.~\ref{eq:GRE},
\begin{equation}\mathcal{G}_{ni,mj}\approx A_{ni} A_{mj}^\star 2\sqrt{J(\omega_{in})J(\omega_{jm})}=M_{ni}M_{mj}^\star,
\label{eq:sqrt}
\end{equation}
where $M_{ij}=A_{ij}\sqrt{2J(\omega_{ji})}$.
The RHS is the positive definite Kossakowski matrix for the GKSL master equation
\begin{equation}
\frac{d\rho}{dt}=-i[H_S+H_L,\rho]+M\rho M^\dagger-\frac{1}{2}\{M^\dagger M,\rho\},
\label{eq:game}
\end{equation}
where $M$ is
\begin{equation}
M=A\circ\sqrt{2J^T},
\end{equation}
and $\sqrt{2J^T}$ is defined to be the matrix with elements $[\sqrt{2J^T}]_{ij}=\sqrt{2J(\omega_{ji})}$.
The explanation of the square-root approximation of spectral densities (Eq.~\ref{eq:sqrt}) is complicated, and is relegated to appendix~\ref{appendix:GAME}.

GAME and ULE have a pleasant feature. If we discard the Lamb-shift at $T=0$K, then the asymptotic state will be
the ground state of the qubit Hamiltonian $H_S$, which is natural. Thus, if the asymptotic state is found to be different from the ground state, this will be due to the unitary effect of the environment expressed through the Lamb-shift. In this context, the Lamb-shift
becomes critical for the existence of the FTES. (By contrast, at $T>0$K,
the asymptotic state is no longer the Gibbs state,~\cite{lee2020comment} and the above statement no longer holds.)

\subsection{Time-Convolutionless (TCL) Master Equation.~\label{sec:TCL}}

The TCL-master equation is an exact master equation obtained using the Nakajima-Zwanzig projection-operator technique.~\cite{BreuerHeinz-Peter1961-2007TToO}
The main assumption is that the time-evolution of the reduced density matrix of the system is invertible. In that case, it is possible to eliminate the integral over the history in the Nakajima-Zwanzig equation. It may be debatable if the assumption is applicable for the evaluation of the asymptotic state. Namely, since the asymptotic state results from an arbitrary initial state, it is clearly not invertible. The asymptotic state is still worth exploring, since the TCL
is more accurate than the Redfield equation, which suffers from the same problem.

The generator of the TCL-equation can be expressed as perturbation series in $\alpha$.~\cite{BreuerHeinz-Peter1961-2007TToO} In the lowest (linear) order of $\alpha$, the TCL equation is equivalent to Eq.~\ref{eq:RE2}. In the literature, the names RE and TCL2 are used interchangeably, but here we only use RE. In the second order of $\alpha$, TCL is named TCL4. Implementation of TCL4 is complicated, and the discussion is given in appendix~\ref{appendix:tcl}.

The TCL4 master equation reads as Eq.~\ref{eq:RE2} with an extra term,
\begin{eqnarray}
\nonumber
\frac{d\rho_{nm}}{dt}&=&\{-i[H_S,\rho]+A\rho \Lambda(t)^\dagger-\rho \Lambda(t)^\dagger A\\
&-&A\Lambda(t)\rho+\Lambda(t)\rho A\}_{nm}\\
&+&\sum\limits_{i,j}\delta\mathcal{G}_{ni,mj}(t)\rho_{ij}.
\label{eq:RE4}
\end{eqnarray}
Tensor $\delta\mathcal{G}_{ni,mj}(t)$ is quadratic with $\alpha$ and time dependent.
It is given by triple integrals of various
four-point bath correlation functions, which can be found in the appendix~\ref{appendix:tcl}.

Since our focus is to study the asymptotic state, we consider Eq.~\ref{eq:RE4} in the Markov limit
$t\to\infty$. We find $\delta\mathcal{G}_{ni,mj}(t)$ converges well for both Ohmic and Super-Ohmic SDs. Calculating the tensor $\delta\mathcal{G}_{ni,mj}(\infty)$ is the hard part,
after which the numerical calculation of system dynamics and the asymptotic state can be determined as easily as with the RE and GAME master equations.

%From this tensor, the Lamb-shift contribution can be also identified in second order with $\alpha$,
%We can identify the unitary contribution or the Lamb shift, analogous to how we identified it in the RE, which leads to
%\begin{equation}
%H_L^{(2)}=\frac{\mathcal{F}^\dagger-\mathcal{F}}{2i},
%\end{equation}
%where $\mathcal{F}$ is obtained as discussed in the appendix~\ref{appendix:tcl}. In contrast to the Lamb-shift in Eq.~\ref{eq:LambShift},
%the second order contribution is independent of $\omega_c$ at $\omega_c\gg \vert\omega\vert$ and it doesn't need the compensating counterterm.

\subsection{Time Evolving Matrix Product Operators (TEMPO).}

The Born-Markov approximation used in the previous sections is usually valid if the coupling between the system and the heat bath is weak. If the coupling increases, at some point the correlations in the heat bath imparted by the system dynamics persist long enough to feed back into the system, altering the system dynamics. This feedback is nonlocal in time and can dramatically change the asymptotic state. In an extreme example, one qubit coupled to Ohmic bath displays the spin-boson quantum phase transition at the critical coupling to the bath.~\cite{Leggett,ulrich,HUR20082208} Even if the coupling is far below critical, the entanglement of the asymptotic state of the system, calculated using exact quantum dynamics, can differ significantly from that calculated by Markov master equations.~\cite{Hartmann_2020}

Since the phenomenon of FTES involves feedback from the bath, there is an immediate concern if the phenomenon is well represented by the Markov master equations. Thus, we
will investigate if the phenomenon persists in non-Markovian dynamics.
In the path-integral formulation of quantum mechanics, the effect of the environment of linear harmonic oscillators can be exactly accounted for by the Feynman-Vernon influence functional~\cite{FEYNMAN1963118}. Its implementation includes the exact quasi adiabatic path integral method (QUAPI),~\cite{Makri1,Makri2,THORWART2004333,Nalbach} where the central quantity is the augmented density tensor (ADT). This carries the probability distribution of possible histories and auto-correlations of the system. The density matrix is obtained by tracing over those histories. Another path integral method commonly used is Monte Carlo~\cite{Komnik}. More recently, tensor network methods have been utilized~\cite{Chin,Prior,Strathearn2018,Strathearn2019,Florian,Maria,Modi,Filippov} to simplify and extend the QUAPI method.

Other methods of exact dynamics include the hierarchical equations of motion,~\cite{Tanimura,Tanimura1,Tanimura2,ZhenHua,Cheng_2015}, the
multiconfiguration time-dependent Hartree (MCTDH) algorithm~\cite{MEYER199073,BECK20001}, and the multilayer formulation of the previous algorithm.~\cite{Haobin,Jie} The Nakajima-Zwanzig equation is also exact,~\cite{Nakajima,Zwanzig} but it is as difficult to solve as the Liouville equation.~\cite{BreuerHeinz-Peter1961-2007TToO}  The efficiency of the method has recently been improved using quantum trajectory based hierarchy of stochastic pure states,~\cite{Suess,Pan-Pan,Hartmann1} including investigation of entanglement between two spins mediated by the bath.~\cite{Hartmann_2020}

In this paper, we apply the time evolving matrix product operators (TEMPO)~\cite{Strathearn2018} technique. The main idea is to compress the ADT in the form of a
matrix product state~\cite{ORUS2014117,cirac2020matrix} using singular-value-decomposition (SVD), which can significantly extend the history time of the approximate path integral. The introduction to the method and the python code we use in most of this paper
are available in Ref.~\cite{Strathearn2018}.
The TEMPO algorithm can also be modified to calculate the process tensor~\cite{Jorgensen2019}. This alternative formulation has been used in both optimizing quantum control procedures~\cite{Fux2021} and in calculating exact bath dynamics.~\cite{Gribben2021}
An open source python package~\cite{TimeEvolvingMPO} improves the performance and includes new approaches.

\section{Heat-Bath as State Stabilizer~\label{sec:stabilizer}}

It is theoretically established that two qubits interacting solely via a common heat bath can exhibit permanent entanglement, depending on the system-bath coupling and bath parameters.~\cite{Braun,McCutcheon_2009,Benatti_2009,Kast_2014,Makri2013,Hartmann_2020}
The entanglement usually results from the unitary transformation that the heat bath applies on the qubit system.~\cite{Majenz,Benatti,Schaller,Rivas,Hartmann,Davidovi__2020}
However, asymptotic entanglement of qubits is typically weak, as dissipative dynamics drives the system close to the ground state, which is not easily entangled.

Our claim is that the unitary effect of the bath can, in addition, create fault tolerant excitations. In the context of Eq.~\ref{eq:game}, an excited state $\vert FT\rangle$ of the Hamiltonian $H_S+H_L$ will be fault tolerant if two conditions are met: 1) the state is annihilated by the generator, e.g., $M\vert FT\rangle =0$; 2) the asymptotic state of the master equation $\rho_\infty$ must be unique.

Under the first condition, $\vert FT\rangle\langle FT\vert$ is a solution of Eq.~\ref{eq:game}, while
under the second condition, the state can recover from errors. Let us assume that an error corrupts $\vert FT\rangle\langle FT\vert$, which can be due to an environment not accounted for by the model.
Since $\vert FT\rangle\langle FT\vert$ is both asymptotic and unique, the corrupted state will spontaneously recover under Eq.~\ref{eq:game}, assuming no additional error will take place during the recovery time.

In general, the asymptotic state need not be unique. For example,  noiseless quantum codes~\cite{Zanardi_1997} or
decoherence free subspaces (DFS)~\cite{Lidar_1998} have degenerate asymptotic states. A state in a DFS is annihilated by the generator. However, the purity of the asymptotic state is not guaranteed, because a convex sum of different asymptotic states is a mixed asymptotic state. %If an extrinsic noise source drives the state out of DFS, the state cannot be corrected by the action of the bath alone.
Noiseless quantum codes and DFS occur if there is a symmetry of the Hamiltonian with respect to
permutations between qubits. All such symmetries are broken in our model, guaranteeing the uniqueness of the asymptotic state.

\subsection{Bath Induced Renormalization~\label{subsec:bir}}

The unitary effect of the heat-bath transforms the two lowest excited states of two weakly detuned qubits into singlet-like and triplet-like superpositions. The singlet-like state exhibits suppressed damping because of the destructive interference
caused by the singlet superposition. The interference is at the heart of suppressing damping
to create a stable excited state. Thus, we begin by finding out
how the bath transforms quantum states of the qubit system.

\begin{figure}
\centering
\includegraphics[width=0.49\textwidth]{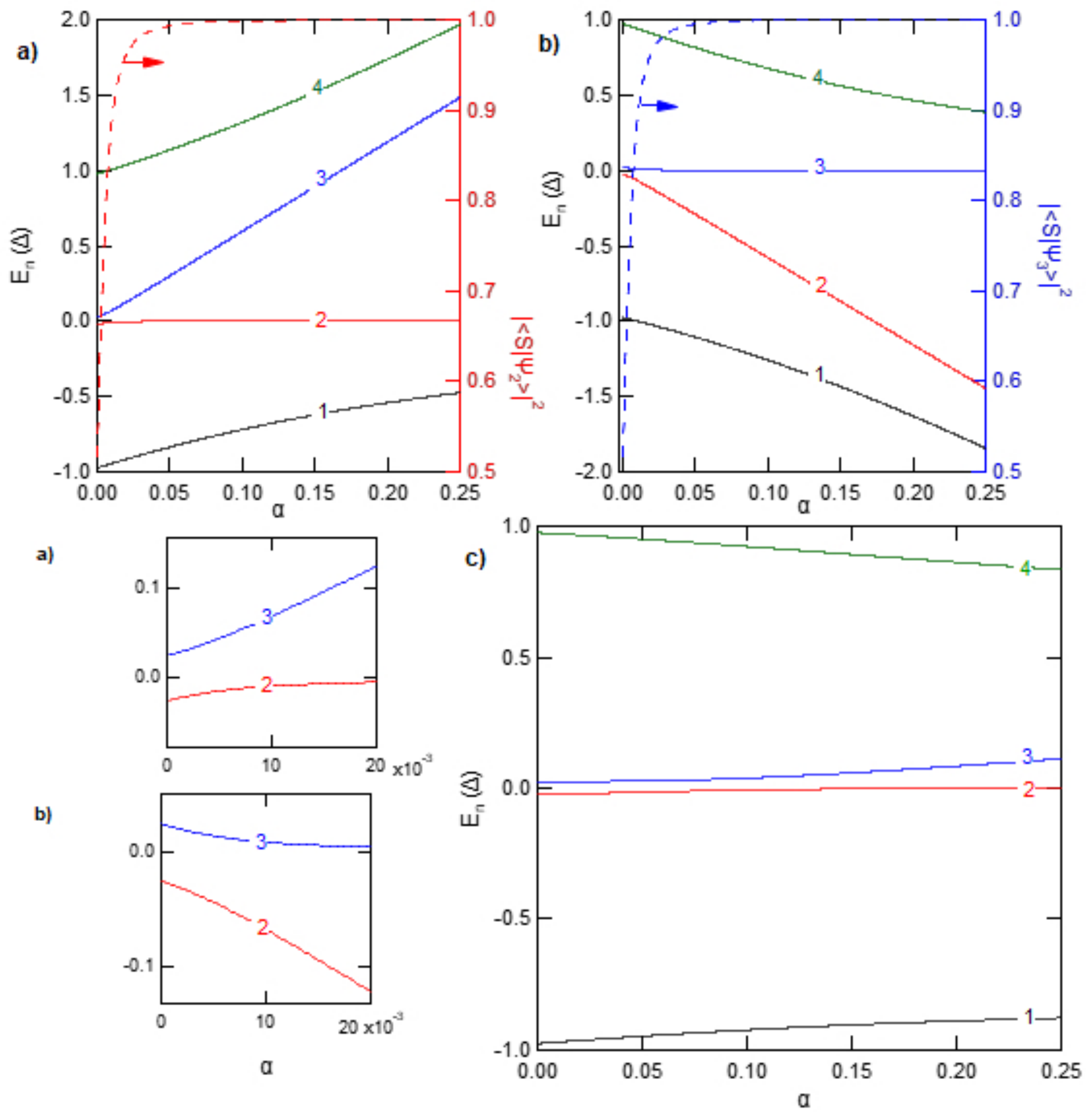}
  \caption{Renormalizations of the qubit energy levels by the Ohmic heat bath.
   {\bf a)} Counterterm only. {\bf b)} Lamb-shift only.  {\bf c)} Both the counterterm and the Lamb-shift.
    Full lines show 2-qubit system energy levels versus $\alpha$, on the left axes. Dashed-lines display singlet fidelity
    of the appropriate (e.g., singlet-like) excited state versus $\alpha$, on the right axis. Insets: zoom-in at avoided level-crossings. $\Delta=1$, $\xi=0.05$, $\eta=0.179$, $\zeta=0$, $\omega_c=10$, $s=1$.}\label{fig:crossings}
\end{figure}

In our recent work,~\cite{Davidovi__2020} we studied bath induced renormalization, or avoided crossing, between two weakly detuned excited levels in a three-level Jaynes-Cummings model. Excluding the counterterm, the avoided crossing energy is $2S(\Delta)$, where $\Delta$ is the drive frequency  of the model. The pair of weakly detuned excited levels is transformed into a singlet-like state $\vert S\rangle'\approx
\vert S\rangle =(\vert 1,0\rangle-\vert 0,1\rangle)/\sqrt{2}$ at higher energy, and a triplet like state $\vert T0\rangle'\approx
\vert T0\rangle=(\vert 1,0\rangle+\vert 0,1\rangle)/\sqrt{2}$ at lower energy.  The avoided crossing energy can be quite large for large $\omega_c$, suggesting that the counterterms should be considered.

Here we investigate two qubit system Hamiltonian renormalizations. If we take the limit $\omega/\omega_c \to 0$ in Eqs.~\ref{eq:SOhm}-\ref{eq:SSOhm},
then $A_f\approx -iS(0)A$, and the Lamb-shift in Eq.~\ref{eq:LambShift} becomes $S(0)A^2=-\alpha\omega_c\Gamma(s)A^2/2$, equal and opposite to the counterterm given by Eq.~\ref{eq:counterterm}. At finite $\omega/\omega_c$, there is incomplete cancellation of the terms, the effects of which have
recently been studied in Ref.~\cite{Hartmann_2020}
\begin{figure}
\centering
\includegraphics[width=0.49\textwidth]{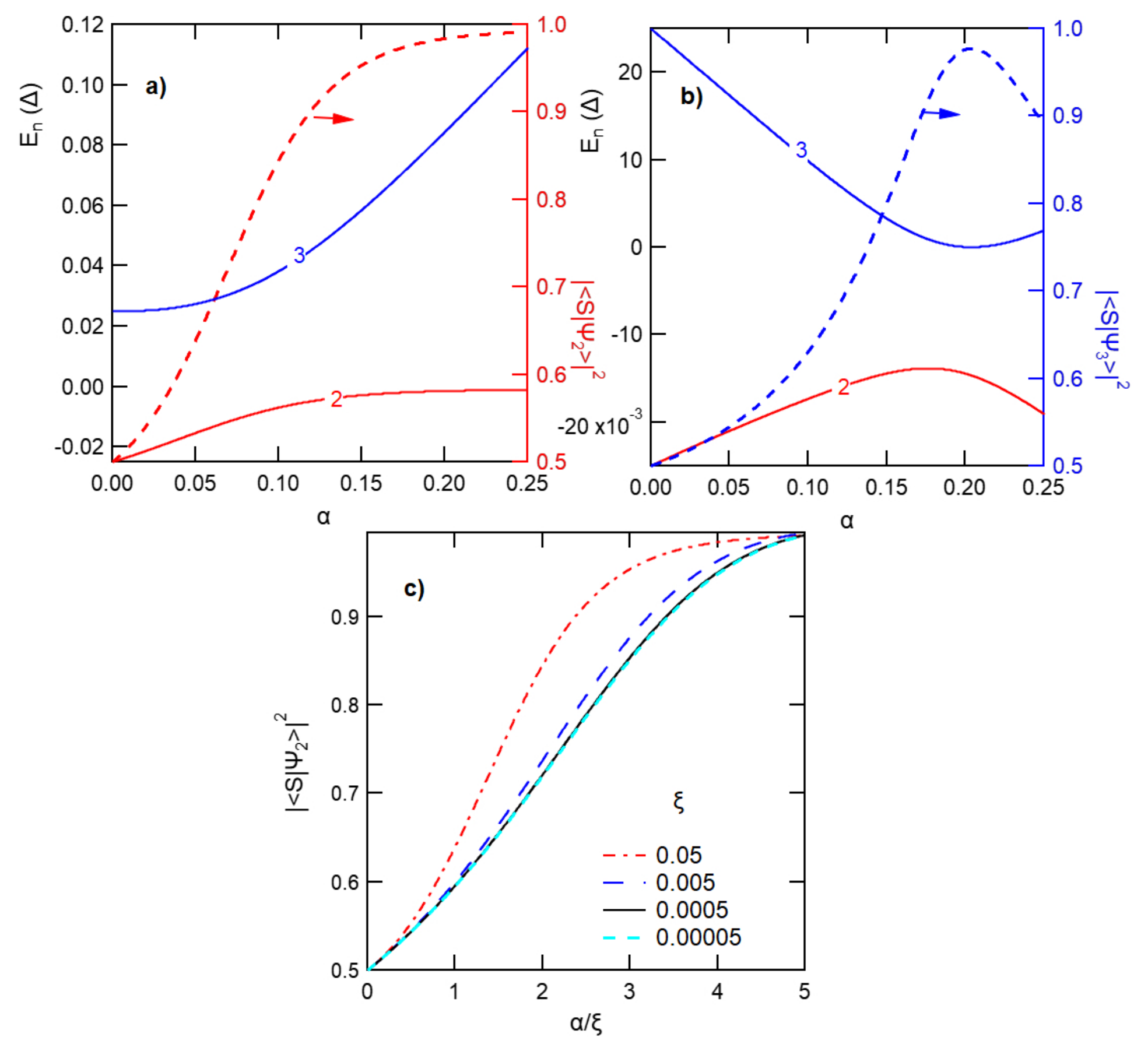}
  \caption{Renormalizations of the qubit energy levels by an Ohmic and a super-Ohmic heat bath.
  {\bf a)} $\Delta=1$, $\xi=0.05$, $\eta=0.179$, $\zeta=0$, $s=1$.
   {\bf b)} $\Delta=1$, $\xi=0.05$, $\eta=0.37$, $\zeta=0$, $s=3$.
   {\bf c)} Singlet fidelity versus scaled damping $\alpha/\xi$, at four values of $\xi$. $s=1$, $\eta = 0.179$, $\zeta= 0$, $s=1$. In all figures, $\omega_c=10$. }\label{fig:crossingsSO}
\end{figure}

Note that in our model,
the renormalized system Hamiltonian is always block-diagonal. In the computational basis, it is
\begin{equation}
\left(
\begin{array}{cccc}
a & 0 & 0 & b \\
0 & c & d & 0 \\
0 & d^\star & e & 0 \\
b^\star & 0 & 0 & f\\
\end{array}
\right).
\label{matrix}
\end{equation}
We focus first on the subspace spanned by the basis $\vert 01\rangle$ and $\vert 10\rangle$.
Without any detuning, the Hamiltonian in the subspace is explicitly given as
\begin{equation}
H_S+H_L=\bigg[q\Delta+\frac{S(\mathcal{E}_-)\mathcal{E}_+-S(\mathcal{E}_+)\mathcal{E}_-}{\Delta\sqrt{q^2+4}}\bigg]\vert T0\rangle\langle T0\vert
\label{eq:zerosplit}
\end{equation}
where $q=\Gamma(s)\alpha\omega_c/2\Delta$ and
\begin{equation}
\mathcal{E}_\pm=\frac{\Delta}{2}\left(q\pm\sqrt{q^2+4}\right)
\end{equation}
are the extremal eigenenergies of $H_c$.
In the singlet-triplet basis, the renormalized Hamiltonian is diagonal, with the singlet-triplet splitting depending on the delicate balance between the counterterm and Lamb-shift.

As an example, for $\alpha=0.2$ and $\omega_c=10\Delta$, we find a singlet-triplet splitting of $-0.067\Delta$ [$0.0108\Delta$] for the Ohmic [super-Ohmic] bath. In comparison, the norm of the counterterm for the Ohmic bath is half that of the super-Ohmic. The larger counterterm with a smaller level repulsion implies that the super-Ohmic bath has a much better cancellation of the counterterm and the Lamb-shift, which is consistent with the cancellation trends studied recently in Ref.~\cite{Hartmann_2020}

The regime at finite system and system-bath detunings is numerically examined.
We determine the eigenvalues and eigenstates
of Hamiltonians $H_S$, $H_S-H_c+H_L$, and $H_S+H_L$.
In Fig.~\ref{fig:crossings}, we plot the renormalized qubit levels versus damping $\alpha$.
In {\bf a)} and {\bf b)}, the renormalizations are only due to the counterterm and Lamb-shift, respectively. The levels shift asymmetrically with increasing $\alpha$, with approximately equal and opposite centers of gravity. The insets on the bottom left show repulsion between weakly detuned-levels with an avoided crossing at $\alpha\approx 0.005$. At larger $\alpha$, the singlet-like level does not shift with $\alpha$. The singlet fidelity,
indicated by the dashed line (right axis), ranges from one-half at small $\alpha$ to one at large $\alpha$.

In comparison, Fig.~\ref{fig:crossings}~{\bf c)} displays the qubit levels renormalized by both the counterterm and the Lamb-shift as a function of $\alpha$. The center of gravity of the spectrum remains close to zero, showing an overall cancellation of the two terms. Despite this, some weak shifts remain.

Figs.~\ref{fig:crossingsSO}~{\bf a)} and~{\bf b)}  display avoided crossings between weakly detuned levels versus damping for the  Ohmic and super-Ohmic bath, respectively, with both terms included. The minimum spacing between levels 2 and 3 is located at $\alpha=0.0673$ and $0.1908$ for~{\bf a)} and~{\bf b)}, respectively. The avoided crossing energy is approximately three times larger for the Ohmic bath, even though in this case the super-Ohmic bath has a 6.7 times larger counterterm norm. This again shows that the counterterm and the Lamb-shift cancel more effectively with large $s$.
Jumping ahead, stronger level repulsion for the Ohmic bath, relative to the super-Ohmic bath, will make it easier to prepare FTES and gives an advantage to slower baths.

If we plot the singlet fidelity of the singlet-like excited state versus scaled parameter $\alpha/\xi$, then the curves will coalesce
at very small $\alpha$, as shown in Fig.~\ref{fig:crossingsSO}~{\bf c)}.
Thus, in the weak coupling regime, singlet fidelity versus $\alpha$ and $\xi$ is a function of only one
variable, $\xi/\alpha$. We will encounter this scaling again and again.

\subsection{Suppression of Singlet Damping by System-Bath Detuning}

Throughout the paper, the initial condition will be the singlet state. Master equations are solved by decomposing the initial vectorized density matrix into a linear combination of the eigenvectors of the superoperator, after which the solutions can be found by exponentiation in a fraction of a second. TEMPO calculations are much slower and can take anywhere between several days to a month per curve, as detailed in appendix~\ref{appendix:TEMPO}.

Figure~\ref{fig:arrayCT}~[\ref{fig:array}] displays singlet fidelity $\langle S\vert\rho\vert S\rangle$ versus time, with~[without] the counterterm, at fixed qubit detuning $\xi$ and varied qubit-bath detuning $\eta$. %[Few TEMPO curves in Figs.~\ref{fig:arrayCT}~{\bf d)} and~\ref{fig:array}~{\bf d)} are incomplete, because the program crashed due to inability of the SVD to converge.]
Overall, different methods lead to qualitatively similar results.
They all exhibit a strong suppression of singlet damping as a function of a bath detuning $\eta$,  which is part of the main result of the paper.

Singlet fidelity decays incoherently in Fig.~\ref{fig:arrayCT}.
As $\eta$ increases in the range $[0,0.35]$, the renormalized energy spacing between singlet and triplet-like states changes, corresponding to the oscillation period in the interval $[60/\Delta,78/\Delta]$. The period is too large compared to the damping and decoherence time, explaining the incoherent singlet decay.
In the absence of the counterterm, the splitting is much larger [compare the spacing between levels 2 and 3 in Fig.~\ref{fig:crossings}~{\bf b)} versus \ref{fig:crossings}~{\bf c)}], leading to a much shorter oscillation period, in quantitative agreement with the  oscillations in Fig.~\ref{fig:array}.
\begin{figure}
\centering
\includegraphics[width=0.45\textwidth]{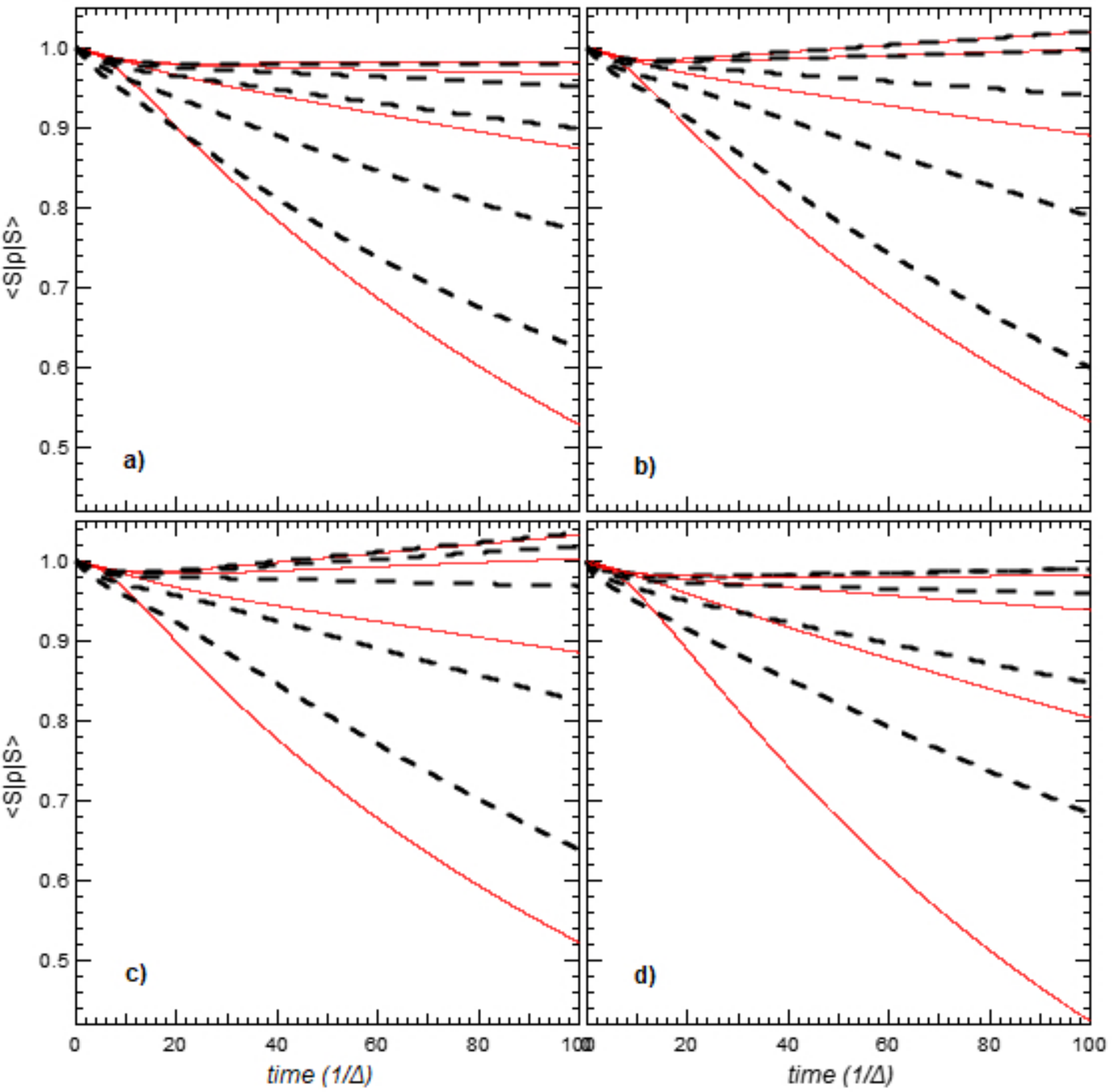}
  \caption{\label{fig:arrayCT}Singlet fidelity versus time for Ohmic heat bath including the counterterm effect. {\bf (a-d)} Numerical results obtained, respectively, using GAME, RE, TCL4, and TEMPO. Full red lines, bottom-to-top: $\eta=0,0.1,0.15,0.179$, respectively. Dashed black lines, top-to-bottom: $\eta=0.19,0.22,0.25,0.3,0.35$, respectively. Other
  parameters: $\Delta=1$, $\xi=0.05$, $\zeta=0.0437$, $\alpha=0.2$, $\omega_c=10$, and $\rho(0)=\vert S\rangle\langle S\vert$.
  TEMPO parameters: time step $\tau_s=0.02$, number of steps stored in the ADT $K_{max}=150$, $\epsilon_{SVD}=10^{-6}$.}
\end{figure}

Depending on the qubit-bath detuning $\eta$, Figs.~\ref{fig:arrayCT}~{\bf b-c)} and~\ref{fig:array}~{\bf b-c)} can display negative rates. That is, the singlet probability can increase with time and exceed one. We checked that this unphysical behavior is due to density matrix positivity violation, which is not that uncommon in the case of the RE and TCL4 master equations. The asymptotic state is not bound in the range of $\eta$ with negative rates.
In comparison, the TEMPO curves in Fig.~\ref{fig:arrayCT} have significant negative curvature when the singlet fidelity increases versus time, suggesting a bound asymptotic state. We also verified density matrix positivity violations in our TEMPO calculations.

TEMPO represents  exact quantum dynamics,~\cite{Fux2021} which must be completely positive. We lack the computational resources needed to maintain positivity of the density matrix in TEMPO. We vary the simulation parameters, such as memory cutoff time, precision of the SVD compression, the time step, and the boundary condition at memory cut-off, and find that positivity violations are very stubborn within reasonable calculation time-frames.

\begin{figure}
\centering
\includegraphics[width=0.45\textwidth]{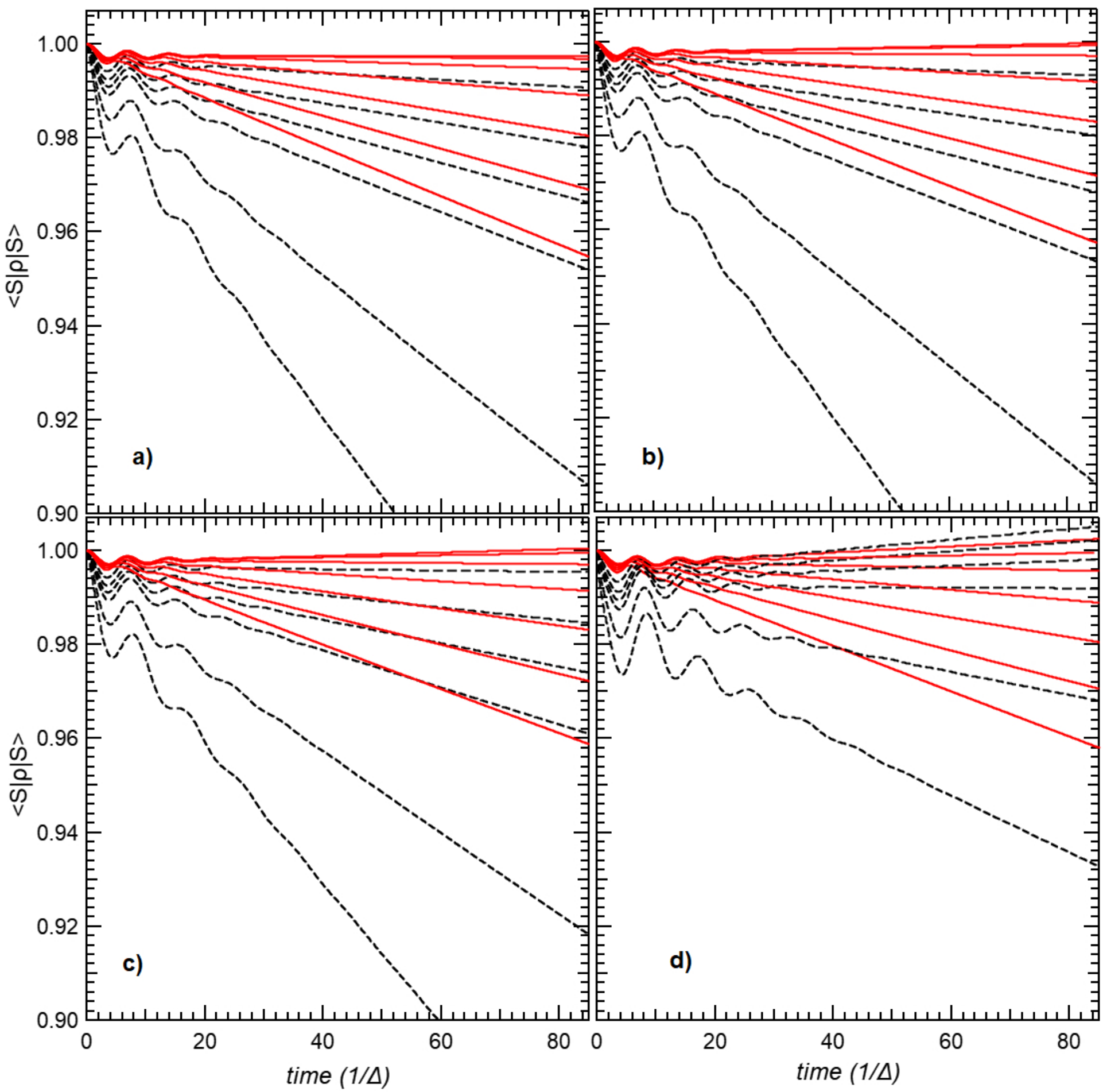}
  \caption{\label{fig:array}The same as Fig.~\ref{fig:arrayCT},  but without the counterterm.  {\bf (a-d)} Full red lines, bottom-to-top: $\eta=0.02,0,-0.02,-0.04,-0.06,-0.075,-0.09$, respectively. Dashed black lines, top-to-bottom: $\eta=-0.13,-0.16,-0.18,-0.2,-0.25,-0.3$, respectively. Other
  parameters: $\Delta=1$, $\xi=0.05$, $\zeta=0$, $\alpha=0.2$, $\omega_c=10$, and $\rho(0)=\vert S\rangle\langle S\vert$. TEMPO parameters: time step $\tau_s=0.02$, number of steps stored in the ADT $K_{max}=150$, compression precision $\epsilon_{SVD}=10^{-6}$.}
\end{figure}
One likely explanation of this problem is that the Ohmic bath has a slow decay of bath correlations. In fact, we find that the problem goes away if we consider the super-Ohmic bath.
With a smaller memory cut-off for the ADT, we still get a positive semidefinite density matrix, thereby making the simulation more reliable. We can even explore singlet decay in the strong coupling regime, where master equations fail and the dynamics is not Markovian, e.g., as follows.

\begin{figure}
\centering
\includegraphics[width=0.49\textwidth]{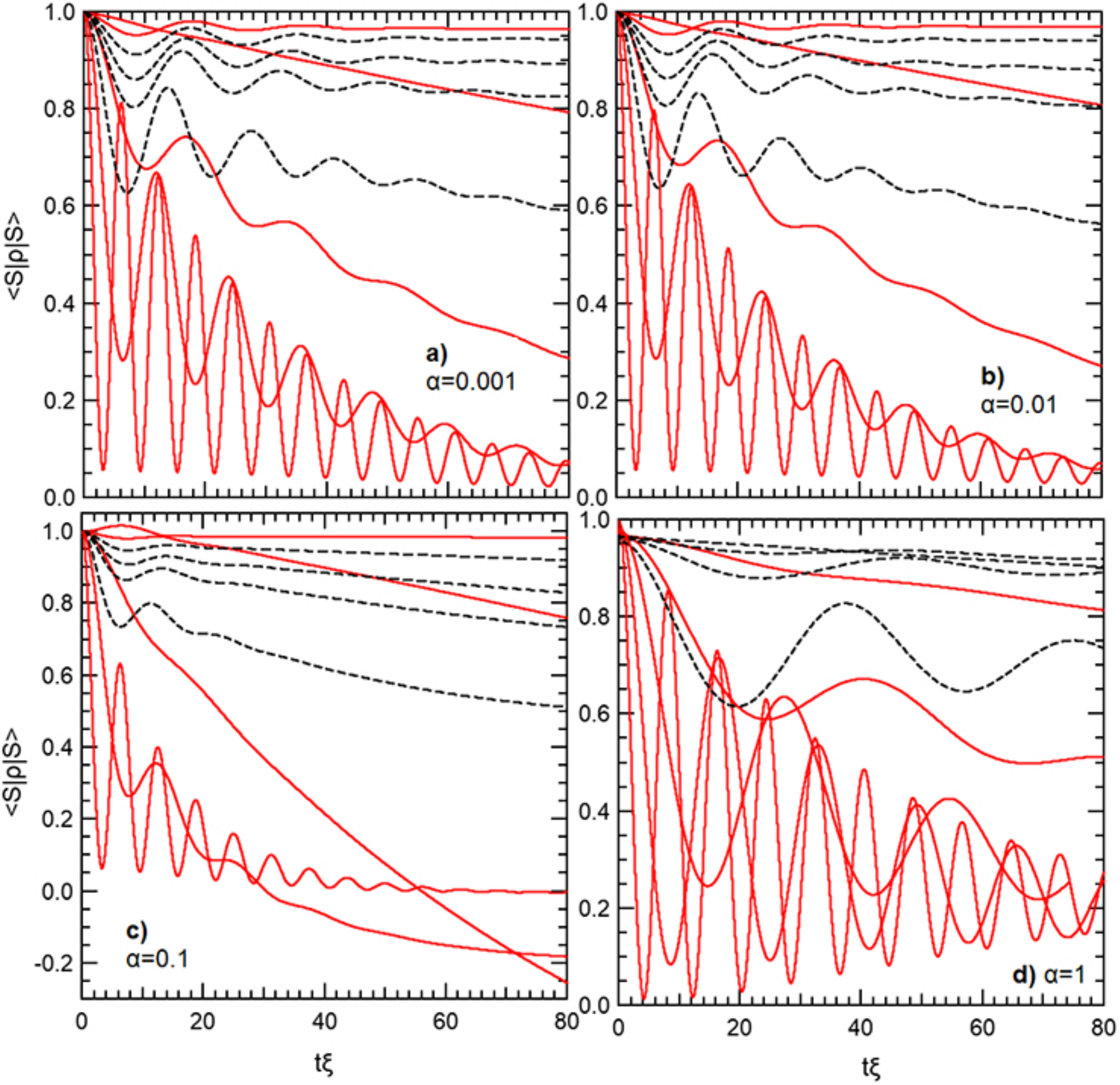}
  \caption{\label{fig:arraySO}Singlet decay for the super-Ohmic bath. {\bf (a-c)} show the RE results, for $\alpha=10^{-3},10^{-2},10^{-1}$, respectively. {\bf d)} shows the exact solution by TEMPO.
 Full red lines, bottom-to-top: $\eta=0,0.20,0.28,0.33,0.36$, respectively. Dashed black lines, top-to-bottom: $\eta=0.37,0.38,0.39,0.42$, respectively. Other parameters: $\alpha=1$. $\Delta=1$, $\xi/\alpha=0.2$, $\zeta=0$, $\omega_c=10$, and $\rho(0)=\vert S\rangle\langle S\vert$. TEMPO: $\tau_s=0.02$, $K_{max}=75$, $\epsilon_{SVD}=10^{-6}$.}
\end{figure}

Fig.~\ref{fig:arraySO}~{\bf a-d)} displays singlet fidelity versus scaled time $\alpha t$ for the super-Ohmic bath.
Between~{\bf a)} and~{\bf d)}, $\alpha$ increases by four orders of magnitude, while $\xi/\alpha=0.2$, and $\zeta$, $\Delta$, and $\omega_c$ are constant. In each sub-figure, $\eta$ changes within the same set of values.

The curves in Fig.~\ref{fig:arraySO}~{\bf a-c)} are obtained by solving the Redfield master equation. For $\alpha=0.001$, Fig.~\ref{fig:arraySO}~{\bf a)},
we also find the solutions of the GAME master equation, and obtain curves indistinguishable from those in the figure.
Since the RE and GAME master equations have equivalent accuracy that scales linearly with the coupling parameter $\alpha$,~\cite{Davidovi__2020}
this indistinguishability confirms correctness of both master equations. In addition, we determine  the solutions of the RE master equation at $\alpha=10^{-4}$,  and also find curves indistinguishable from those in~{\bf a)}.  This implies scaling of qubit dynamics in the weak coupling limit,
similar to the invariance in the singlet fidelity in Fig.~\ref{fig:crossingsSO}~{\bf c)}.

Comparing Figs.~\ref{fig:arraySO}~{\bf a)} and~{\bf b)}, we see that the curves are slightly different. They no longer scale, which is attributed to the non-negligible coupling to the bath in~{\bf b)}. With another order of magnitude increase in damping,  Fig.~\ref{fig:arraySO}~{\bf c)} displays unphysical singlet fidelity, larger than one or less than zero.
In this regime, the weak-coupling approximation to reduced quantum dynamics is failing. With further increase in orders of magnitude of $\alpha$ not shown, the
RE master equation is very unstable.

On the other hand, the TEMPO simulation is stable at $\alpha=1$ and exhibits a positive definite density matrix and singlet damping as shown in Fig.~\ref{fig:arraySO}~{\bf d)}. The damping curves in the weak and strong coupling regimes are overall very similar.
For example, the singlet damping rate is strongly suppressed near the same value of $\eta$ in~{\bf a)} and~{\bf d)} ($\eta\approx 0.37$).
At $\eta=0$, the oscillation periods measured in the damping curves are $6.11/\xi$ and $8.08/\xi$  in ~{\bf a)} and~{\bf d)}, respectively. We compare those periods with the detuning of the renormalized qubit Hamiltonian, and obtain the respective numbers of $6.12/\xi$ and $7.32/\xi$. The discrepancy between the frequencies in the strong coupling regime will be the subject of future work.
Regardless, the main conclusion of this section is that the strong suppression versus bath detuning in singlet damping remains valid in the strong coupling regime.

We conclude that the suppression of the singlet damping rate is overall weakly dependent on the coupling strength $\alpha$ if $\xi/\alpha$ is constant. It behooves that the suppression of damping of the singlet-like excited state persists in the strong-coupling regime. Phenomenological theory of quantum error correction is based on the operator-sum representation of quantum
operations,~\cite{Nielsen} which is correct regardless of the strength of the coupling.

\subsection{Cancelation of "Singlet" Damping}

Consistency of the solutions between the four different methods is encouraging. Out of these methods, only the geometric-arithmetic master equation has an easily calculable and positive semidefinite asymptotic density matrix. We thus focus on the GAME method.

\begin{figure}
\centering
\includegraphics[width=0.49\textwidth]{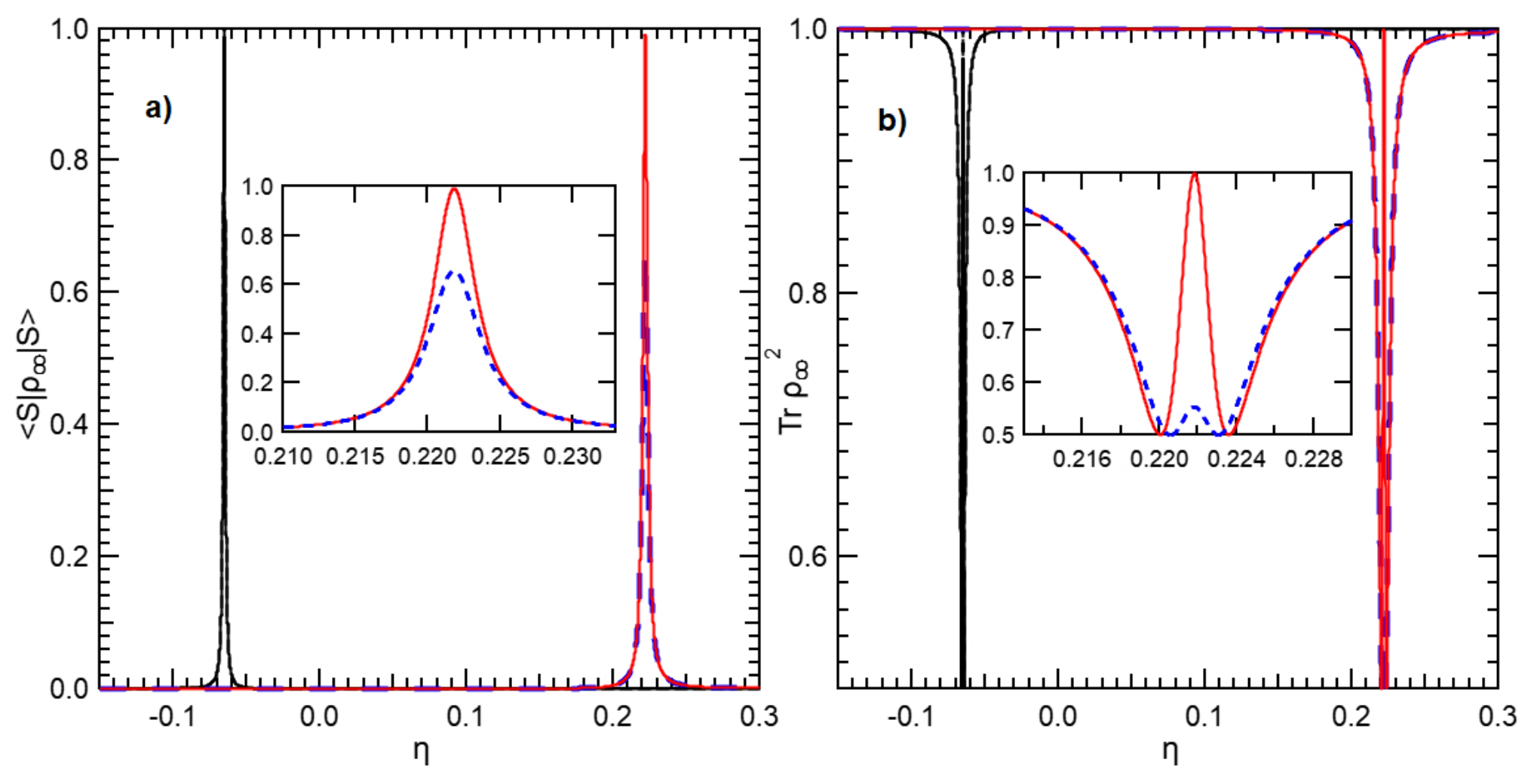}
  \caption{{\bf a)} and {\bf b)}: Singlet fidelity $\langle S\vert \rho_\infty\vert S\rangle$ and purity $Tr\,\rho_\infty^2$ of the asymptotic state, respectively, for  $\alpha=0.004$.
Full and barely visible dashed black line: $\zeta=7.5\times 10^{-5}$ and $\zeta=0$, respectively, with no counterterm. Full red and dashed blue line: $\zeta=1.368\times 10^{-3}$ and $\zeta=0$, respectively, with counterterm included. Insets: Zoom-in peak fidelity and purity, with counterterm included. $\xi=0.001$, $s=1$, $\omega_c=10$.\label{Fig:peaks1}}
\end{figure}

 The asymptotic state is determined as the eigenstate of the superoperator with eigenvalue zero.
 Fig.~\ref{Fig:peaks1} displays the singlet fidelity and purity of the asymptotic state as a function of bath detuning $\eta$ at qubit detuning $\xi=0.001$ and damping $\alpha=0.004$. Dashed lines are calculated using $\zeta=0$.
The black curves with peaks at negative $\eta$ correspond to the system Hamiltonian
without the counterterm, and vice versa for the red and blue curves with peaks at positive $\eta$. The insets zoom-in
the peaks with the counterterm.

One observes from the insets in Fig.~\ref{Fig:peaks1} that if $\zeta=0$, the maximum singlet fidelity and state purity are only $0.66$ and $0.55$, respectively.
So, we optimize both $\eta$ and $\zeta$ to maximize the asymptotic state purity. This can be done efficiently using standard optimization methods.
The full black ($\zeta=7.5\times 10^{-5}$) and red ($\zeta=-1.36\times 10^{-3}$) lines show singlet fidelity and state purity versus $\eta$ at the optimum $\zeta$. The optimization significantly increases the maximum singlet fidelity and purity if the counterterm is included.
The asymptotic state purity reaches one, within the precision of at least $10^{-8}$, limited by the numerical accuracy
of the optimizer. We thus consider the asymptotic state to be pure. We verify that the asymptotic state is an excited eigenstate of the qubit Hamiltonian.

\begin{figure}
\centering
\includegraphics[width=0.49\textwidth]{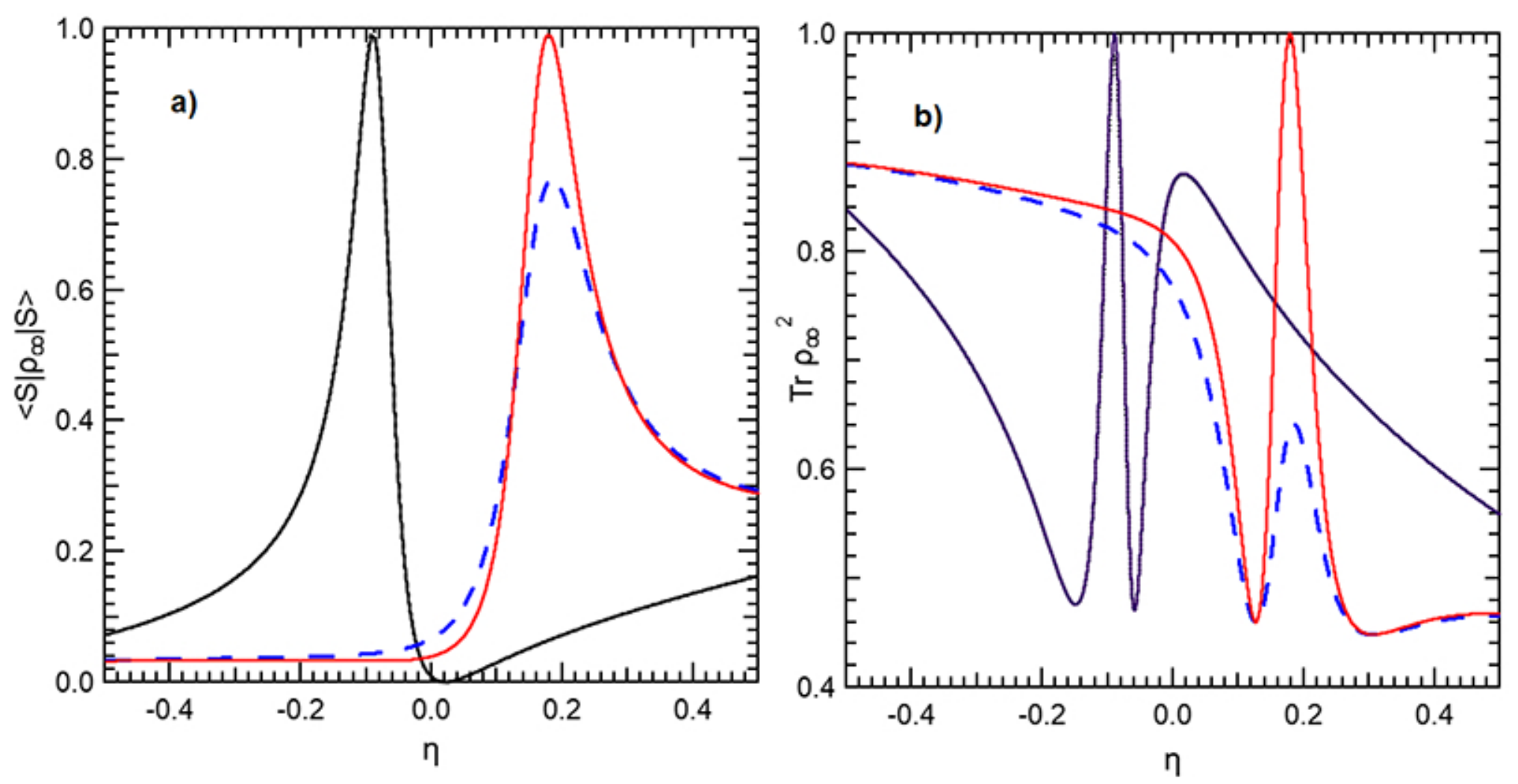}
  \caption{{\bf a)} and {\bf b)}: Singlet fidelity $\langle S\vert \rho_\infty\vert S\rangle$ and purity $Tr\,\rho_\infty^2$ of the asymptotic state, respectively, for $\alpha=0.2$.
Full and barely visible dashed black line: $\zeta=3.6\times 10^{-3}$ and $\zeta=0$, respectively, with no counterterm. Full red and dashed blue line: $\zeta=0.0437$ and $\zeta=0$, respectively, with counterterm included. $\xi=0.05$, $s=1$, $\omega_c=10$.\label{Fig:peaks2}}
\end{figure}
For a weak system-bath coupling, the FTES occurs within a narrow range of system-environment parameters. This range increases with increasing damping and detuning, as shown by the broader peaks in Fig.~\ref{Fig:peaks2}~{\bf a)} and~{\bf b)}. At the maximum, the purity is again larger than $1-10^{-8}$, and the asymptotic state is
the singlet-like excited eigenstate of the qubit Hamiltonian.

\subsection{FTES Recovery}

In QEC, the faster an unknown quantum state recovers from an error transient, the higher the state fidelity. Similarly, one figure of merit of FTES is the recovery rate, which can be obtained from the smallest magnitude nonzero eigenvalue of the superoperator of the quantum dynamical semigroup.

%While the eigenvectors of the superoperators versus system parameters can be discontinuous, which poses problems finding the asymptotic states of Re2 and RE4 as in Fig.~\ref{fig:arrayCT} and~\ref{fig:array}, the eigenvalues are always continuous. Thus, we can compare the recovery rates determined using GAME, RE2, RE4, and TEMPO methods, all on the same graph. (By contrast, it is hard to compare asymptotic states when some are diverging.)

\begin{figure}
\centering
\includegraphics[width=0.45\textwidth]{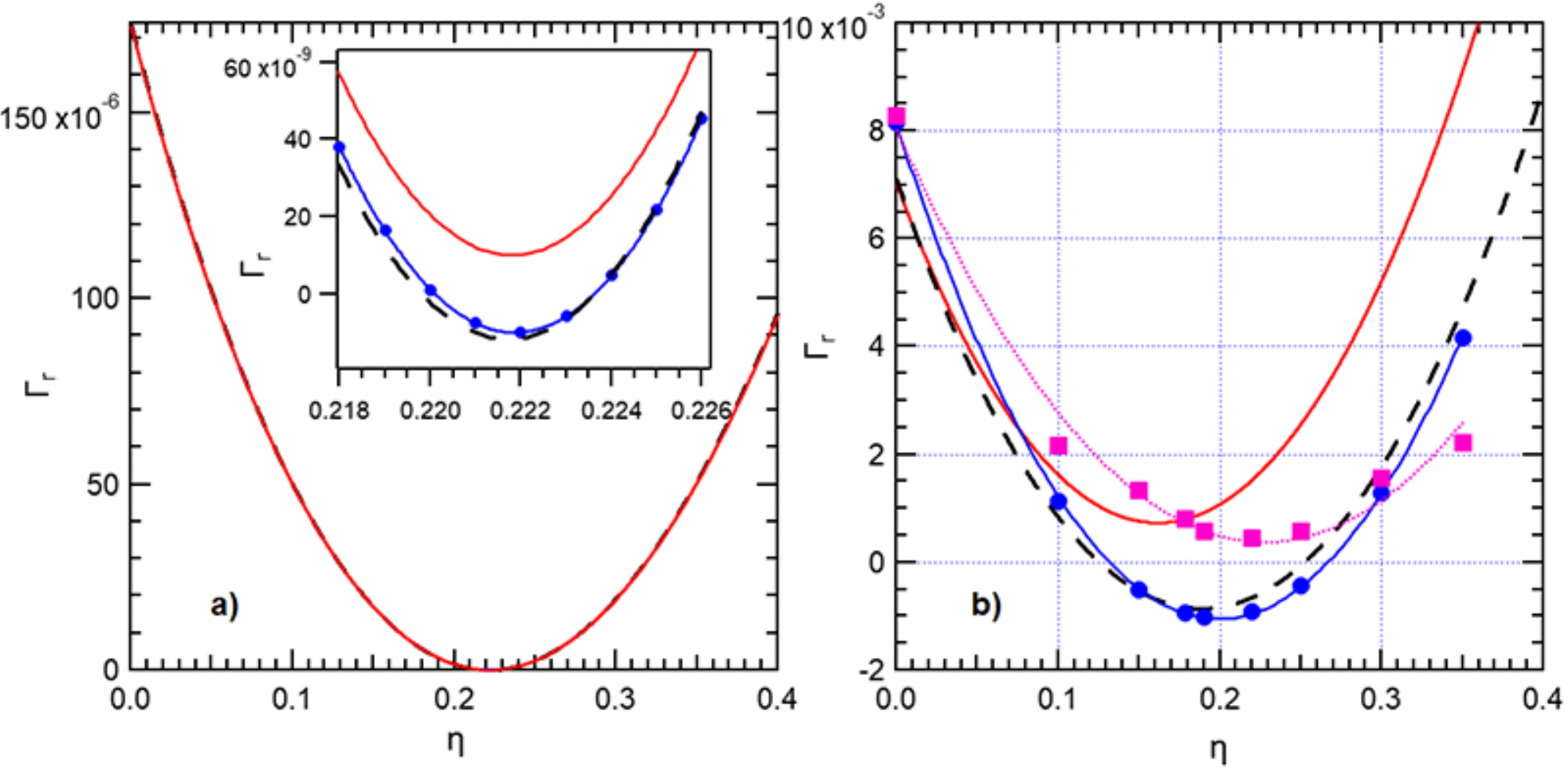}
  \caption{FTES recovery rate $\Gamma_r$.
  {\bf a)} $\alpha=0.004$. Full-red line:  GAME-values. Thick dashed black line: RE values. Inset: Zoom in at minimum rate. Blue circles and line: TCL4 values and best parabolic fit. $\zeta=1.36\times 10^{-3}$, $\xi=0.001$.
  {\bf b)} $\alpha=0.2$. Full-red line: GAME-values. Thick dashed black line: RE values. Blue circles and line: TCL4 values and best parabolic fit. Pink squares and line: TEMPO-values and best parabolic fit.
 $\zeta=0.0437$, $\xi=0.05$.
All $\Gamma_r$ are in units of drive frequency $\Delta$. $s=1$, $\omega_c=10$, counterterm included.\label{fig:recovery}}
\end{figure}

Figure~\ref{fig:recovery}~{\bf a)} displays the smallest relaxation rate ($\Gamma_r$) versus bath detuning $\eta$ at $\alpha=0.004$.
Looking at the larger range of $\eta$, there is virtually no difference between GAME and RE curves. However, the inset zooms in at the minimum and shows differences. Critically, the relaxation rates of RE and TCL4 become negative, and the asymptotic state diverges in the neighborhood of the FTES identified by GAME. In the latter approximation, $\Gamma_r$ is positive, guaranteed by the contracting semigroup.~\cite{AlickiBook} Since RE and GAME have comparable accuracy with respect to exact quantum dynamics, we take the recovery rate of the FTES to be only an estimate, with an error comparable to the estimate.

In the case of stronger damping, Figure~\ref{fig:recovery}~{\bf b)} includes the recovery rate obtained from TEMPO simulations, where $\Gamma_r$ is obtained by fitting singlet fidelity versus time to the exponential, as discussed in appendix~\ref{appendix:TEMPO}. The magnitudes of the minima are comparable between the four methods, confirming the validity of the GAME recovery rate within the factor of order one. The ability to estimate the recovery rate easily makes GAME the preferred method to search for and study FTES.
%It remains an open question if $\Gamma_r$ is identically zero in exact quantum dynamics. If so, the recovery
%rate would need to be calculated for the required purity of the asymptotic state. But this is not necessary in context of the GAME, because
%$\Gamma_r$ is nonzero at fault tolerance point where the state purity is $1$.

\subsection{Wrapping it up}

\begin{figure}
\centering
\includegraphics[width=0.45\textwidth]{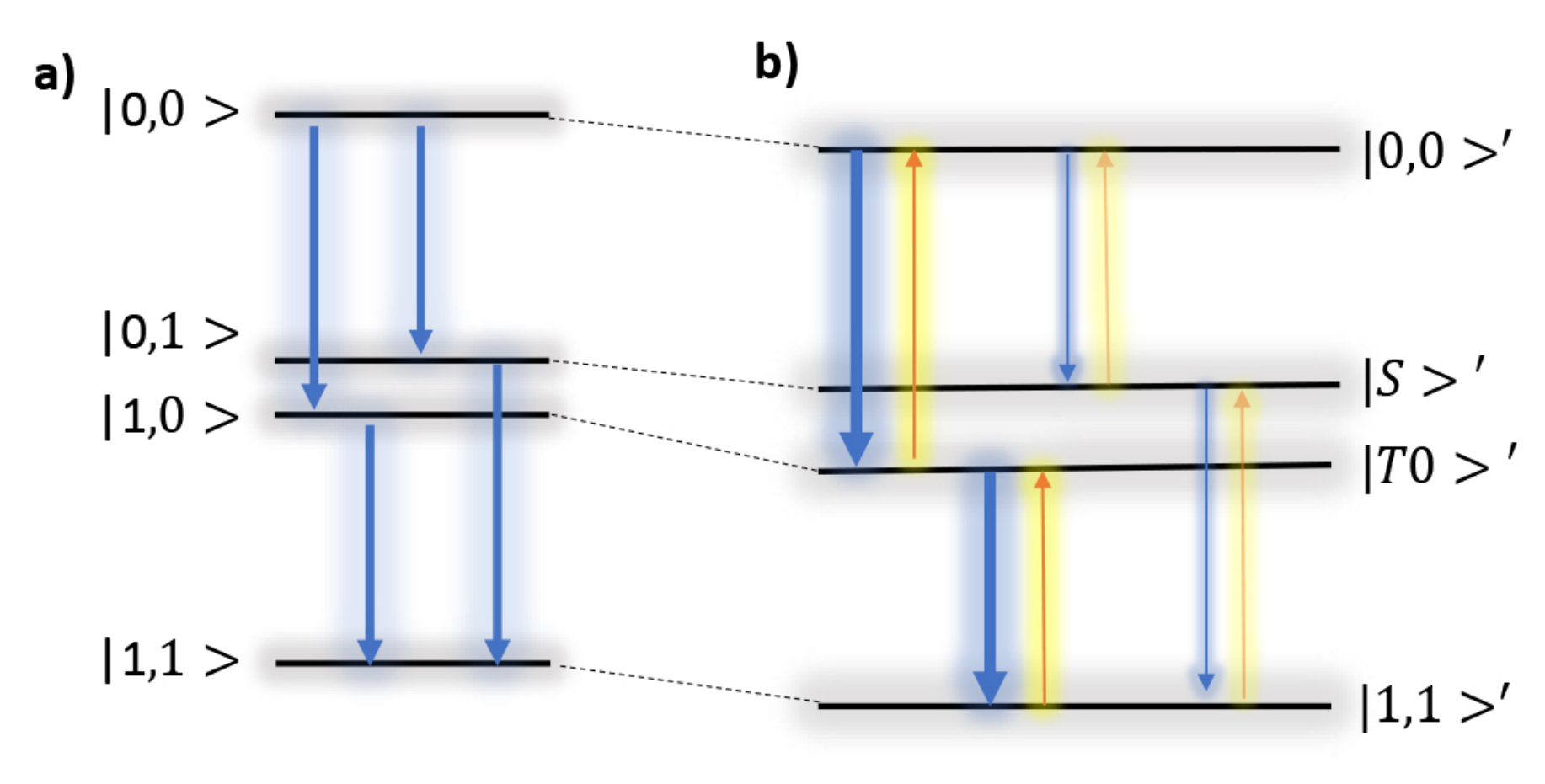}
  \caption{Dynamics of the FTES $\vert S\rangle'$. {\bf a)} Energy levels of a two qubit system in the absence of a unitary transformation by the heat-bath. The arrows represent the relaxation process due to a dissipative coupling to the bath at zero temperature. The rates are proportional to the thickness of the arrows and given by the Fermi-golden rule, where only energy-decreasing transitions are allowed. {\bf b)} Energy levels after the unitary transformation. The top-most and the lowest levels are weakly renormalized. The weakly detuned levels, $\vert 0,1\rangle$ and $\vert 1,0\rangle$, are strongly renormalized into singlet and triplet like states $\vert S\rangle'$ and $\vert T0\rangle'$. The triplet relaxation rate is enhanced, while that of the singlet is suppressed. Energy increasing processes are weakly allowed, due to the admixing.}\label{fig:FTdynamics}
\end{figure}

We proceed to qualitatively explain the dynamics of FTES.
The sketch in Fig.~\ref{fig:FTdynamics}~{\bf a)} displays four energy levels of two weakly detuned qubits, without taking into account any unitary effect induced by the heat bath.
We assume that the Fermi-golden rule relaxation rates (equal to $2J$ and indicated by the blue arrows) are much larger than the detuning $\xi\Delta$ between states $\vert 0,1\rangle$ and $\vert 1,0\rangle$, but much smaller than $\Delta$. In terms of the GKSL-equation~\ref{eq:game}, all the information about these rates is in the generator $M$.

In the presence of the unitary contribution from the bath,
levels $\vert 0,1\rangle$ and $\vert 1,0\rangle$ are strongly admixed, since their relaxation rate was assumed to be much larger than the detuning. On the other hand, the matrix~\ref{matrix} admixes levels $\vert 0,0\rangle$ and $\vert 1,1\rangle$ weakly, because their spacing is much larger than the relaxation rates.

We consider the GKSL equation in the
basis of the renormalized states, which are the eigenstates of $H_S+H_L$. The effect of the unitary transformation on the generator is
\begin{equation}
M=\left(
  \begin{array}{cccc}
    0 & 0 & 0 & 0 \\
    \bullet & 0 & 0 & 0 \\
    \bullet & 0 & 0 & 0 \\
    0 & \bullet & \bullet & 0 \\
  \end{array}
\right)
\to
\left(
  \begin{array}{cccc}
    0 & \bullet & \bullet & 0 \\
    \bullet & 0 & 0 & \bullet \\
    \bullet & 0 & 0 & \bullet \\
    0 & \bullet & \bullet & 0 \\
  \end{array}
\right),
\end{equation}
while that on the damping term is
\begin{equation}
M^\dagger M=\left(
  \begin{array}{cccc}
    \bullet & 0 & 0 & 0 \\
    0 & \bullet & \bullet & 0 \\
    0 & \bullet & \bullet & 0 \\
    0 & 0 & 0 & 0 \\
  \end{array}
\right)
\to
\left(
  \begin{array}{cccc}
    \bullet & 0 & 0 & \bullet \\
    0 & \bullet & \bullet & 0 \\
    0 & \bullet & \bullet & 0 \\
    \bullet & 0 & 0 & \bullet \\
  \end{array}
\right),
\label{eq:dampreno}
\end{equation}
where full circles indicate nonzero matrix elements.
The arrows in Fig.~\ref{fig:FTdynamics}~{\bf b)} sketch the relaxation rates in the renormalized basis: the triplet-like relaxation rate is approximately doubled, while the singlet-like one is strongly suppressed relative to the Fermi-golden rule, due to constructive and destructive interference, respectively.

Additionally, the rates of the reverse processes (yellow arrows) are nonzero (but weak). This can be seen, for example,
by the nonzero matrix element $(4,4)$
on the RHS of Eq.~\ref{eq:dampreno}, which indicates a nonzero loss rate of the renormalized ground state $\vert 1,1\rangle'$.

If we tweak the system-bath interaction Hamiltonian through parameters ($\eta,\zeta$) we can make the singlet relaxation rate be the bottleneck rate. As a result, the asymptotic state of the qubits will not be far from the singlet-like excitation. In the regime of fault tolerance, the bottleneck rate is precisely zero, and the asymptotic state is the pure "singlet" excitation. In that regime, the recovery rate is given by a nontrivial combination of various rates indicated in picture~\ref{fig:FTdynamics}~{\bf b)}, which we obtain by finding the smallest magnitude nonzero eigenvalue of the superoperator of the GAME master equation.

\section{Scaling Properties of the FTES~\label{sec:scaling}}

Here, we vary the system-environment parameters $\xi,\eta,\zeta,\alpha$ and $s$, and identify the parameter range that exhibits FTES, e.g.,  where the asymptotic state $\rho_\infty$ has a fidelity of one to an excited eigenstate of the renormalized qubit Hamiltonian. We also explore the parameter range where the FTES has high singlet-fidelity. All results in this section are obtained by finding the asymptotic solutions of the GAME master equation and maximizing its purity versus system parameters, so that $\text{Tr}\rho_\infty^2>1-10^{-8}$.

\subsection{The effects of qubit detuning on FTES.}

\begin{figure}
\centering
\includegraphics[width=0.45\textwidth]{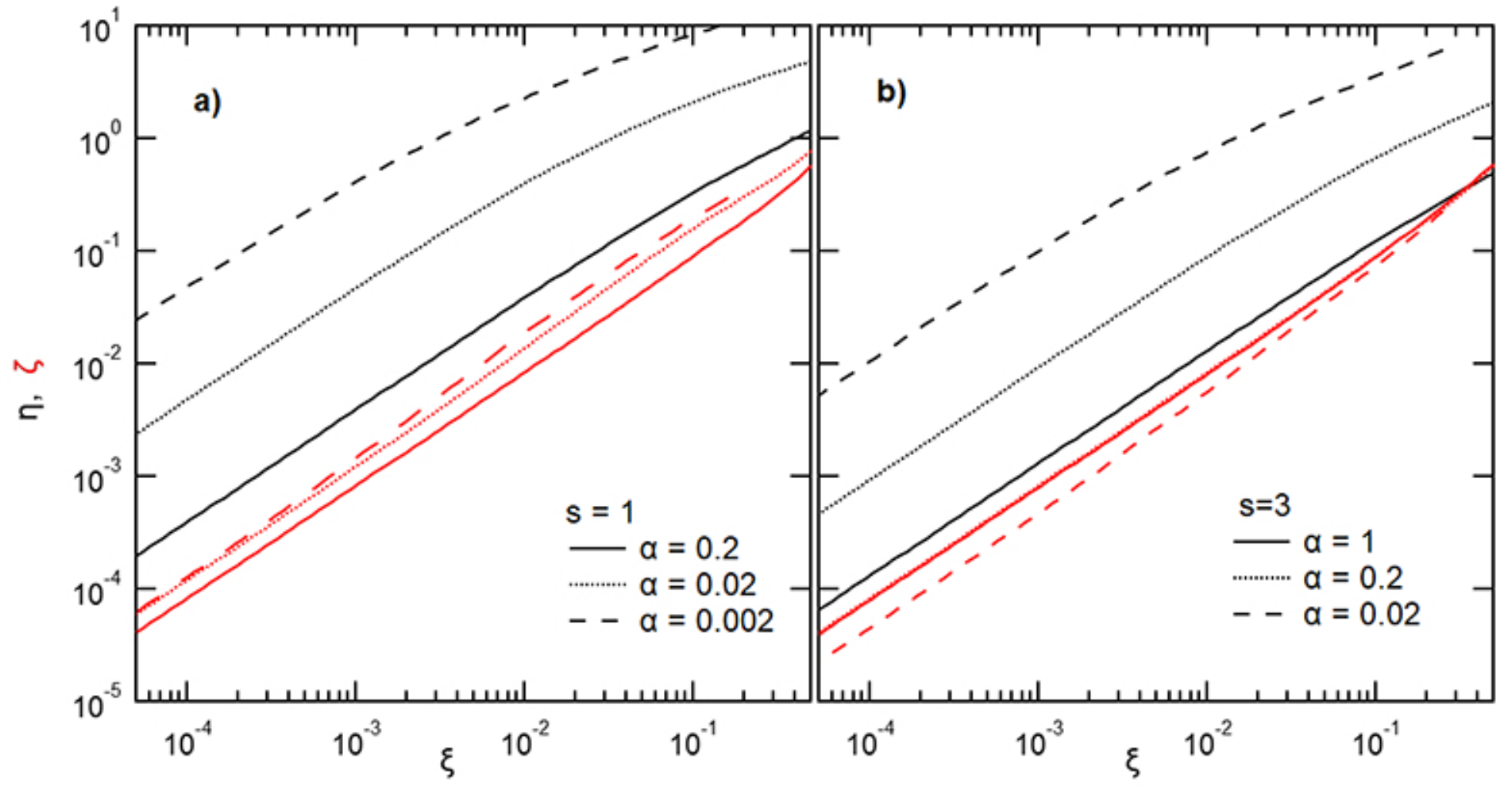}
  \caption{System-bath coupling parameters $\eta$ and $\zeta$ versus system detuning $\xi$, at the FTES, for Ohmic~{\bf a}) and super-Ohmic~{\bf b}) heat baths. The top three (black) curves show $\eta$. The bottom three (red) curves show $\zeta$. The dashes correspond to different dampings $\alpha$, as displayed in the legend. $\omega_c=10$.\label{fig:FTparams1}}
\end{figure}

Here, we vary the detuning $\xi$ at fixed values of $\alpha$, $s$, and $\omega_c$. For each $\xi$, we optimize $\eta$ and $\zeta$ to find the FTES, which leads to the unique value of  $\eta$ and $\zeta$.
The optimum $\eta$ and $\zeta$ versus $\xi$ are displayed in Figs.~\ref{fig:FTparams1}~{\bf a)} and~{\bf b)} for the Ohmic and super-Ohmic baths, respectively.
Note that here $\eta$ can be larger than $1$, which means there's strong asymmetry between the couplings of the qubits to the environment,
where the damping of one qubit is enhanced by $(1+\eta)$.

Let us first examine the regime of weak detuning, which we define as the range where $\xi<0.2$, $\eta<0.2$, and $\zeta<0.2$.
At small $\xi$, $\eta$ and $\zeta$ increase linearly with $\xi$, as shown in Figs.~\ref{fig:FTparams1}~{\bf a)} and~{\bf b)}. The slope  $\eta$ versus $\xi$ is proportional to $1/\alpha$. The property $\eta\approx\text{const}\, \xi/\alpha$ is consistent
with the analysis of the suppression of singlet damping in Fig.~\ref{fig:arraySO}.% where $\eta$ changed negligibly, while
%$\alpha$ and $\xi$ changed by three orders of magnitude with the constraint $\xi/\alpha=0.2$.

\begin{figure}
\centering
\includegraphics[width=0.45\textwidth]{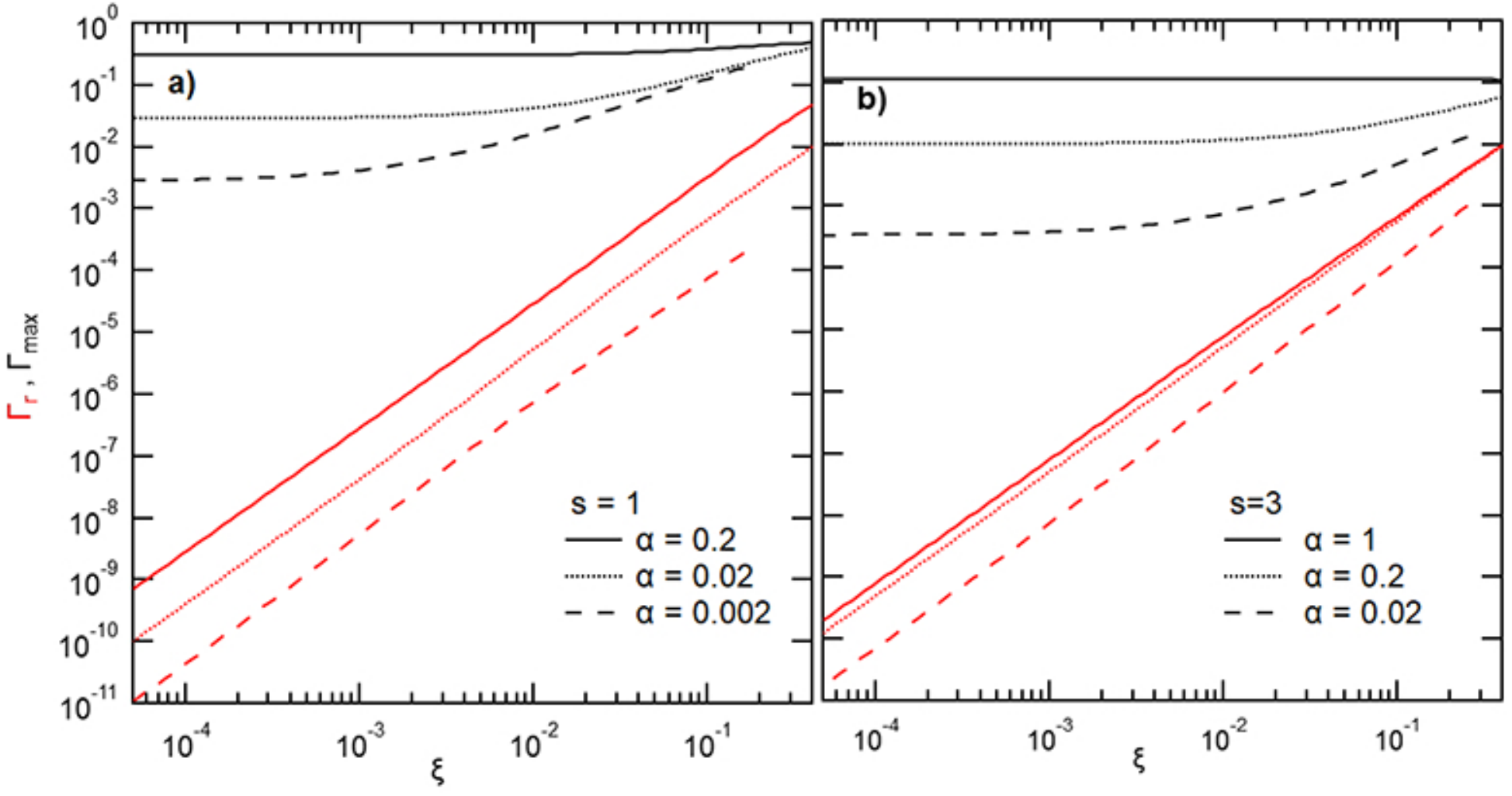}
  \caption{Maximum and minimum relaxation rate versus system detuning for Ohmic ({\bf a}) and super-Ohmic ({\bf b}) heat baths. The top three (black) curves show the maximum relaxation rate. The bottom three (red) curves show the recovery (e.g., minimum relaxation) rate. The dashes correspond to different dampings $\alpha$, as displayed in the legends. $\omega_c=10$ and all frequencies are in units of the drive frequency $\Delta$. \label{fig:FTrecovery1}}
\end{figure}The recovery rate and the maximum qubit relaxation rate are obtained by diagonalizing the superoperator, finding the eigenvalues with the largest and smallest real part, taking the real part, and multiplying by $-1$. The rates are displayed in Fig.~\ref{fig:FTrecovery1}.
The recovery rate for the Ohmic bath scales as $\Gamma_r\simeq\alpha\xi^2$ over the entire range in Fig.~\ref{fig:FTrecovery1}~{\bf a)}. By contrast, $\Gamma_{max}\simeq \alpha$ and is independent of $\xi$ at small $\xi$, with a crossover to a more complicated behavior at high $\xi$. The complicated behavior will be discussed in the next section. The quadratic scaling of $\Gamma_r$ with $\xi$ is the general property of decoherence-free proximity spaces.~\cite{Lidar_1998} Namely, if the permutational symmetry of the qubits is broken by a weak inhomogeneity (e.g., the detuning), the relaxation rate out of the decoherence free subspace scales quadratically with the inhomogeneity.

Note that the master equation is likely valid in most of the parameter range in Fig.~\ref{fig:FTrecovery1} since the maximum relaxation rate is at worst somewhat smaller than the drive $\Delta$. The highest recovery rate is found at the highest damping. For the Ohmic bath, $\Gamma_r$ is in the $10^{-2}\Delta$ range, which is a rapid error correction rate. Namely, if there is an extrinsic noise source that drives the qubit-state out of its asymptotic state at the rate $\Gamma_{error}$, then the state fidelity will be approximately $1-\exp (-\Gamma_r/\Gamma_{error})$. For example, if $\Gamma_r=0.01\Delta$ and  $\Gamma_{error}=10^{-3}\Delta$, then the FTES fidelity will be $>0.9999$.

\subsection{The effects of damping on FTES.}
\begin{figure}
\centering
\includegraphics[width=0.45\textwidth]{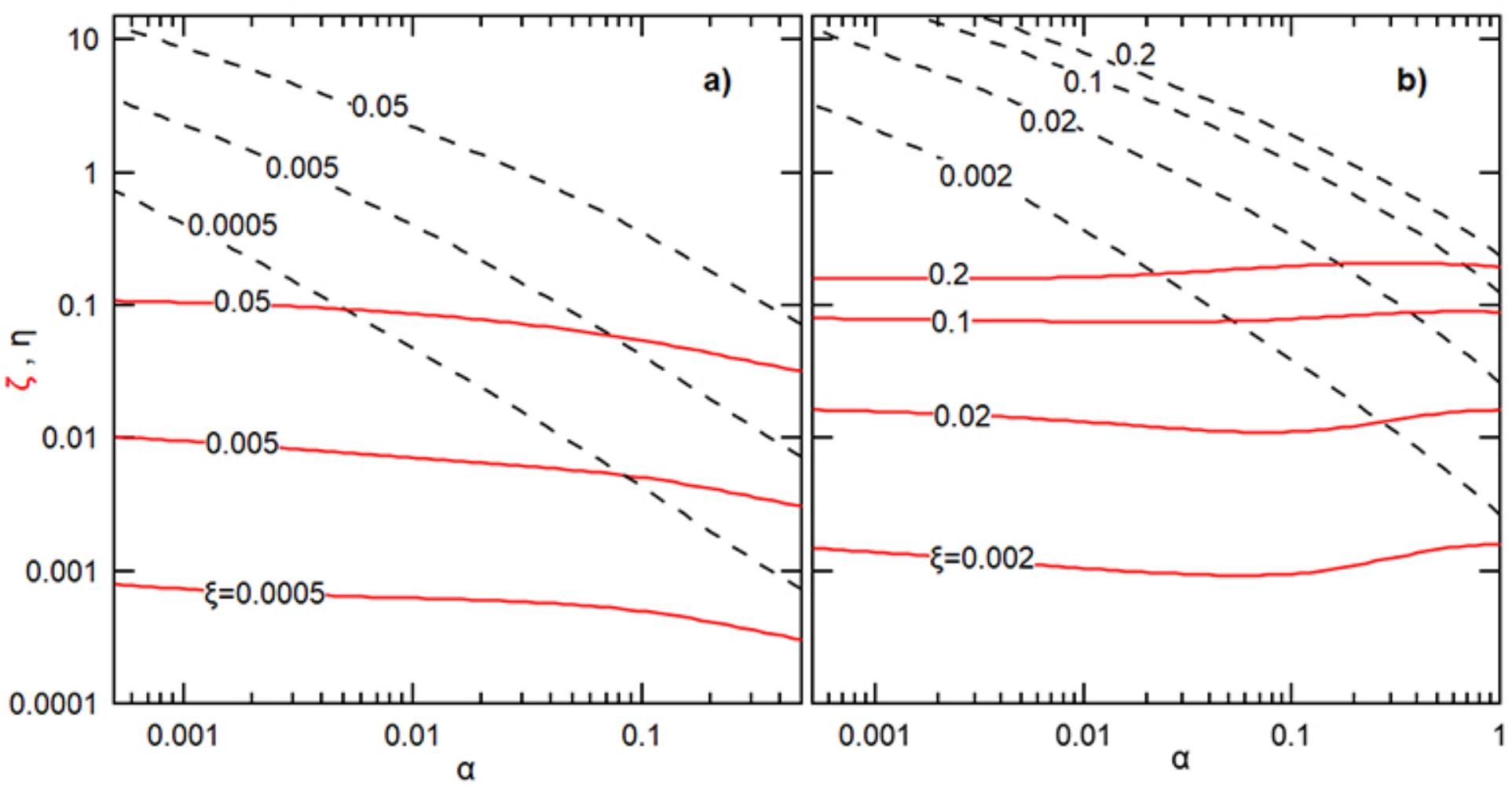}
  \caption{System-bath coupling parameters $\eta$ and $\zeta$ at fault tolerance versus damping $\alpha$ at fixed $\xi$, for Ohmic ({\bf a}) and super-Ohmic ({\bf b}) heat baths. The top three (black-dashed) curves show $\eta$. The bottom three (red-full) curves show $\zeta$. $\omega_c=10$.\label{fig:FTparams2}}
\end{figure}
\begin{figure}
\centering
\includegraphics[width=0.45\textwidth]{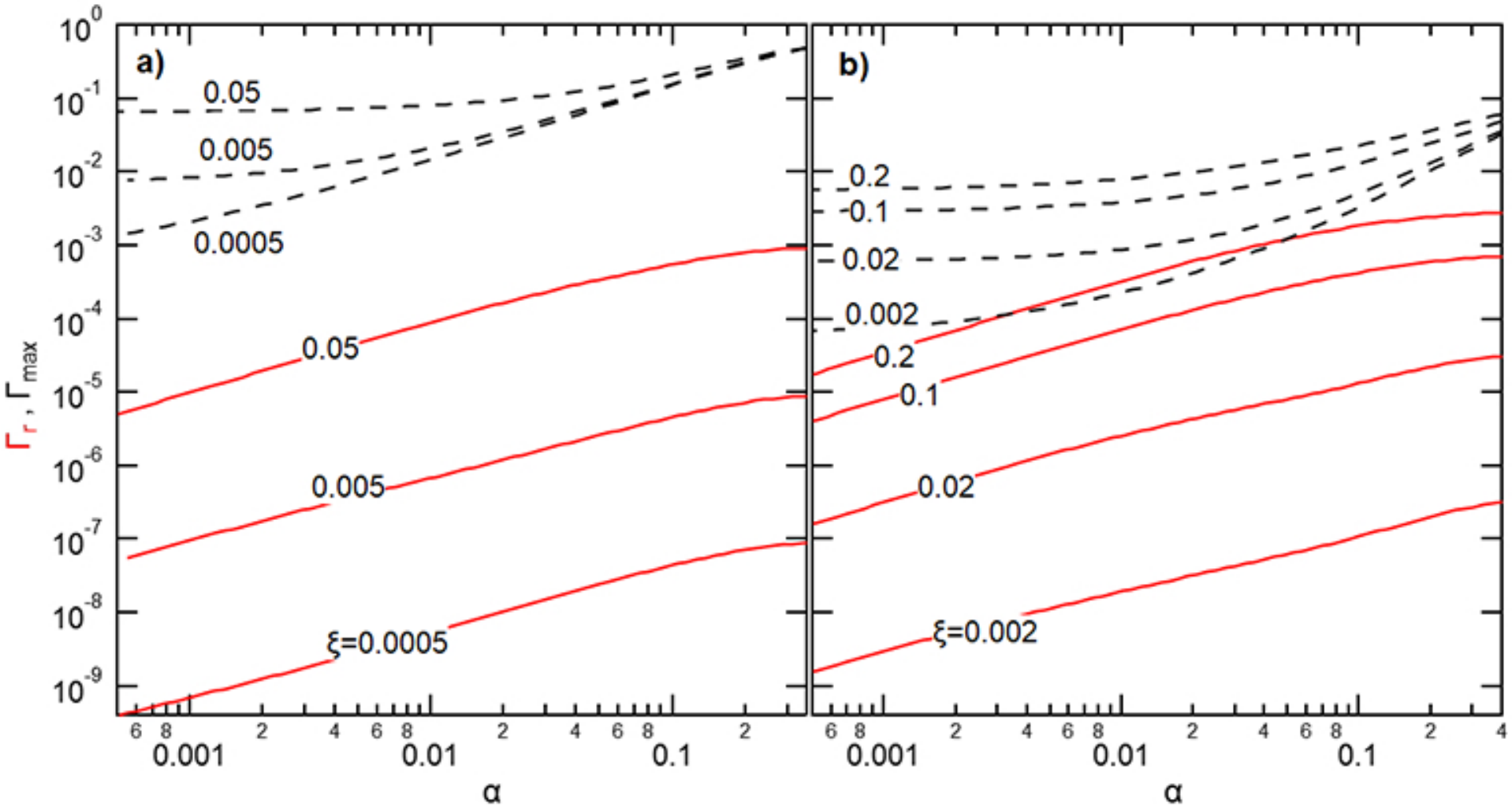}
  \caption{Maximum and minimum relaxation rate versus damping, for Ohmic ({\bf a}) and super-Ohmic ({\bf b}) heat baths. The top three (black-dashed) curves show maximum relaxation rate. The bottom three (red-full) curves show the recovery (e.g., the minimum relaxation) rate. $\omega_c=10$ and all frequencies in unit of drive frequency $\Delta$.\label{fig:FTrecovery2}}
\end{figure}

We obtain further insights by following the fault-tolerant excited state as a function of damping at a fixed  system detuning $\xi$. The bath coupling
parameters $\eta$ and $\zeta$ are obtained as explained in the previous section, and follow the curves in Fig.~\ref{fig:FTparams2}. The bath detuning $\eta$ exceeds one at small $\alpha$. As discussed in the previous section, FTES follows the constraint $\alpha\eta\simeq\xi$. So, for fixed $\xi$, $\eta$ increases as $1/\alpha$ and can become larger than 1.

The recovery and the maximum relaxation rate versus $\alpha$ are shown in Fig.~\ref{fig:FTrecovery2}. As in the previous section, $\Gamma_r\simeq\alpha\xi^2$, while $\Gamma_{max}$ has a crossover from $\simeq\alpha$ at large $\alpha$ to $\simeq$const at small $\alpha$.  The flattening of $\Gamma_{max}$ versus $\alpha$ at small $\alpha$ is also related to the constraint
$\eta\simeq\xi/\alpha$. If $\alpha$ is small, $\eta$ can be larger than one, and the damping rate of one qubit will be enhanced to $(1+\eta)\alpha\approx\eta\alpha\propto\xi$. This is constant in this figure, thereby explaining why the rate $\Gamma_{max}$ becomes flat at small $\alpha$. In comparison, the recovery rate is not affected by large $\eta$.
The following table summarizes the various scalings of the FTES.

\centerline{
\begin{tabular}{|c|c|c|}
  \hline
  & $\eta<1$ & $\eta>1$ \\
  \hline
  $\eta$ & $\xi/\alpha$ & $\xi/\alpha$ \\
  \hline
  $\Gamma_{max}$ & $\alpha$ & $\text{const}$ \\
  \hline
  $\Gamma_r$ & $\alpha\xi^2$ & $\alpha\xi^2$ \\
  \hline
\end{tabular}
}

\subsection{Singlet fidelity of the FTES.}

\begin{figure}
\centering
\includegraphics[width=0.25\textwidth]{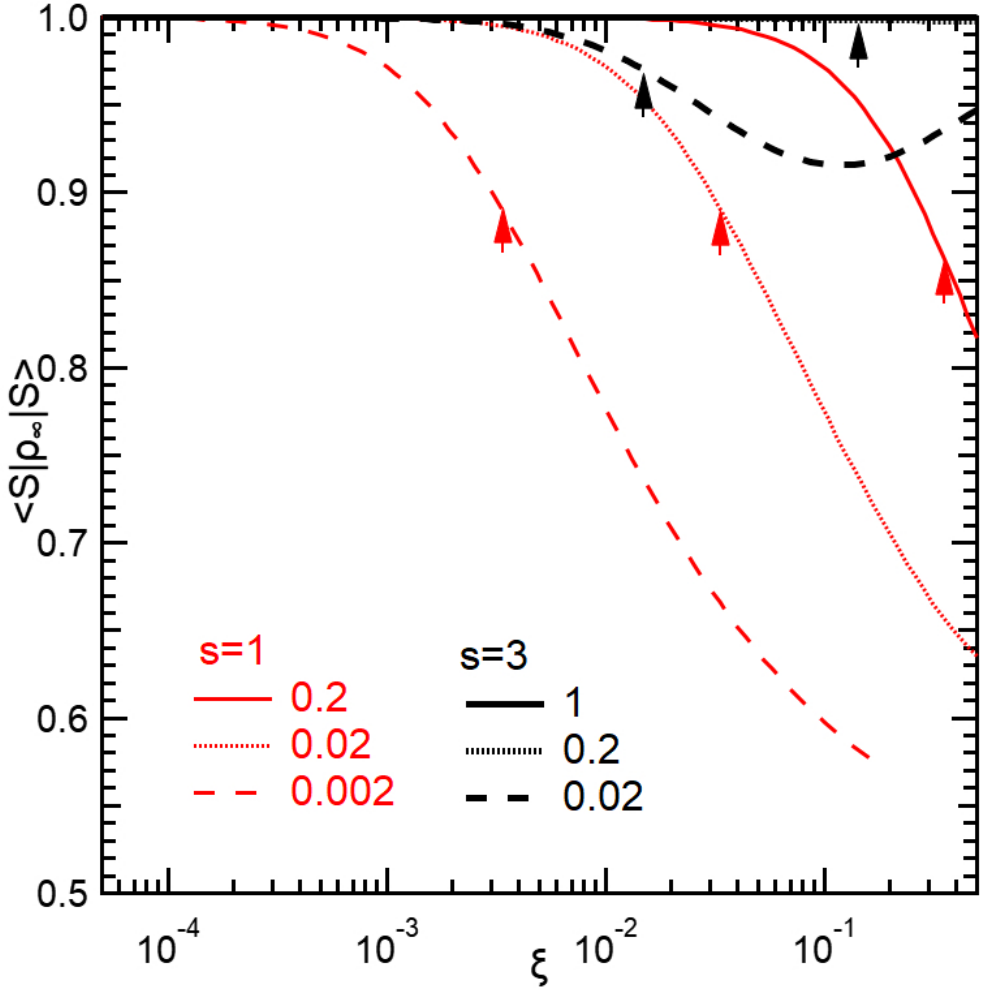}
  \caption{Singlet fidelity of the FTES versus system detuning $\xi$. The higher the damping, the higher the characteristic system detuning, above which the FTES becomes non-universal and not singlet like. \label{fig:fidelity}}
\end{figure}

High singlet fidelity would be important if the goal were to prepare and preserve singlet states as resource for quantum information processing.
Fig.~\ref{fig:fidelity} displays singlet fidelity of the FTES versus system detuning.
The arrows in the figure indicate the points where the bath-coupling detuning $\eta$ is one. That is, $\eta>1$ to the right of the arrows.
We conclude that the condition for significant singlet fidelity of the FTES is that the detuning of the qubit-coupling to the heat bath is significantly less than 1, e.g., that the couplings are close to symmetric. In that regime, the microscopic details of the Hamiltonian and spectral density have only a weak effect on the FTES.

\section{Discussion~\label{discussion}}

In this paper we have presented a theoretical argument for the fault-tolerant excited state (FTES) under the umbrella of a unified system-environment Hamiltonian. The system  is two qubits coupled to a common bath of linear harmonic oscillators. The FTES is an excited state of qubits that behaves like the ground state. It is excited because its energy is higher than the lowest eigenenergy of the qubit Hamiltonian. It behaves like the ground state in that the coupling to the environment at zero temperature produces equilibrium, where the reduced qubit state is pure.

The FTES is not the Gibbs state at an effective temperature, and exemplifies break-down of the eigenstate thermalization hypothesis.~\cite{Deutsch_2018,Tarun} There is a universality aspect to the FTES, in that over a wide range of system-environment parameters
in the regime of strong repulsion between weakly detuned qubit levels caused by the unitary effect of the heat bath, the FTES
has high singlet fidelity and therefore weakly depends on those parameters. This is analogous to
the universality of the Boltzmann distribution, which is also independent of system-environment parameters over a significant range.

The eigenstate thermalization hypothesis can break down if the system is integrable.~\cite{Deutsch_2018}
Since the bath of linear harmonic oscillators is integrable, this may be one reason that the FTES is in violation of the hypothesis. A possible next step of this research could be to examine if the FTES can occur with an anharmonic heat bath, which is not integrable.

We have shown that the FTES is the consequence of the GKSL master equation if two conditions are met: first, an excited eigenstate of the qubit Hamiltonian is annihilated by the dissipative generator of the equation; second, the asymptotic state of the equation is unique. A counterargument for the existence of the FTES would thus be the counterargument for the validity of the GKSL master equation, which would be challenging to find.

A peculiar feature of our model in the weak system-bath coupling regime is that the slower the bath, the less sensitive the FTES with respect to
changes in the model parameters. This will make it easier to experimentally find the parameter range for a FTES if the bath is slower. The reason is that slower baths have a weaker compensation by the counterterm, leading to stronger net unitary
effect on the reduced qubit system.

An interesting question is what kind of entangled qubit states can be created if several qubits are coupled to the same environment. We have found fault-tolerant manifolds of excited states in systems with four and six qubits, which we will publish in future. Thus, fault tolerant excitations can be extended to multiple qubits. This leads to a topic worth exploring if FTES can be used to prepare entangled states as  initial states for quantum computation. This capability would bypass the need for the large number of unitary gates necessary for state preparation. However, experimental tuning of the environmental couplings for the FTES will be very challenging as well.

The question that remains to be answered is if the FTES can occur in non-Markovian quantum dynamics. Our simulations based on TEMPO are consistent with the master equations, but only on a short time scale compared to the recovery time of the FTES. We do not have the computational resources to determine if non-Markovian dynamics can produce a
FTES. With better computational resources and improvements in the algorithm, a study of this question could be doable.
Alternatively, note that FTES shares some properties with decoherence-free-subspaces, as discussed in the text. The latter are valid for exact quantum dynamics. It will be worth exploring if FTES can be understood in terms of some decoherence free proximity space,
which in turn would imply the exactness of FTES for arbitrary strengths of system-bath coupling from the exactness of decoherence-free-subspace.

FTES is not QEC, at least not yet. QEC protects unknown quantum states, while FTES is a known quantum state. Expanding the approach presented in this paper could lead to a Hamiltonian model of QEC.

The authors would like to thank Gerald Fux for discussion and adding a feature to the open source package on our request, and Brayden Ware for bringing to our attention the eigenstate thermalization hypothesis. This research was supported
by DOE contract DE-FG02-06ER46281. Additional support from the Georgia Tech Quantum Alliance (GTQA), a center
funded by the Georgia Tech Institute of Electronics and Nanotechnology was used to develop exact numerical
methods on entangled low-dimensional magnetic system.

\section{Appendix 1. Approximation in Eq.~\ref{eq:sqrt}\label{appendix:GAME}}
To approximate a master equation well,  it is neither necessary nor sufficient that the coefficients of the before/after equations be very close.
Such is the case with the approximation in Eq.~\ref{eq:sqrt}. The accuracy of GAME stems from the coarse-graining properties of master equations, rather than any similarity in coefficients.

To see how coarse-graining works, we first vectorize the master equation. A matrix like $\rho$
is replaced by a column vector $\vert\rho\rangle$ made up of the columns of the matrix appended one after the other.
In this format, $\mathcal{G}$ becomes a Hermitian matrix, $\mathcal{G}_{ni,mj}=
\mathcal{G}_{mj,ni}^\star$,
while the ME is
\begin{equation}
\frac{d\vert\rho\rangle}{dt}=\mathcal{R}\vert\rho\rangle.
\end{equation}
Note that here the generator $\mathcal{R}$ is more general than in Eq.~\ref{Eq:LindGens}. For example, it may be the generator of GAME.
The generator is often referred to as a superoperator, and can be obtained from the master equation as described, for example, in Ref.~\cite{Havel_2003}.
Unlike  $\mathcal{G}$, $\mathcal{R}$ need not be Hermitian.

A coarse-grained master equation is obtained by time averaging the coefficients in the interaction picture. In particular, the coarse-grained superoperator is
\begin{equation}
\tilde{\mathcal{R}}=\frac{1}{T_c}\int\limits_{-T_c/2}^{T_c/2}e^{-\mathcal{R}_0\tau}\mathcal{R}e^{\mathcal{R}_0\tau}d\tau,
\end{equation}
where
$\mathcal{R}_0=-i[H_S,\bullet]$ is the qubit free Liouvillian, and $\mathcal{R}_{int}=e^{-\mathcal{R}_0\tau}\mathcal{R}e^{\mathcal{R}_0\tau}$ is the superoperator in the interaction picture. In terms of matrix elements, coarse-graining amounts to
$\tilde{\mathcal{R}}_{nm,ij}=\mathcal{R}_{nm,ij}\text{sinc}[(\omega_{nm}-\omega_{ij})T_c/2]$, where $\text{sinc}(x)=\sin(x)/x$.
\begin{figure}
\centering
\includegraphics[width=0.45\textwidth]{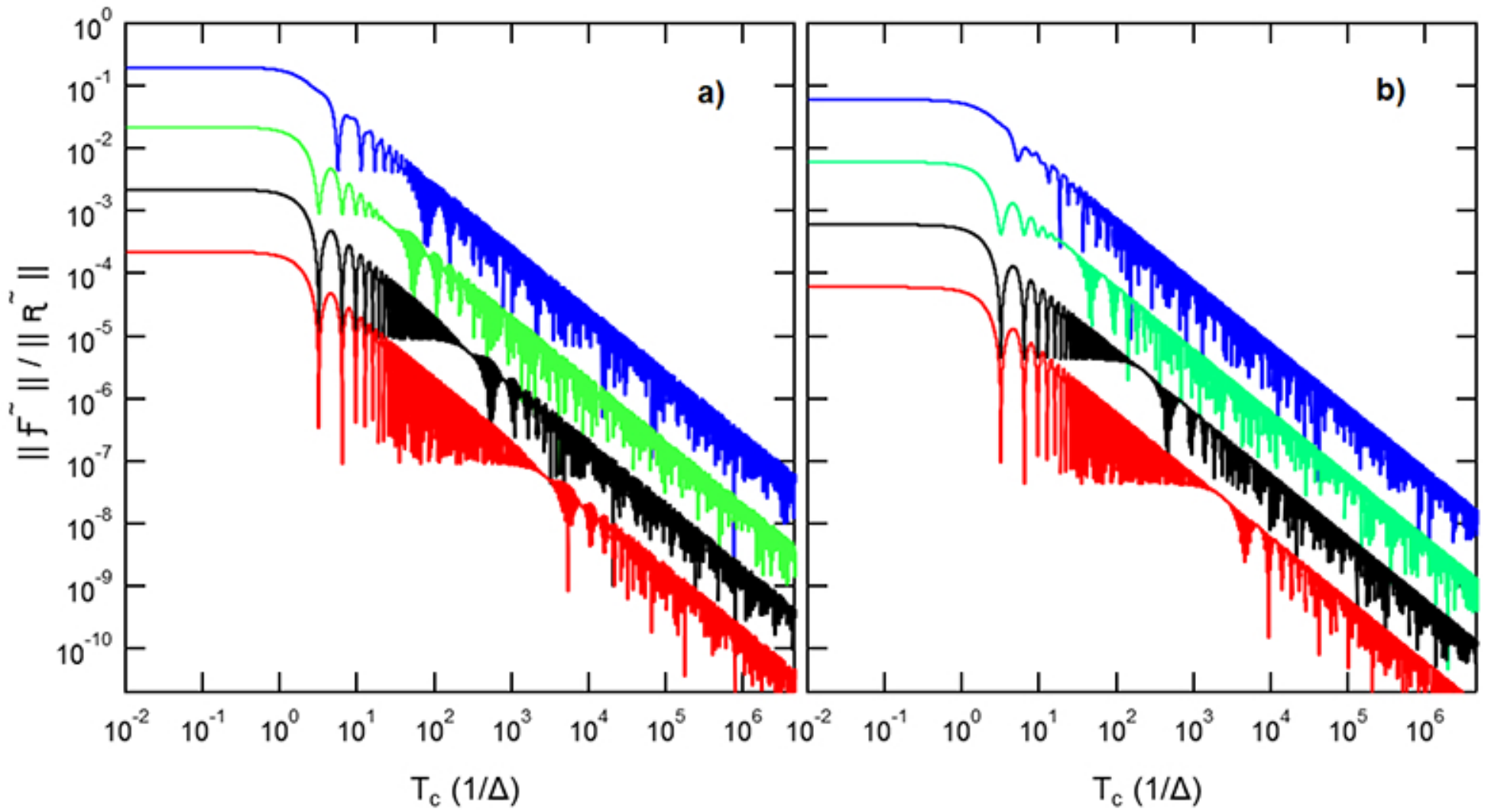}
  \caption{Conjoining of superoperators of two coarse-grained master equations versus coarse-graining time $T_c$. Lines, top-to bottom: $\alpha=2\times 10^{-k}$, $k=1,2,3,4$. {\bf a)}
$s=1$, $\xi=\alpha/4$, $\eta=0.1788$, $\zeta=-0.04349$.
{\bf b)} $s=3$, $\xi=\alpha/5$, $\eta=0.37$, $\zeta=0$.
$\tilde{\mathcal{F}}$ is the Frobenius distance between the superoperators of the RE and GAME master equations.
Schr\"odinger picture. $\Delta=1$, $\omega_c=10$.\label{fig:coarsegrain}}
\end{figure}

The Frobenius distance between coarse-grained superoperators of RE and GAME decreases inversely with coarse-graining time,
as shown in Fig.~\ref{fig:coarsegrain}. In two qubit system, we see that the onset coarse-grain time for the inverse dependence is of order inverse drive frequency $1/\Delta$.

On a time scale much shorter than the relaxation time, the dynamics of the density matrix in the interaction picture, under Markov ME, is approximately the same as the effect of coarse-graining.  To see this, note that the dynamics is
\begin{equation}
\varrho(t/2)=\varrho(-t/2)+\int\limits_{-t/2}^{t/2} d\tau \mathcal{R}_{int}(\tau)\varrho(\tau).
\label{eq:dynamics}
\end{equation}
If $\tau$ is much smaller than the relaxation time, we can make the approximation $\varrho(\tau)\approx \varrho(0)$, and  Eq.~\ref{eq:dynamics} becomes
\begin{equation}
\varrho(t/2)=\varrho(-t/2)+t\tilde{\mathcal{R}}\varrho(0).
\label{eq:CGdynamics}
\end{equation}

Just as the distance between coarse-grained superoperators decreases above the onset time in Fig.~\ref{fig:coarsegrain}, the dynamics under the RE and GAME expressed through Eq.~\ref{eq:CGdynamics} become closer versus time, validating the approximation given by Eq. \ref{eq:sqrt}. Further details can be found in Ref.~\cite{Davidovi__2020} Our numerical studies find accuracy comparable to that of the RE Master equation.~\cite{Davidovi__2020} Equivalent accuracy is analytically established for ULE.~\cite{Nathan}

\begin{widetext}
\section{Appendix 2. TCL4 Master Equation\label{appendix:tcl}}
We closely follow the derivation of the TCL4 master equation given by Breuer, Kappler and Petruccione in
Ref.~\cite{Breuer_1999}. The master equation is obtained as a perturbative expansion of the Nakajima-Zwanzig equation in $\alpha$. Note that $\alpha$ in this article is
not the same as in that reference. The $\alpha$ that we use is quadratic versus $\alpha$ from Ref.~\cite{Breuer_1999}.
In the lowest order, the perturbative expansion gives the time dependent Redfield master equation~\ref{eq:RE2}.
As discussed in Sec.~\ref{sec:RED}, the Markov limit of the equation emerges by replacing the timed spectral density with its limit at $t\to\infty$.

The next term in the perturbative expansion is second order in $\alpha$. In the interaction picture, it is given by Eq. 29 in Ref.~\cite{Breuer_1999},
\begin{eqnarray}
  \label{eq:K_4_exp}
   \lefteqn{ \left(\frac{d\varrho}{dt}\right)_2=\int_0^tdt_1\int_0^{t_1}dt_2\int_0^{t_2}
  dt_3}\nonumber\\
  &&\times\Big\{\langle 02\rangle\langle 13\rangle\left[\hat 0,\left[\hat 1,\hat 2\right]\hat 3\varrho\right]
   - \langle 02\rangle\langle 31\rangle\left[\hat 0,\left[\hat 1,\hat 2\right]\varrho \hat 3\right]
  - \langle 20\rangle\langle 13\rangle\left[\hat 0,\hat 3\varrho\left[\hat 1,\hat 2\right]\right]
   + \langle 20\rangle\langle 31\rangle\left[\hat 0,\varrho \hat 3\left[\hat 1,\hat 2\right]\right]\nonumber\\
 &&  + \langle 03\rangle\langle
  12\rangle\left(\left[\hat 0,\left[\hat 3,\hat 2\right]\varrho \hat 1\right]+\left[\hat 0,\left[\hat 1\hat 2,\hat 3\right]\varrho\right]\right)
  + \langle 30\rangle\langle 21\rangle\left(\left[\hat 0,\hat 1\varrho\left[\hat 3,\hat 2\right]\right]+\left[\hat 0,\varrho\left[\hat 2\hat 1,\hat 3\right]\right]\right)\nonumber\\
 &&  - \langle 03\rangle\langle 21\rangle\left[\hat 0,\left[\hat 1,\hat 3\right]\varrho\hat 2\right]
   - \langle 30 \rangle\langle12\rangle\left[\hat 0,\hat 2\varrho\left[\hat 1,\hat 3\right]\right]\Big\}.
   \label{eq:monstercorrelation}
\end{eqnarray}
Here the short-hand notion is introduced,
\begin{eqnarray}
\nonumber
\hat{i}&=&A(t_i),\,\,i\doteq 0,1,2,3\\
\nonumber
\langle ij\rangle&=&C(t_i-t_j),\,\,i,j\doteq 0,1,2,3.
\end{eqnarray}

Evaluating the expression~\ref{eq:monstercorrelation} is immensely complicated. For completeness, we present the final result for $\delta\mathcal{G}_{in,jm}(t)$ in Eq.~\ref{eq:RE4}:
\begin{eqnarray}
\nonumber
\delta\mathcal{G}_{ni,mj}(t)&=&\sum\limits_{a,b}A_{ni}A_{ja}A_{ab}A_{bm}\int\limits_0^td\tau e^{i\omega_{mj}t+i\omega_{jb}\tau}H_{bm,ja}(t,\tau)-
e^{i\omega_{mj}t+i(\omega_{ja}+\omega_{bm})\tau})H_{ab,ja}(t,\tau)\\
\nonumber
&-&\sum\limits_{a,b,k}\delta_{ni}A_{ja}A_{ab}A_{bk}A_{km}\int\limits_0^td\tau
[e^{i\omega_{kj}t+i\omega_{jb}\tau}H_{bk,ja}(t,\tau)-e^{i\omega_{kj}t+i(\omega_{ja}+\omega_{bk})\tau}H_{ab,ja}(t,\tau)]\\
\nonumber
&-&\sum\limits_{a,b}A_{na}A_{ab}A_{bi}A_{jm}\int\limits_0^td\tau e^{i(\omega_{ij}+\omega_{ma})t+i\omega_{jm}\tau}
[e^{i\omega_{ab}\tau}F_{bi,jm}(t,\tau)-e^{i\omega_{bi}\tau}F_{ab,jm}(t,\tau)]\\
\nonumber
&+&\sum\limits_{a,b}A_{nb}A_{bi}A_{ja}A_{am}\int\limits_0^t\tau e^{i(\omega_{ij}+\omega_{an})t+i\omega_{ja}\tau}
[e^{i\omega_{nb}\tau}F_{bi,ja}(t,\tau)-e^{i\omega_{bi}\tau}F_{nb,ja}(t,\tau)]\\
\nonumber
&+&\sum\limits_{a,b}A_{na}A_{ai}A_{jb}A_{bm}\int\limits_0^td\tau
[e^{i(\omega_{ma}+\omega_{ib})t+i\omega_{ai}\tau}I_{bm,jb}(t,\tau)-
e^{i(\omega_{ba}+\omega_{ij})t+i\omega_{ai}\tau}I_{jb,bm}(t,\tau)]
\\
\nonumber
&-&\sum\limits_{a,b}A_{ni}A_{jb}A_{ba}A_{am}\int\limits_0^td\tau
[e^{i(\omega_{an}+\omega_{ib})t+i\omega_{ni}\tau}I_{ba,jb}(t,\tau)-
e^{i(\omega_{bn}+\omega_{ij})t+i\omega_{ni}\tau}I_{jb,ba}(t,\tau)
]
\\
\nonumber
&+&\sum\limits_{a,b}A_{ni}A_{jb}A_{ba}A_{am}\int\limits_0^td\tau
[e^{i\omega_{aj}t+i\omega_{ba}\tau}I_{jb,am}(t,\tau)
  - e^{i\omega_{aj}t+i\omega_{jb}\tau}I_{ba,am}(t,\tau)]\\
  \nonumber
&-&\sum\limits_{a,b,k}\delta_{ni}A_{jk}A_{ka}A_{ab}A_{bm}\int\limits_0^td\tau
[e^{i\omega_{aj}t+i\omega_{ka}\tau}I_{jk,ab}(t,\tau)-
e^{i\omega_{aj}t+i\omega_{jk}\tau} I_{ka,ab}(t,\tau)]\\
\nonumber
&-&\sum\limits_{a,b}A_{na}A_{ai}A_{jb}A_{bm}\int\limits_0^td\tau
\bigg [e^{i(\omega_{ba}+\omega_{ij})t+i\omega_{jb}\tau}Y_{ai,bm}(t,\tau)-
e^{i(\omega_{ma}+\omega_{ib})t+i\omega_{bm}\tau}Y_{ai,jb}(t,\tau)\bigg]
\\
\nonumber
&+&\sum\limits_{a,b}A_{ni}A_{jb}A_{ba}A_{am}\int\limits_0^td\tau
\bigg [e^{i(\omega_{bn}+\omega_{ij})t+i\omega_{jb}\tau}Y_{ni,ba}(t,\tau)-
e^{i(\omega_{an}+\omega_{ib})t+i\omega_{ba}\tau}Y_{ni,jb}(t,\tau)\bigg].
\label{eq:monsterformula}
\end{eqnarray}
%while for the Lamb-shift,
%\begin{eqnarray}
%
%\mathcal{F}_{jm}&=&\sum\limits_{a,b,k}A_{ja}A_{ab}A_{bk}A_{km}\int\limits_0^tdt_1\Big[
%e^{i\omega_{kj}t+i\omega_{jb}t_1}H_{bk,ja}(t,t_1)-e^{i\omega_{kj}t+i(\omega_{ja}+\omega_{bk})t_1}H_{ab,ja}(t,t_1)\\
%\nonumber
%&+&
%e^{i\omega_{bj}t+i\omega_{ab}t_1}I_{ja,bk}(t,t_1)-
%e^{i\omega_{bj}t+i\omega_{ja}t_1} I_{ab,bk}(t,t_1)\Big].
%\end{eqnarray}

Here we introduce the bath correlation functions
\begin{eqnarray}
\label{corF}
F_{ab,jm}(t,\tau)&=&\int\limits_0^{\tau}du C(t-u)e^{i\omega_{ab}u}\big[\Gamma_{jm}(\tau)-\Gamma_{jm}(\tau-u)\big]^\star\\
\label{corH}
H_{ab,jm}(t,\tau)&=&\int\limits_0^{\tau}du C(u-t)e^{i\omega_{ab}u}\big[\Gamma_{jm}(\tau)-\Gamma_{jm}(\tau-u)\big]^\star\\
\label{corI}
I_{ab,jm}(t,\tau)&=&\int\limits_0^{\tau}du C(u-\tau)e^{i\omega_{ab}u}\big[\Gamma_{jm}(t)-\Gamma_{jm}(t-u)\big]^\star\\
\label{corY}
Y_{ab,jm}(t,\tau)&=&\int\limits_0^{\tau}du C(\tau-u)e^{i\omega_{ab}u}\big[\Gamma_{jm}(t)-\Gamma_{jm}(t-u)\big]^\star,
\end{eqnarray}
where $\Gamma_{jm}(t)=\Gamma(\omega_{jm},t)$ is the timed spectral density given by Eqs.~\ref{eq:sdtOhmic} or~\ref{eq:sdtSuperOhmic}.
We must also add the Hermitian-conjugate, e.g., substitute $\delta \mathcal{G}_{ni,mj}(t)\to \delta \mathcal{G}_{ni,mj}(t)+\delta \mathcal{G}_{mj,ni}(t)^\star$ before inserting in Eq.~\ref{eq:RE4}.

The most time consuming step is to find the correlation functions~\ref{corF}-\ref{corY}. $I$ and $Y$ decay more slowly with time than $F$ and $H$ and require longer $t$ to reach the
Markov limit ($t\to\infty$). For the Ohmic heat bath, $I$ and $Y$ decay inversely with $t-t_1$, and calculating the integrals is time consuming.

First we determine the timed spectral densities analytically using Eqs.~\ref{eq:sdtOhmic} or~\ref{eq:sdtSuperOhmic},
and apply Simpson's rule with time increment $dt=0.04$ and $n_0=2^{17}$ time-steps to find $\delta G_{in,jm}(t)$, where $t=n_0dt\approx  52420/\omega_c$. This leads to the accuracy of approximately five significant digits. For $F$ and $H$, $n_0=2^{14}$ time steps are sufficient to reach that accuracy. For super-Ohmic baths, the integrals converge much faster, and
we use $n_0=2^{14}$ time steps with $dt=0.02$. To verify the accuracy of our TCL4 implementation, we reproduced the low temperature results from Fig. 10.15 in Ref.~\cite{BreuerHeinz-Peter1961-2007TToO}.

An interesting question is should there be a second or higher order counterterms in order to cancel any remaining $\omega_c$ dependence of $\delta\mathcal{G}$ in the Markov limit?
Such higher order counterterms have been implied by the dots in the sum given in Eq. 4.6 in Ref.~\cite{Alicki1}, but no analysis of the higher order terms were given. The derivation of the most general coupling between the system and bath of linear harmonic oscillators leads to the counterterm being linear with $\alpha$.~\cite{CALDEIRA1983374}.

So let us examine the  second order in $\alpha$ terms of the TCL4 master equation, and find out if they depend on $\omega_c$ for $\omega_c\gg\Delta$. Eqs.~\ref{corF}-\ref{corY} show
that we integrate over difference between two timed spectral densities at two different times. In the Markov limit $t\to\infty$,  the terms in Eqs.~\ref{eq:sdtOhmic} and~\ref{eq:sdtSuperOhmic} linear with $\omega_c$ effectively cancel after taking the integrals over both $\tau$ and $u$. Indeed, we calculate the Frobenius distance between $\delta\mathcal{G}$ at cutoff frequencies $10^{k}\omega_c$ and
 $10^{k+1}\omega_c$, for $k=0,1,2,3$, and find that the distance decreases inversely with the cutoff frequency. Thus, in the Markov limit, $\delta\mathcal{G}$ is independent of $\omega_c$ at $\omega_c\gg\Delta$.
 For the same reason, we expect that all higher order terms in the ME expansion of ordered cumulants~\cite{BreuerHeinz-Peter1961-2007TToO} are also independent of $\omega_c$ under those conditions, which is consistent with the absence of higher order counterterms.

The Lamb-shift portion independent of $\omega_c$ for $\omega_c\gg\Delta$ has higher order terms with $\alpha$.
In particular, we can identify the unitary contribution, or the Lamb-shift, in Eq.~\ref{eq:monsterformula} analogously to how we identified it in the RE, which leads to
\begin{equation}
H_L^{(2)}=\frac{\mathcal{F}^\dagger-\mathcal{F}}{2i},
\end{equation}
where
\begin{equation}
\mathcal{F}=\langle 20\rangle\langle 31\rangle\hat{3}[\hat{1},\hat{2}]\hat{0}+\langle 30\rangle\langle 21\rangle[\hat{2}\hat{1},\hat{3}]\hat{0},
\end{equation}
with the matrix elements
\begin{eqnarray}
\mathcal{F}_{jm}&=&\sum\limits_{a,b,k}A_{ja}A_{ab}A_{bk}A_{km}\int\limits_0^tdt_1\Big[
e^{i\omega_{kj}t+i\omega_{jb}t_1}H_{bk,ja}(t,t_1)-e^{i\omega_{kj}t+i(\omega_{ja}+\omega_{bk})t_1}H_{ab,ja}(t,t_1)\\
\nonumber
&+&
e^{i\omega_{bj}t+i\omega_{ab}t_1}I_{ja,bk}(t,t_1)-
e^{i\omega_{bj}t+i\omega_{ja}t_1} I_{ab,bk}(t,t_1)\Big].
\end{eqnarray}
\end{widetext}

\section{Appendix 3. TEMPO Simulation Parameters\label{appendix:TEMPO}}

Throughout this paper we use the TEMPO software package downloaded from Ref.~\cite{Strathearn2018}. We made some modifications, such as updating the system-bath coupling Hamiltonian according to Eq.~\ref{eq:A} and including the counterterm in Eq.~\ref{eq:counterterm}. There are three sources of errors: 1) The Trotter splitting error, which is due to the finite time step $\tau_{s}$. In all simulations presented here, the time step is $0.02/\Delta$; 2) The finite memory approximation, originally introduced by Makri and Makarov,~\cite{Makri1,Makri2} limits the time over which the histories of the system are stored. This error is measured here by the finite number of time steps $K_{max}$ in the ADT. That is, the various histories of the system are stored over the last $K_{max}$ time steps. In our simulations, $K_{max}=150$ for the Ohmic bath and $K_{max}=75$ for the super-Ohmic bath; 3) Precision parameter of the SVD-compression. In all our simulations, the precision is $\epsilon_{SVD}=10^{-6}$, which means that in the tensor network representation of the ADT, we neglect the singular vectors with singular values $<10^{-6}$.

We have varied $\tau_{s}$, $K_{max}$, and $\epsilon$  until the numerically obtained quantum dynamics is roughly converged. Here, that means that changing the parameters in the direction of higher accuracy, by factors of two, will change a curve in Figs.~\ref{fig:array}-\ref{fig:arraySO} by approximately five percent.
However, the array of curves in Figs.~\ref{fig:array}-\ref{fig:arraySO} remain very similar as the accuracy is increased,
unambiguously confirming the suppression of singlet damping.
The positivity violations in the density matrix
can lead to singlet fidelity exceeding one near fault tolerance conditions due to these errors.

\begin{figure}
\centering
\includegraphics[width=0.45\textwidth]{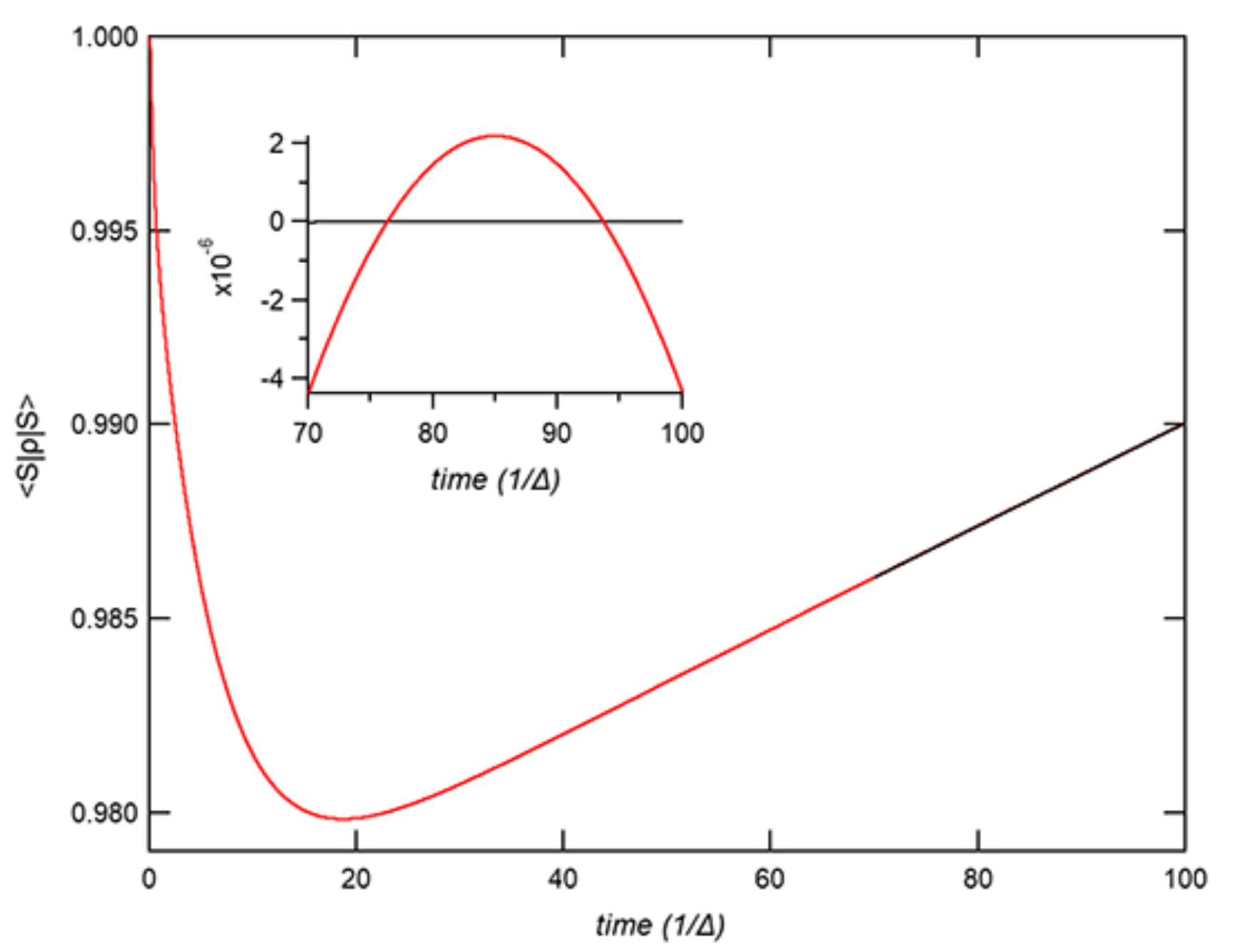}
  \caption{Tempo simulation: Red line: Singlet fidelity versus time, for $\eta=0.22$ in Fig.~\ref{fig:arrayCT}~{\bf d)}.
  Black line:best fit to exponential, in the interval $t>70/\Delta$. Inset: Difference between best fit and TEMPO simulation, for linear (red) and exponential fit (black).
  \label{fig:fitting}}
\end{figure}

Fig.~\ref{fig:fitting} shows singlet fidelity versus time at $\eta=0.22$ from Fig.~\ref{fig:arrayCT}~{\bf d)}. The data between $70/\Delta$ and $100/\Delta$ is fit to an exponential with offset and a linear function. The differences between best fits and data are shown in the inset. The exponential fit is much better than the linear, showing that we can reliably infer the curvature from which we estimate the recovery time. The latter is $T_r=2286/\Delta$, an unfeasible long time for TEMPO, while the asymptotic state singlet fidelity is $1.29$. Thus, while the solution appears to be bound, the fidelity exceeding one confirms a positivity violation, which we have also confirmed by the negative eigenvalues of the density matrix.

\nocite{apsrev41Control}
\bibliography{master}

\begin{thebibliography}{87}%
\makeatletter
\providecommand \@ifxundefined [1]{%
 \@ifx{#1\undefined}
}%
\providecommand \@ifnum [1]{%
 \ifnum #1\expandafter \@firstoftwo
 \else \expandafter \@secondoftwo
 \fi
}%
\providecommand \@ifx [1]{%
 \ifx #1\expandafter \@firstoftwo
 \else \expandafter \@secondoftwo
 \fi
}%
\providecommand \natexlab [1]{#1}%
\providecommand \enquote  [1]{``#1''}%
\providecommand \bibnamefont  [1]{#1}%
\providecommand \bibfnamefont [1]{#1}%
\providecommand \citenamefont [1]{#1}%
\providecommand \href@noop [0]{\@secondoftwo}%
\providecommand \href [0]{\begingroup \@sanitize@url \@href}%
\providecommand \@href[1]{\@@startlink{#1}\@@href}%
\providecommand \@@href[1]{\endgroup#1\@@endlink}%
\providecommand \@sanitize@url [0]{\catcode `\\12\catcode `\$12\catcode
  `\&12\catcode `\#12\catcode `\^12\catcode `\_12\catcode `\%12\relax}%
\providecommand \@@startlink[1]{}%
\providecommand \@@endlink[0]{}%
\providecommand \url  [0]{\begingroup\@sanitize@url \@url }%
\providecommand \@url [1]{\endgroup\@href {#1}{\urlprefix }}%
\providecommand \urlprefix  [0]{URL }%
\providecommand \Eprint [0]{\href }%
\providecommand \doibase [0]{http://dx.doi.org/}%
\providecommand \selectlanguage [0]{\@gobble}%
\providecommand \bibinfo  [0]{\@secondoftwo}%
\providecommand \bibfield  [0]{\@secondoftwo}%
\providecommand \translation [1]{[#1]}%
\providecommand \BibitemOpen [0]{}%
\providecommand \bibitemStop [0]{}%
\providecommand \bibitemNoStop [0]{.\EOS\space}%
\providecommand \EOS [0]{\spacefactor3000\relax}%
\providecommand \BibitemShut  [1]{\csname bibitem#1\endcsname}%
\let\auto@bib@innerbib\@empty
%</preamble>
\bibitem [{\citenamefont {Nielsen}\ and\ \citenamefont
  {Chuang}(2011)}]{Nielsen}%
  \BibitemOpen
  \bibfield  {author} {\bibinfo {author} {\bibfnamefont {M.~A.}\ \bibnamefont
  {Nielsen}}\ and\ \bibinfo {author} {\bibfnamefont {I.~L.}\ \bibnamefont
  {Chuang}},\ }\href@noop {} {\emph {\bibinfo {title} {Quantum Computation and
  Quantum Information: 10th Anniversary Edition}}},\ \bibinfo {edition} {10th}\
  ed.\ (\bibinfo  {publisher} {Cambridge University Press},\ \bibinfo {address}
  {USA},\ \bibinfo {year} {2011})\BibitemShut {NoStop}%
\bibitem [{\citenamefont {Shor}(1995)}]{Shor}%
  \BibitemOpen
  \bibfield  {author} {\bibinfo {author} {\bibfnamefont {P.~W.}\ \bibnamefont
  {Shor}},\ }\href {\doibase 10.1103/PhysRevA.52.R2493} {\bibfield  {journal}
  {\bibinfo  {journal} {Phys. Rev. A}\ }\textbf {\bibinfo {volume} {52}},\
  \bibinfo {pages} {R2493} (\bibinfo {year} {1995})}\BibitemShut {NoStop}%
\bibitem [{\citenamefont {Steane}(1996)}]{Steane}%
  \BibitemOpen
  \bibfield  {author} {\bibinfo {author} {\bibfnamefont {A.~M.}\ \bibnamefont
  {Steane}},\ }\href {\doibase 10.1103/PhysRevLett.77.793} {\bibfield
  {journal} {\bibinfo  {journal} {Phys. Rev. Lett.}\ }\textbf {\bibinfo
  {volume} {77}},\ \bibinfo {pages} {793} (\bibinfo {year} {1996})}\BibitemShut
  {NoStop}%
\bibitem [{\citenamefont {Wiseman}(1994)}]{Wiseman}%
  \BibitemOpen
  \bibfield  {author} {\bibinfo {author} {\bibfnamefont {H.~M.}\ \bibnamefont
  {Wiseman}},\ }\href {\doibase 10.1103/PhysRevA.49.2133} {\bibfield  {journal}
  {\bibinfo  {journal} {Phys. Rev. A}\ }\textbf {\bibinfo {volume} {49}},\
  \bibinfo {pages} {2133} (\bibinfo {year} {1994})}\BibitemShut {NoStop}%
\bibitem [{\citenamefont {Goetsch}\ \emph {et~al.}(1996)\citenamefont
  {Goetsch}, \citenamefont {Tombesi},\ and\ \citenamefont {Vitali}}]{Vitali}%
  \BibitemOpen
  \bibfield  {author} {\bibinfo {author} {\bibfnamefont {P.}~\bibnamefont
  {Goetsch}}, \bibinfo {author} {\bibfnamefont {P.}~\bibnamefont {Tombesi}}, \
  and\ \bibinfo {author} {\bibfnamefont {D.}~\bibnamefont {Vitali}},\ }\href
  {\doibase 10.1103/PhysRevA.54.4519} {\bibfield  {journal} {\bibinfo
  {journal} {Phys. Rev. A}\ }\textbf {\bibinfo {volume} {54}},\ \bibinfo
  {pages} {4519} (\bibinfo {year} {1996})}\BibitemShut {NoStop}%
\bibitem [{\citenamefont {Wiseman}\ and\ \citenamefont
  {Milburn}(2009)}]{wiseman_milburn_2009}%
  \BibitemOpen
  \bibfield  {author} {\bibinfo {author} {\bibfnamefont {H.~M.}\ \bibnamefont
  {Wiseman}}\ and\ \bibinfo {author} {\bibfnamefont {G.~J.}\ \bibnamefont
  {Milburn}},\ }\href {\doibase 10.1017/CBO9780511813948} {\emph {\bibinfo
  {title} {Quantum Measurement and Control}}}\ (\bibinfo  {publisher}
  {Cambridge University Press},\ \bibinfo {year} {2009})\BibitemShut {NoStop}%
\bibitem [{\citenamefont {Ahn}\ \emph {et~al.}(2002)\citenamefont {Ahn},
  \citenamefont {Doherty},\ and\ \citenamefont {Landahl}}]{Ahn}%
  \BibitemOpen
  \bibfield  {author} {\bibinfo {author} {\bibfnamefont {C.}~\bibnamefont
  {Ahn}}, \bibinfo {author} {\bibfnamefont {A.~C.}\ \bibnamefont {Doherty}}, \
  and\ \bibinfo {author} {\bibfnamefont {A.~J.}\ \bibnamefont {Landahl}},\
  }\href {\doibase 10.1103/physreva.65.042301} {\bibfield  {journal} {\bibinfo
  {journal} {Physical Review A}\ }\textbf {\bibinfo {volume} {65}} (\bibinfo
  {year} {2002}),\ 10.1103/physreva.65.042301}\BibitemShut {NoStop}%
\bibitem [{\citenamefont {Lloyd}(2000)}]{Lloyd}%
  \BibitemOpen
  \bibfield  {author} {\bibinfo {author} {\bibfnamefont {S.}~\bibnamefont
  {Lloyd}},\ }\href {\doibase 10.1103/PhysRevA.62.022108} {\bibfield  {journal}
  {\bibinfo  {journal} {Phys. Rev. A}\ }\textbf {\bibinfo {volume} {62}},\
  \bibinfo {pages} {022108} (\bibinfo {year} {2000})}\BibitemShut {NoStop}%
\bibitem [{\citenamefont {Nelson}\ \emph {et~al.}(2000)\citenamefont {Nelson},
  \citenamefont {Weinstein}, \citenamefont {Cory},\ and\ \citenamefont
  {Lloyd}}]{Nelson}%
  \BibitemOpen
  \bibfield  {author} {\bibinfo {author} {\bibfnamefont {R.~J.}\ \bibnamefont
  {Nelson}}, \bibinfo {author} {\bibfnamefont {Y.}~\bibnamefont {Weinstein}},
  \bibinfo {author} {\bibfnamefont {D.}~\bibnamefont {Cory}}, \ and\ \bibinfo
  {author} {\bibfnamefont {S.}~\bibnamefont {Lloyd}},\ }\href {\doibase
  10.1103/PhysRevLett.85.3045} {\bibfield  {journal} {\bibinfo  {journal}
  {Phys. Rev. Lett.}\ }\textbf {\bibinfo {volume} {85}},\ \bibinfo {pages}
  {3045} (\bibinfo {year} {2000})}\BibitemShut {NoStop}%
\bibitem [{\citenamefont {Alicki}\ \emph {et~al.}(2006)\citenamefont {Alicki},
  \citenamefont {Lidar},\ and\ \citenamefont {Zanardi}}]{Alicki_2006}%
  \BibitemOpen
  \bibfield  {author} {\bibinfo {author} {\bibfnamefont {R.}~\bibnamefont
  {Alicki}}, \bibinfo {author} {\bibfnamefont {D.~A.}\ \bibnamefont {Lidar}}, \
  and\ \bibinfo {author} {\bibfnamefont {P.}~\bibnamefont {Zanardi}},\ }\href
  {\doibase 10.1103/physreva.73.052311} {\bibfield  {journal} {\bibinfo
  {journal} {Physical Review A}\ }\textbf {\bibinfo {volume} {73}} (\bibinfo
  {year} {2006}),\ 10.1103/physreva.73.052311}\BibitemShut {NoStop}%
\bibitem [{\citenamefont {Mohseni}\ and\ \citenamefont
  {Lidar}(2005)}]{Mohseni}%
  \BibitemOpen
  \bibfield  {author} {\bibinfo {author} {\bibfnamefont {M.}~\bibnamefont
  {Mohseni}}\ and\ \bibinfo {author} {\bibfnamefont {D.~A.}\ \bibnamefont
  {Lidar}},\ }\href {\doibase 10.1103/PhysRevLett.94.040507} {\bibfield
  {journal} {\bibinfo  {journal} {Phys. Rev. Lett.}\ }\textbf {\bibinfo
  {volume} {94}},\ \bibinfo {pages} {040507} (\bibinfo {year}
  {2005})}\BibitemShut {NoStop}%
\bibitem [{\citenamefont {Terhal}\ and\ \citenamefont
  {Burkard}(2005)}]{Barbara}%
  \BibitemOpen
  \bibfield  {author} {\bibinfo {author} {\bibfnamefont {B.~M.}\ \bibnamefont
  {Terhal}}\ and\ \bibinfo {author} {\bibfnamefont {G.}~\bibnamefont
  {Burkard}},\ }\href {\doibase 10.1103/PhysRevA.71.012336} {\bibfield
  {journal} {\bibinfo  {journal} {Phys. Rev. A}\ }\textbf {\bibinfo {volume}
  {71}},\ \bibinfo {pages} {012336} (\bibinfo {year} {2005})}\BibitemShut
  {NoStop}%
\bibitem [{\citenamefont {Aliferis}\ \emph {et~al.}(2006)\citenamefont
  {Aliferis}, \citenamefont {Gottesman},\ and\ \citenamefont
  {Preskill}}]{Aliferis}%
  \BibitemOpen
  \bibfield  {author} {\bibinfo {author} {\bibfnamefont {P.}~\bibnamefont
  {Aliferis}}, \bibinfo {author} {\bibfnamefont {D.}~\bibnamefont {Gottesman}},
  \ and\ \bibinfo {author} {\bibfnamefont {J.}~\bibnamefont {Preskill}},\
  }\href@noop {} {\bibfield  {journal} {\bibinfo  {journal} {Quantum Info.
  Comput.}\ }\textbf {\bibinfo {volume} {6}},\ \bibinfo {pages} {97–165}
  (\bibinfo {year} {2006})}\BibitemShut {NoStop}%
\bibitem [{\citenamefont {Aharonov}\ \emph {et~al.}(2006)\citenamefont
  {Aharonov}, \citenamefont {Kitaev},\ and\ \citenamefont
  {Preskill}}]{Aharonov}%
  \BibitemOpen
  \bibfield  {author} {\bibinfo {author} {\bibfnamefont {D.}~\bibnamefont
  {Aharonov}}, \bibinfo {author} {\bibfnamefont {A.}~\bibnamefont {Kitaev}}, \
  and\ \bibinfo {author} {\bibfnamefont {J.}~\bibnamefont {Preskill}},\ }\href
  {\doibase 10.1103/PhysRevLett.96.050504} {\bibfield  {journal} {\bibinfo
  {journal} {Phys. Rev. Lett.}\ }\textbf {\bibinfo {volume} {96}},\ \bibinfo
  {pages} {050504} (\bibinfo {year} {2006})}\BibitemShut {NoStop}%
\bibitem [{\citenamefont {Sarovar}\ \emph {et~al.}(2004)\citenamefont
  {Sarovar}, \citenamefont {Ahn}, \citenamefont {Jacobs},\ and\ \citenamefont
  {Milburn}}]{Sarovar_2004}%
  \BibitemOpen
  \bibfield  {author} {\bibinfo {author} {\bibfnamefont {M.}~\bibnamefont
  {Sarovar}}, \bibinfo {author} {\bibfnamefont {C.}~\bibnamefont {Ahn}},
  \bibinfo {author} {\bibfnamefont {K.}~\bibnamefont {Jacobs}}, \ and\ \bibinfo
  {author} {\bibfnamefont {G.~J.}\ \bibnamefont {Milburn}},\ }\href {\doibase
  10.1103/PhysRevA.69.052324} {\bibfield  {journal} {\bibinfo  {journal} {Phys.
  Rev. A}\ }\textbf {\bibinfo {volume} {69}},\ \bibinfo {pages} {052324}
  (\bibinfo {year} {2004})}\BibitemShut {NoStop}%
\bibitem [{\citenamefont {Sarovar}\ and\ \citenamefont
  {Milburn}(2005)}]{Sarovar_2005}%
  \BibitemOpen
  \bibfield  {author} {\bibinfo {author} {\bibfnamefont {M.}~\bibnamefont
  {Sarovar}}\ and\ \bibinfo {author} {\bibfnamefont {G.~J.}\ \bibnamefont
  {Milburn}},\ }\href {\doibase 10.1103/PhysRevA.72.012306} {\bibfield
  {journal} {\bibinfo  {journal} {Phys. Rev. A}\ }\textbf {\bibinfo {volume}
  {72}},\ \bibinfo {pages} {012306} (\bibinfo {year} {2005})}\BibitemShut
  {NoStop}%
\bibitem [{\citenamefont {Khodjasteh}\ and\ \citenamefont
  {Lidar}(2005)}]{Khodjasteh}%
  \BibitemOpen
  \bibfield  {author} {\bibinfo {author} {\bibfnamefont {K.}~\bibnamefont
  {Khodjasteh}}\ and\ \bibinfo {author} {\bibfnamefont {D.~A.}\ \bibnamefont
  {Lidar}},\ }\href {\doibase 10.1103/PhysRevLett.95.180501} {\bibfield
  {journal} {\bibinfo  {journal} {Phys. Rev. Lett.}\ }\textbf {\bibinfo
  {volume} {95}},\ \bibinfo {pages} {180501} (\bibinfo {year}
  {2005})}\BibitemShut {NoStop}%
\bibitem [{\citenamefont {Mozgunov}\ and\ \citenamefont
  {Lidar}(2020)}]{mozgunov}%
  \BibitemOpen
  \bibfield  {author} {\bibinfo {author} {\bibfnamefont {E.}~\bibnamefont
  {Mozgunov}}\ and\ \bibinfo {author} {\bibfnamefont {D.}~\bibnamefont
  {Lidar}},\ }\href {\doibase https://doi.org/10.22331/q-2020-02-06-227} {\
  \textbf {\bibinfo {volume} {4}},\ \bibinfo {pages} {227} (\bibinfo {year}
  {2020})},\ \Eprint {http://arxiv.org/abs/1908.01095} {1908.01095 [Quantum]}
  \BibitemShut {NoStop}%
\bibitem [{\citenamefont {Davidović}(2020)}]{Davidovi__2020}%
  \BibitemOpen
  \bibfield  {author} {\bibinfo {author} {\bibfnamefont {D.}~\bibnamefont
  {Davidović}},\ }\href {\doibase 10.22331/q-2020-09-21-326} {\bibfield
  {journal} {\bibinfo  {journal} {Quantum}\ }\textbf {\bibinfo {volume} {4}},\
  \bibinfo {pages} {326} (\bibinfo {year} {2020})}\BibitemShut {NoStop}%
\bibitem [{\citenamefont {Nathan}\ and\ \citenamefont {Rudner}(2020)}]{Nathan}%
  \BibitemOpen
  \bibfield  {author} {\bibinfo {author} {\bibfnamefont {F.}~\bibnamefont
  {Nathan}}\ and\ \bibinfo {author} {\bibfnamefont {M.~S.}\ \bibnamefont
  {Rudner}},\ }\href {\doibase 10.1103/PhysRevB.102.115109} {\bibfield
  {journal} {\bibinfo  {journal} {Phys. Rev. B}\ }\textbf {\bibinfo {volume}
  {102}},\ \bibinfo {pages} {115109} (\bibinfo {year} {2020})}\BibitemShut
  {NoStop}%
\bibitem [{\citenamefont {Strathearn}\ \emph {et~al.}(2018)\citenamefont
  {Strathearn}, \citenamefont {Kirton}, \citenamefont {Kilda}, \citenamefont
  {Keeling},\ and\ \citenamefont {Lovett}}]{Strathearn2018}%
  \BibitemOpen
  \bibfield  {author} {\bibinfo {author} {\bibfnamefont {A.}~\bibnamefont
  {Strathearn}}, \bibinfo {author} {\bibfnamefont {P.}~\bibnamefont {Kirton}},
  \bibinfo {author} {\bibfnamefont {D.}~\bibnamefont {Kilda}}, \bibinfo
  {author} {\bibfnamefont {J.}~\bibnamefont {Keeling}}, \ and\ \bibinfo
  {author} {\bibfnamefont {B.~W.}\ \bibnamefont {Lovett}},\ }\href {\doibase
  10.1038/s41467-018-05617-3} {\bibfield  {journal} {\bibinfo  {journal} {Nat.
  Commun.}\ }\textbf {\bibinfo {volume} {9}},\ \bibinfo {pages} {3322}
  (\bibinfo {year} {2018})}\BibitemShut {NoStop}%
\bibitem [{\citenamefont {Caldeira}\ and\ \citenamefont
  {Leggett}(1983)}]{CALDEIRA1983374}%
  \BibitemOpen
  \bibfield  {author} {\bibinfo {author} {\bibfnamefont {A.}~\bibnamefont
  {Caldeira}}\ and\ \bibinfo {author} {\bibfnamefont {A.}~\bibnamefont
  {Leggett}},\ }\href {\doibase https://doi.org/10.1016/0003-4916(83)90202-6}
  {\bibfield  {journal} {\bibinfo  {journal} {Annals of Physics}\ }\textbf
  {\bibinfo {volume} {149}},\ \bibinfo {pages} {374} (\bibinfo {year}
  {1983})}\BibitemShut {NoStop}%
\bibitem [{\citenamefont {Weiss}(2012)}]{ulrich}%
  \BibitemOpen
  \bibfield  {author} {\bibinfo {author} {\bibfnamefont {U.}~\bibnamefont
  {Weiss}},\ }\href@noop {} {} (\bibinfo {year} {2012})\BibitemShut {NoStop}%
\bibitem [{\citenamefont {Breuer}\ and\ \citenamefont
  {Petruccione}(2007)}]{BreuerHeinz-Peter1961-2007TToO}%
  \BibitemOpen
  \bibfield  {author} {\bibinfo {author} {\bibfnamefont {H.-P.}\ \bibnamefont
  {Breuer}}\ and\ \bibinfo {author} {\bibfnamefont {F.}~\bibnamefont
  {Petruccione}},\ }\href@noop {} {\enquote {\bibinfo {title} {The theory of
  open quantum systems},}\ } (\bibinfo {year} {2007})\BibitemShut {NoStop}%
\bibitem [{\citenamefont {Hartmann}\ and\ \citenamefont
  {Strunz}(2020{\natexlab{a}})}]{Hartmann_2020}%
  \BibitemOpen
  \bibfield  {author} {\bibinfo {author} {\bibfnamefont {R.}~\bibnamefont
  {Hartmann}}\ and\ \bibinfo {author} {\bibfnamefont {W.~T.}\ \bibnamefont
  {Strunz}},\ }\href {\doibase 10.22331/q-2020-10-22-347} {\bibfield  {journal}
  {\bibinfo  {journal} {Quantum}\ }\textbf {\bibinfo {volume} {4}},\ \bibinfo
  {pages} {347} (\bibinfo {year} {2020}{\natexlab{a}})}\BibitemShut {NoStop}%
\bibitem [{\citenamefont {Alicki}\ and\ \citenamefont
  {Lendi}(2014)}]{AlickiBook}%
  \BibitemOpen
  \bibfield  {author} {\bibinfo {author} {\bibfnamefont {R.}~\bibnamefont
  {Alicki}}\ and\ \bibinfo {author} {\bibfnamefont {K.}~\bibnamefont {Lendi}},\
  }\href@noop {} {\emph {\bibinfo {title} {Quantum Dynamical Semigroups and
  Applications}}}\ (\bibinfo  {publisher} {Springer Publishing Company,
  Incorporated},\ \bibinfo {year} {2014})\BibitemShut {NoStop}%
\bibitem [{\citenamefont {Gorini}\ \emph {et~al.}(1976)\citenamefont {Gorini},
  \citenamefont {Kossakowski},\ and\ \citenamefont {Sudarshan}}]{Gorini}%
  \BibitemOpen
  \bibfield  {author} {\bibinfo {author} {\bibfnamefont {V.}~\bibnamefont
  {Gorini}}, \bibinfo {author} {\bibfnamefont {A.}~\bibnamefont {Kossakowski}},
  \ and\ \bibinfo {author} {\bibfnamefont {E.~C.~G.}\ \bibnamefont
  {Sudarshan}},\ }\href {\doibase 10.1063/1.522979} {\bibfield  {journal}
  {\bibinfo  {journal} {Journal of Mathematical Physics}\ }\textbf {\bibinfo
  {volume} {17}},\ \bibinfo {pages} {821} (\bibinfo {year} {1976})},\ \Eprint
  {http://arxiv.org/abs/https://aip.scitation.org/doi/pdf/10.1063/1.522979}
  {https://aip.scitation.org/doi/pdf/10.1063/1.522979} \BibitemShut {NoStop}%
\bibitem [{\citenamefont {Lindblad}(1976)}]{lindblad1976}%
  \BibitemOpen
  \bibfield  {author} {\bibinfo {author} {\bibfnamefont {G.}~\bibnamefont
  {Lindblad}},\ }\href {https://projecteuclid.org:443/euclid.cmp/1103899849}
  {\bibfield  {journal} {\bibinfo  {journal} {Comm. Math. Phys.}\ }\textbf
  {\bibinfo {volume} {48}},\ \bibinfo {pages} {119} (\bibinfo {year}
  {1976})}\BibitemShut {NoStop}%
\bibitem [{\citenamefont {Davies}(1974)}]{davies1974}%
  \BibitemOpen
  \bibfield  {author} {\bibinfo {author} {\bibfnamefont {E.~B.}\ \bibnamefont
  {Davies}},\ }\href {https://projecteuclid.org:443/euclid.cmp/1103860160}
  {\bibfield  {journal} {\bibinfo  {journal} {Comm. Math. Phys.}\ }\textbf
  {\bibinfo {volume} {39}},\ \bibinfo {pages} {91} (\bibinfo {year}
  {1974})}\BibitemShut {NoStop}%
\bibitem [{\citenamefont {Chen}\ and\ \citenamefont
  {Lidar}(2020)}]{chen2020hoqst}%
  \BibitemOpen
  \bibfield  {author} {\bibinfo {author} {\bibfnamefont {H.}~\bibnamefont
  {Chen}}\ and\ \bibinfo {author} {\bibfnamefont {D.~A.}\ \bibnamefont
  {Lidar}},\ }\href@noop {} {\enquote {\bibinfo {title} {Hoqst: Hamiltonian
  open quantum system toolkit},}\ } (\bibinfo {year} {2020}),\ \Eprint
  {http://arxiv.org/abs/2011.14046} {arXiv:2011.14046 [quant-ph]} \BibitemShut
  {NoStop}%
\bibitem [{\citenamefont {Majenz}\ \emph {et~al.}(2013)\citenamefont {Majenz},
  \citenamefont {Albash}, \citenamefont {Breuer},\ and\ \citenamefont
  {Lidar}}]{Majenz}%
  \BibitemOpen
  \bibfield  {author} {\bibinfo {author} {\bibfnamefont {C.}~\bibnamefont
  {Majenz}}, \bibinfo {author} {\bibfnamefont {T.}~\bibnamefont {Albash}},
  \bibinfo {author} {\bibfnamefont {H.-P.}\ \bibnamefont {Breuer}}, \ and\
  \bibinfo {author} {\bibfnamefont {D.~A.}\ \bibnamefont {Lidar}},\ }\href
  {\doibase 10.1103/PhysRevA.88.012103} {\bibfield  {journal} {\bibinfo
  {journal} {Phys. Rev. A}\ }\textbf {\bibinfo {volume} {88}},\ \bibinfo
  {pages} {012103} (\bibinfo {year} {2013})}\BibitemShut {NoStop}%
\bibitem [{\citenamefont {Schaller}\ and\ \citenamefont
  {Brandes}(2008)}]{Schaller}%
  \BibitemOpen
  \bibfield  {author} {\bibinfo {author} {\bibfnamefont {G.}~\bibnamefont
  {Schaller}}\ and\ \bibinfo {author} {\bibfnamefont {T.}~\bibnamefont
  {Brandes}},\ }\href {\doibase 10.1103/PhysRevA.78.022106} {\bibfield
  {journal} {\bibinfo  {journal} {Phys. Rev. A}\ }\textbf {\bibinfo {volume}
  {78}},\ \bibinfo {pages} {022106} (\bibinfo {year} {2008})}\BibitemShut
  {NoStop}%
\bibitem [{\citenamefont {Albash}\ \emph {et~al.}(2012)\citenamefont {Albash},
  \citenamefont {Boixo}, \citenamefont {Lidar},\ and\ \citenamefont
  {Zanardi}}]{Albash_2012}%
  \BibitemOpen
  \bibfield  {author} {\bibinfo {author} {\bibfnamefont {T.}~\bibnamefont
  {Albash}}, \bibinfo {author} {\bibfnamefont {S.}~\bibnamefont {Boixo}},
  \bibinfo {author} {\bibfnamefont {D.~A.}\ \bibnamefont {Lidar}}, \ and\
  \bibinfo {author} {\bibfnamefont {P.}~\bibnamefont {Zanardi}},\ }\href
  {\doibase 10.1088/1367-2630/14/12/123016} {\bibfield  {journal} {\bibinfo
  {journal} {New Journal of Physics}\ }\textbf {\bibinfo {volume} {14}},\
  \bibinfo {pages} {123016} (\bibinfo {year} {2012})}\BibitemShut {NoStop}%
\bibitem [{\citenamefont {Vogt}\ \emph {et~al.}(2013)\citenamefont {Vogt},
  \citenamefont {Jeske},\ and\ \citenamefont {Cole}}]{Vogt}%
  \BibitemOpen
  \bibfield  {author} {\bibinfo {author} {\bibfnamefont {N.}~\bibnamefont
  {Vogt}}, \bibinfo {author} {\bibfnamefont {J.}~\bibnamefont {Jeske}}, \ and\
  \bibinfo {author} {\bibfnamefont {J.~H.}\ \bibnamefont {Cole}},\ }\href
  {\doibase 10.1103/PhysRevB.88.174514} {\bibfield  {journal} {\bibinfo
  {journal} {Phys. Rev. B}\ }\textbf {\bibinfo {volume} {88}},\ \bibinfo
  {pages} {174514} (\bibinfo {year} {2013})}\BibitemShut {NoStop}%
\bibitem [{\citenamefont {Tscherbul}\ and\ \citenamefont
  {Brumer}(2015)}]{Tscherbul}%
  \BibitemOpen
  \bibfield  {author} {\bibinfo {author} {\bibfnamefont {T.~V.}\ \bibnamefont
  {Tscherbul}}\ and\ \bibinfo {author} {\bibfnamefont {P.}~\bibnamefont
  {Brumer}},\ }\href {\doibase 10.1063/1.4908130} {\bibfield  {journal}
  {\bibinfo  {journal} {The Journal of Chemical Physics}\ }\textbf {\bibinfo
  {volume} {142}},\ \bibinfo {pages} {104107} (\bibinfo {year} {2015})},\
  \Eprint {http://arxiv.org/abs/https://doi.org/10.1063/1.4908130}
  {https://doi.org/10.1063/1.4908130} \BibitemShut {NoStop}%
\bibitem [{\citenamefont {Kir\ifmmode~\check{s}\else \v{s}\fi{}anskas}\ \emph
  {et~al.}(2018)\citenamefont {Kir\ifmmode~\check{s}\else \v{s}\fi{}anskas},
  \citenamefont {Francki\'e},\ and\ \citenamefont {Wacker}}]{perlind}%
  \BibitemOpen
  \bibfield  {author} {\bibinfo {author} {\bibfnamefont {G.}~\bibnamefont
  {Kir\ifmmode~\check{s}\else \v{s}\fi{}anskas}}, \bibinfo {author}
  {\bibfnamefont {M.}~\bibnamefont {Francki\'e}}, \ and\ \bibinfo {author}
  {\bibfnamefont {A.}~\bibnamefont {Wacker}},\ }\href {\doibase
  10.1103/PhysRevB.97.035432} {\bibfield  {journal} {\bibinfo  {journal} {Phys.
  Rev. B}\ }\textbf {\bibinfo {volume} {97}},\ \bibinfo {pages} {035432}
  (\bibinfo {year} {2018})}\BibitemShut {NoStop}%
\bibitem [{\citenamefont {Lee}\ and\ \citenamefont
  {Yeo}(2020)}]{lee2020comment}%
  \BibitemOpen
  \bibfield  {author} {\bibinfo {author} {\bibfnamefont {J.~S.}\ \bibnamefont
  {Lee}}\ and\ \bibinfo {author} {\bibfnamefont {J.}~\bibnamefont {Yeo}},\
  }\href@noop {} {\enquote {\bibinfo {title} {Comment on "universal lindblad
  equation for open quantum systems"},}\ } (\bibinfo {year} {2020}),\ \Eprint
  {http://arxiv.org/abs/2011.00735} {arXiv:2011.00735 [quant-ph]} \BibitemShut
  {NoStop}%
\bibitem [{\citenamefont {Leggett}\ \emph {et~al.}(1987)\citenamefont
  {Leggett}, \citenamefont {Chakravarty}, \citenamefont {Dorsey}, \citenamefont
  {Fisher}, \citenamefont {Garg},\ and\ \citenamefont {Zwerger}}]{Leggett}%
  \BibitemOpen
  \bibfield  {author} {\bibinfo {author} {\bibfnamefont {A.~J.}\ \bibnamefont
  {Leggett}}, \bibinfo {author} {\bibfnamefont {S.}~\bibnamefont
  {Chakravarty}}, \bibinfo {author} {\bibfnamefont {A.~T.}\ \bibnamefont
  {Dorsey}}, \bibinfo {author} {\bibfnamefont {M.~P.~A.}\ \bibnamefont
  {Fisher}}, \bibinfo {author} {\bibfnamefont {A.}~\bibnamefont {Garg}}, \ and\
  \bibinfo {author} {\bibfnamefont {W.}~\bibnamefont {Zwerger}},\ }\href
  {\doibase 10.1103/RevModPhys.59.1} {\bibfield  {journal} {\bibinfo  {journal}
  {Rev. Mod. Phys.}\ }\textbf {\bibinfo {volume} {59}},\ \bibinfo {pages} {1}
  (\bibinfo {year} {1987})}\BibitemShut {NoStop}%
\bibitem [{\citenamefont {Hur}(2008)}]{HUR20082208}%
  \BibitemOpen
  \bibfield  {author} {\bibinfo {author} {\bibfnamefont {K.~L.}\ \bibnamefont
  {Hur}},\ }\href {\doibase https://doi.org/10.1016/j.aop.2007.12.003}
  {\bibfield  {journal} {\bibinfo  {journal} {Annals of Physics}\ }\textbf
  {\bibinfo {volume} {323}},\ \bibinfo {pages} {2208} (\bibinfo {year}
  {2008})}\BibitemShut {NoStop}%
\bibitem [{\citenamefont {Feynman}\ and\ \citenamefont
  {Vernon}(1963)}]{FEYNMAN1963118}%
  \BibitemOpen
  \bibfield  {author} {\bibinfo {author} {\bibfnamefont {R.}~\bibnamefont
  {Feynman}}\ and\ \bibinfo {author} {\bibfnamefont {F.}~\bibnamefont
  {Vernon}},\ }\href {\doibase https://doi.org/10.1016/0003-4916(63)90068-X}
  {\bibfield  {journal} {\bibinfo  {journal} {Annals of Physics}\ }\textbf
  {\bibinfo {volume} {24}},\ \bibinfo {pages} {118} (\bibinfo {year}
  {1963})}\BibitemShut {NoStop}%
\bibitem [{\citenamefont {Makri}\ and\ \citenamefont
  {Makarov}(1995{\natexlab{a}})}]{Makri1}%
  \BibitemOpen
  \bibfield  {author} {\bibinfo {author} {\bibfnamefont {N.}~\bibnamefont
  {Makri}}\ and\ \bibinfo {author} {\bibfnamefont {D.~E.}\ \bibnamefont
  {Makarov}},\ }\href {\doibase 10.1063/1.469508} {\bibfield  {journal}
  {\bibinfo  {journal} {The Journal of Chemical Physics}\ }\textbf {\bibinfo
  {volume} {102}},\ \bibinfo {pages} {4600} (\bibinfo {year}
  {1995}{\natexlab{a}})},\ \Eprint
  {http://arxiv.org/abs/https://doi.org/10.1063/1.469508}
  {https://doi.org/10.1063/1.469508} \BibitemShut {NoStop}%
\bibitem [{\citenamefont {Makri}\ and\ \citenamefont
  {Makarov}(1995{\natexlab{b}})}]{Makri2}%
  \BibitemOpen
  \bibfield  {author} {\bibinfo {author} {\bibfnamefont {N.}~\bibnamefont
  {Makri}}\ and\ \bibinfo {author} {\bibfnamefont {D.~E.}\ \bibnamefont
  {Makarov}},\ }\href {\doibase 10.1063/1.469509} {\bibfield  {journal}
  {\bibinfo  {journal} {The Journal of Chemical Physics}\ }\textbf {\bibinfo
  {volume} {102}},\ \bibinfo {pages} {4611} (\bibinfo {year}
  {1995}{\natexlab{b}})},\ \Eprint
  {http://arxiv.org/abs/https://doi.org/10.1063/1.469509}
  {https://doi.org/10.1063/1.469509} \BibitemShut {NoStop}%
\bibitem [{\citenamefont {Thorwart}\ \emph {et~al.}(2004)\citenamefont
  {Thorwart}, \citenamefont {Paladino},\ and\ \citenamefont
  {Grifoni}}]{THORWART2004333}%
  \BibitemOpen
  \bibfield  {author} {\bibinfo {author} {\bibfnamefont {M.}~\bibnamefont
  {Thorwart}}, \bibinfo {author} {\bibfnamefont {E.}~\bibnamefont {Paladino}},
  \ and\ \bibinfo {author} {\bibfnamefont {M.}~\bibnamefont {Grifoni}},\ }\href
  {\doibase https://doi.org/10.1016/j.chemphys.2003.10.007} {\bibfield
  {journal} {\bibinfo  {journal} {Chemical Physics}\ }\textbf {\bibinfo
  {volume} {296}},\ \bibinfo {pages} {333 } (\bibinfo {year} {2004})},\
  \bibinfo {note} {the Spin-Boson Problem: From Electron Transfer to Quantum
  Computing ... to the 60th Birthday of Professor Ulrich Weiss}\BibitemShut
  {NoStop}%
\bibitem [{\citenamefont {Nalbach}\ and\ \citenamefont
  {Thorwart}(2010)}]{Nalbach}%
  \BibitemOpen
  \bibfield  {author} {\bibinfo {author} {\bibfnamefont {P.}~\bibnamefont
  {Nalbach}}\ and\ \bibinfo {author} {\bibfnamefont {M.}~\bibnamefont
  {Thorwart}},\ }\href {\doibase 10.1103/PhysRevB.81.054308} {\bibfield
  {journal} {\bibinfo  {journal} {Phys. Rev. B}\ }\textbf {\bibinfo {volume}
  {81}},\ \bibinfo {pages} {054308} (\bibinfo {year} {2010})}\BibitemShut
  {NoStop}%
\bibitem [{\citenamefont {M\"uhlbacher}\ \emph {et~al.}(2005)\citenamefont
  {M\"uhlbacher}, \citenamefont {Ankerhold},\ and\ \citenamefont
  {Komnik}}]{Komnik}%
  \BibitemOpen
  \bibfield  {author} {\bibinfo {author} {\bibfnamefont {L.}~\bibnamefont
  {M\"uhlbacher}}, \bibinfo {author} {\bibfnamefont {J.}~\bibnamefont
  {Ankerhold}}, \ and\ \bibinfo {author} {\bibfnamefont {A.}~\bibnamefont
  {Komnik}},\ }\href {\doibase 10.1103/PhysRevLett.95.220404} {\bibfield
  {journal} {\bibinfo  {journal} {Phys. Rev. Lett.}\ }\textbf {\bibinfo
  {volume} {95}},\ \bibinfo {pages} {220404} (\bibinfo {year}
  {2005})}\BibitemShut {NoStop}%
\bibitem [{\citenamefont {Chin}\ \emph {et~al.}(2010)\citenamefont {Chin},
  \citenamefont {Rivas}, \citenamefont {Huelga},\ and\ \citenamefont
  {Plenio}}]{Chin}%
  \BibitemOpen
  \bibfield  {author} {\bibinfo {author} {\bibfnamefont {A.~W.}\ \bibnamefont
  {Chin}}, \bibinfo {author} {\bibfnamefont {n.}~\bibnamefont {Rivas}},
  \bibinfo {author} {\bibfnamefont {S.~F.}\ \bibnamefont {Huelga}}, \ and\
  \bibinfo {author} {\bibfnamefont {M.~B.}\ \bibnamefont {Plenio}},\ }\href
  {\doibase 10.1063/1.3490188} {\bibfield  {journal} {\bibinfo  {journal}
  {Journal of Mathematical Physics}\ }\textbf {\bibinfo {volume} {51}},\
  \bibinfo {pages} {092109} (\bibinfo {year} {2010})},\ \Eprint
  {http://arxiv.org/abs/https://doi.org/10.1063/1.3490188}
  {https://doi.org/10.1063/1.3490188} \BibitemShut {NoStop}%
\bibitem [{\citenamefont {Prior}\ \emph {et~al.}(2010)\citenamefont {Prior},
  \citenamefont {Chin}, \citenamefont {Huelga},\ and\ \citenamefont
  {Plenio}}]{Prior}%
  \BibitemOpen
  \bibfield  {author} {\bibinfo {author} {\bibfnamefont {J.}~\bibnamefont
  {Prior}}, \bibinfo {author} {\bibfnamefont {A.~W.}\ \bibnamefont {Chin}},
  \bibinfo {author} {\bibfnamefont {S.~F.}\ \bibnamefont {Huelga}}, \ and\
  \bibinfo {author} {\bibfnamefont {M.~B.}\ \bibnamefont {Plenio}},\ }\href
  {\doibase 10.1103/PhysRevLett.105.050404} {\bibfield  {journal} {\bibinfo
  {journal} {Phys. Rev. Lett.}\ }\textbf {\bibinfo {volume} {105}},\ \bibinfo
  {pages} {050404} (\bibinfo {year} {2010})}\BibitemShut {NoStop}%
\bibitem [{\citenamefont {Strathearn}(2020)}]{Strathearn2019}%
  \BibitemOpen
  \bibfield  {author} {\bibinfo {author} {\bibfnamefont {A.}~\bibnamefont
  {Strathearn}},\ }\href {\doibase 10.1007/978-3-030-54975-6} {\emph {\bibinfo
  {title} {{Modelling Non-Markovian Quantum Systems Using Tensor Networks}}}},\
  Springer Theses\ (\bibinfo  {publisher} {Springer International Publishing},\
  \bibinfo {address} {Cham},\ \bibinfo {year} {2020})\BibitemShut {NoStop}%
\bibitem [{\citenamefont {Schröder}\ \emph {et~al.}(2019)\citenamefont
  {Schröder}, \citenamefont {Turban}, \citenamefont {Musser}, \citenamefont
  {Hine},\ and\ \citenamefont {Chin}}]{Florian}%
  \BibitemOpen
  \bibfield  {author} {\bibinfo {author} {\bibfnamefont {F.~A. Y.~N.}\
  \bibnamefont {Schröder}}, \bibinfo {author} {\bibfnamefont {D.~H.~P.}\
  \bibnamefont {Turban}}, \bibinfo {author} {\bibfnamefont {A.~J.}\
  \bibnamefont {Musser}}, \bibinfo {author} {\bibfnamefont {N.~D.~M.}\
  \bibnamefont {Hine}}, \ and\ \bibinfo {author} {\bibfnamefont {A.~W.}\
  \bibnamefont {Chin}},\ }\href {\doibase
  https://doi.org/10.1038/s41467-019-09039-7} {\bibfield  {journal} {\bibinfo
  {journal} {Nature Communications}\ }\textbf {\bibinfo {volume} {10}},\
  \bibinfo {pages} {1062} (\bibinfo {year} {2019})}\BibitemShut {NoStop}%
\bibitem [{\citenamefont {Wall}\ \emph {et~al.}(2016)\citenamefont {Wall},
  \citenamefont {Safavi-Naini},\ and\ \citenamefont {Rey}}]{Maria}%
  \BibitemOpen
  \bibfield  {author} {\bibinfo {author} {\bibfnamefont {M.~L.}\ \bibnamefont
  {Wall}}, \bibinfo {author} {\bibfnamefont {A.}~\bibnamefont {Safavi-Naini}},
  \ and\ \bibinfo {author} {\bibfnamefont {A.~M.}\ \bibnamefont {Rey}},\ }\href
  {\doibase 10.1103/PhysRevA.94.053637} {\bibfield  {journal} {\bibinfo
  {journal} {Phys. Rev. A}\ }\textbf {\bibinfo {volume} {94}},\ \bibinfo
  {pages} {053637} (\bibinfo {year} {2016})}\BibitemShut {NoStop}%
\bibitem [{\citenamefont {Pollock}\ \emph {et~al.}(2018)\citenamefont
  {Pollock}, \citenamefont {Rodr\'{\i}guez-Rosario}, \citenamefont
  {Frauenheim}, \citenamefont {Paternostro},\ and\ \citenamefont
  {Modi}}]{Modi}%
  \BibitemOpen
  \bibfield  {author} {\bibinfo {author} {\bibfnamefont {F.~A.}\ \bibnamefont
  {Pollock}}, \bibinfo {author} {\bibfnamefont {C.}~\bibnamefont
  {Rodr\'{\i}guez-Rosario}}, \bibinfo {author} {\bibfnamefont {T.}~\bibnamefont
  {Frauenheim}}, \bibinfo {author} {\bibfnamefont {M.}~\bibnamefont
  {Paternostro}}, \ and\ \bibinfo {author} {\bibfnamefont {K.}~\bibnamefont
  {Modi}},\ }\href {\doibase 10.1103/PhysRevA.97.012127} {\bibfield  {journal}
  {\bibinfo  {journal} {Phys. Rev. A}\ }\textbf {\bibinfo {volume} {97}},\
  \bibinfo {pages} {012127} (\bibinfo {year} {2018})}\BibitemShut {NoStop}%
\bibitem [{\citenamefont {Luchnikov}\ \emph {et~al.}(2019)\citenamefont
  {Luchnikov}, \citenamefont {Vintskevich}, \citenamefont {Ouerdane},\ and\
  \citenamefont {Filippov}}]{Filippov}%
  \BibitemOpen
  \bibfield  {author} {\bibinfo {author} {\bibfnamefont {I.~A.}\ \bibnamefont
  {Luchnikov}}, \bibinfo {author} {\bibfnamefont {S.~V.}\ \bibnamefont
  {Vintskevich}}, \bibinfo {author} {\bibfnamefont {H.}~\bibnamefont
  {Ouerdane}}, \ and\ \bibinfo {author} {\bibfnamefont {S.~N.}\ \bibnamefont
  {Filippov}},\ }\href {\doibase 10.1103/PhysRevLett.122.160401} {\bibfield
  {journal} {\bibinfo  {journal} {Phys. Rev. Lett.}\ }\textbf {\bibinfo
  {volume} {122}},\ \bibinfo {pages} {160401} (\bibinfo {year}
  {2019})}\BibitemShut {NoStop}%
\bibitem [{\citenamefont {Tanimura}\ and\ \citenamefont
  {Kubo}(1989)}]{Tanimura}%
  \BibitemOpen
  \bibfield  {author} {\bibinfo {author} {\bibfnamefont {Y.}~\bibnamefont
  {Tanimura}}\ and\ \bibinfo {author} {\bibfnamefont {R.}~\bibnamefont
  {Kubo}},\ }\href {\doibase 10.1143/JPSJ.58.1199} {\bibfield  {journal}
  {\bibinfo  {journal} {Journal of the Physical Society of Japan}\ }\textbf
  {\bibinfo {volume} {58}},\ \bibinfo {pages} {1199} (\bibinfo {year}
  {1989})},\ \Eprint
  {http://arxiv.org/abs/https://doi.org/10.1143/JPSJ.58.1199}
  {https://doi.org/10.1143/JPSJ.58.1199} \BibitemShut {NoStop}%
\bibitem [{\citenamefont {Tanimura}(2006)}]{Tanimura1}%
  \BibitemOpen
  \bibfield  {author} {\bibinfo {author} {\bibfnamefont {Y.}~\bibnamefont
  {Tanimura}},\ }\href {\doibase 10.1143/JPSJ.75.082001} {\bibfield  {journal}
  {\bibinfo  {journal} {Journal of the Physical Society of Japan}\ }\textbf
  {\bibinfo {volume} {75}},\ \bibinfo {pages} {082001} (\bibinfo {year}
  {2006})},\ \Eprint
  {http://arxiv.org/abs/https://doi.org/10.1143/JPSJ.75.082001}
  {https://doi.org/10.1143/JPSJ.75.082001} \BibitemShut {NoStop}%
\bibitem [{\citenamefont {Tanimura}(2014)}]{Tanimura2}%
  \BibitemOpen
  \bibfield  {author} {\bibinfo {author} {\bibfnamefont {Y.}~\bibnamefont
  {Tanimura}},\ }\href {\doibase 10.1063/1.4890441} {\bibfield  {journal}
  {\bibinfo  {journal} {The Journal of Chemical Physics}\ }\textbf {\bibinfo
  {volume} {141}},\ \bibinfo {pages} {044114} (\bibinfo {year} {2014})},\
  \Eprint {http://arxiv.org/abs/https://doi.org/10.1063/1.4890441}
  {https://doi.org/10.1063/1.4890441} \BibitemShut {NoStop}%
\bibitem [{\citenamefont {Li}\ \emph {et~al.}(2012)\citenamefont {Li},
  \citenamefont {Tong}, \citenamefont {Zheng}, \citenamefont {Hou},
  \citenamefont {Wei}, \citenamefont {Hu},\ and\ \citenamefont
  {Yan}}]{ZhenHua}%
  \BibitemOpen
  \bibfield  {author} {\bibinfo {author} {\bibfnamefont {Z.}~\bibnamefont
  {Li}}, \bibinfo {author} {\bibfnamefont {N.}~\bibnamefont {Tong}}, \bibinfo
  {author} {\bibfnamefont {X.}~\bibnamefont {Zheng}}, \bibinfo {author}
  {\bibfnamefont {D.}~\bibnamefont {Hou}}, \bibinfo {author} {\bibfnamefont
  {J.}~\bibnamefont {Wei}}, \bibinfo {author} {\bibfnamefont {J.}~\bibnamefont
  {Hu}}, \ and\ \bibinfo {author} {\bibfnamefont {Y.}~\bibnamefont {Yan}},\
  }\href {\doibase 10.1103/PhysRevLett.109.266403} {\bibfield  {journal}
  {\bibinfo  {journal} {Phys. Rev. Lett.}\ }\textbf {\bibinfo {volume} {109}},\
  \bibinfo {pages} {266403} (\bibinfo {year} {2012})}\BibitemShut {NoStop}%
\bibitem [{\citenamefont {Cheng}\ \emph {et~al.}(2015)\citenamefont {Cheng},
  \citenamefont {Hou}, \citenamefont {Wang}, \citenamefont {Li}, \citenamefont
  {Wei},\ and\ \citenamefont {Yan}}]{Cheng_2015}%
  \BibitemOpen
  \bibfield  {author} {\bibinfo {author} {\bibfnamefont {Y.}~\bibnamefont
  {Cheng}}, \bibinfo {author} {\bibfnamefont {W.}~\bibnamefont {Hou}}, \bibinfo
  {author} {\bibfnamefont {Y.}~\bibnamefont {Wang}}, \bibinfo {author}
  {\bibfnamefont {Z.}~\bibnamefont {Li}}, \bibinfo {author} {\bibfnamefont
  {J.}~\bibnamefont {Wei}}, \ and\ \bibinfo {author} {\bibfnamefont
  {Y.}~\bibnamefont {Yan}},\ }\href {\doibase 10.1088/1367-2630/17/3/033009}
  {\bibfield  {journal} {\bibinfo  {journal} {New Journal of Physics}\ }\textbf
  {\bibinfo {volume} {17}},\ \bibinfo {pages} {033009} (\bibinfo {year}
  {2015})}\BibitemShut {NoStop}%
\bibitem [{\citenamefont {Meyer}\ \emph {et~al.}(1990)\citenamefont {Meyer},
  \citenamefont {Manthe},\ and\ \citenamefont {Cederbaum}}]{MEYER199073}%
  \BibitemOpen
  \bibfield  {author} {\bibinfo {author} {\bibfnamefont {H.-D.}\ \bibnamefont
  {Meyer}}, \bibinfo {author} {\bibfnamefont {U.}~\bibnamefont {Manthe}}, \
  and\ \bibinfo {author} {\bibfnamefont {L.}~\bibnamefont {Cederbaum}},\ }\href
  {\doibase https://doi.org/10.1016/0009-2614(90)87014-I} {\bibfield  {journal}
  {\bibinfo  {journal} {Chemical Physics Letters}\ }\textbf {\bibinfo {volume}
  {165}},\ \bibinfo {pages} {73 } (\bibinfo {year} {1990})}\BibitemShut
  {NoStop}%
\bibitem [{\citenamefont {Beck}\ \emph {et~al.}(2000)\citenamefont {Beck},
  \citenamefont {Jäckle}, \citenamefont {Worth},\ and\ \citenamefont
  {Meyer}}]{BECK20001}%
  \BibitemOpen
  \bibfield  {author} {\bibinfo {author} {\bibfnamefont {M.}~\bibnamefont
  {Beck}}, \bibinfo {author} {\bibfnamefont {A.}~\bibnamefont {Jäckle}},
  \bibinfo {author} {\bibfnamefont {G.}~\bibnamefont {Worth}}, \ and\ \bibinfo
  {author} {\bibfnamefont {H.-D.}\ \bibnamefont {Meyer}},\ }\href {\doibase
  https://doi.org/10.1016/S0370-1573(99)00047-2} {\bibfield  {journal}
  {\bibinfo  {journal} {Physics Reports}\ }\textbf {\bibinfo {volume} {324}},\
  \bibinfo {pages} {1 } (\bibinfo {year} {2000})}\BibitemShut {NoStop}%
\bibitem [{\citenamefont {Wang}\ and\ \citenamefont {Thoss}(2003)}]{Haobin}%
  \BibitemOpen
  \bibfield  {author} {\bibinfo {author} {\bibfnamefont {H.}~\bibnamefont
  {Wang}}\ and\ \bibinfo {author} {\bibfnamefont {M.}~\bibnamefont {Thoss}},\
  }\href {\doibase 10.1063/1.1580111} {\bibfield  {journal} {\bibinfo
  {journal} {The Journal of Chemical Physics}\ }\textbf {\bibinfo {volume}
  {119}},\ \bibinfo {pages} {1289} (\bibinfo {year} {2003})},\ \Eprint
  {http://arxiv.org/abs/https://doi.org/10.1063/1.1580111}
  {https://doi.org/10.1063/1.1580111} \BibitemShut {NoStop}%
\bibitem [{\citenamefont {Zheng}\ \emph {et~al.}(2016)\citenamefont {Zheng},
  \citenamefont {Xie}, \citenamefont {Jiang},\ and\ \citenamefont {Lan}}]{Jie}%
  \BibitemOpen
  \bibfield  {author} {\bibinfo {author} {\bibfnamefont {J.}~\bibnamefont
  {Zheng}}, \bibinfo {author} {\bibfnamefont {Y.}~\bibnamefont {Xie}}, \bibinfo
  {author} {\bibfnamefont {S.}~\bibnamefont {Jiang}}, \ and\ \bibinfo {author}
  {\bibfnamefont {Z.}~\bibnamefont {Lan}},\ }\href {\doibase
  10.1021/acs.jpcc.5b09921} {\bibfield  {journal} {\bibinfo  {journal} {The
  Journal of Physical Chemistry C}\ }\textbf {\bibinfo {volume} {120}},\
  \bibinfo {pages} {1375} (\bibinfo {year} {2016})},\ \Eprint
  {http://arxiv.org/abs/https://doi.org/10.1021/acs.jpcc.5b09921}
  {https://doi.org/10.1021/acs.jpcc.5b09921} \BibitemShut {NoStop}%
\bibitem [{\citenamefont {Nakajima}(1958)}]{Nakajima}%
  \BibitemOpen
  \bibfield  {author} {\bibinfo {author} {\bibfnamefont {S.}~\bibnamefont
  {Nakajima}},\ }\href {\doibase 10.1143/PTP.20.948} {\bibfield  {journal}
  {\bibinfo  {journal} {Progress of Theoretical Physics}\ }\textbf {\bibinfo
  {volume} {20}},\ \bibinfo {pages} {948} (\bibinfo {year} {1958})},\ \Eprint
  {http://arxiv.org/abs/https://academic.oup.com/ptp/article-pdf/20/6/948/5440766/20-6-948.pdf}
  {https://academic.oup.com/ptp/article-pdf/20/6/948/5440766/20-6-948.pdf}
  \BibitemShut {NoStop}%
\bibitem [{\citenamefont {Zwanzig}(1960)}]{Zwanzig}%
  \BibitemOpen
  \bibfield  {author} {\bibinfo {author} {\bibfnamefont {R.}~\bibnamefont
  {Zwanzig}},\ }\href {\doibase 10.1063/1.1731409} {\bibfield  {journal}
  {\bibinfo  {journal} {The Journal of Chemical Physics}\ }\textbf {\bibinfo
  {volume} {33}},\ \bibinfo {pages} {1338} (\bibinfo {year} {1960})},\ \Eprint
  {http://arxiv.org/abs/https://doi.org/10.1063/1.1731409}
  {https://doi.org/10.1063/1.1731409} \BibitemShut {NoStop}%
\bibitem [{\citenamefont {Suess}\ \emph {et~al.}(2014)\citenamefont {Suess},
  \citenamefont {Eisfeld},\ and\ \citenamefont {Strunz}}]{Suess}%
  \BibitemOpen
  \bibfield  {author} {\bibinfo {author} {\bibfnamefont {D.}~\bibnamefont
  {Suess}}, \bibinfo {author} {\bibfnamefont {A.}~\bibnamefont {Eisfeld}}, \
  and\ \bibinfo {author} {\bibfnamefont {W.~T.}\ \bibnamefont {Strunz}},\
  }\href {\doibase 10.1103/PhysRevLett.113.150403} {\bibfield  {journal}
  {\bibinfo  {journal} {Phys. Rev. Lett.}\ }\textbf {\bibinfo {volume} {113}},\
  \bibinfo {pages} {150403} (\bibinfo {year} {2014})}\BibitemShut {NoStop}%
\bibitem [{\citenamefont {Zhang}\ and\ \citenamefont
  {Eisfeld}(2016)}]{Pan-Pan}%
  \BibitemOpen
  \bibfield  {author} {\bibinfo {author} {\bibfnamefont {P.-P.}\ \bibnamefont
  {Zhang}}\ and\ \bibinfo {author} {\bibfnamefont {A.}~\bibnamefont
  {Eisfeld}},\ }\href {\doibase 10.1021/acs.jpclett.6b02111} {\bibfield
  {journal} {\bibinfo  {journal} {The Journal of Physical Chemistry Letters}\
  }\textbf {\bibinfo {volume} {7}},\ \bibinfo {pages} {4488} (\bibinfo {year}
  {2016})},\ \bibinfo {note} {pMID: 27775345},\ \Eprint
  {http://arxiv.org/abs/https://doi.org/10.1021/acs.jpclett.6b02111}
  {https://doi.org/10.1021/acs.jpclett.6b02111} \BibitemShut {NoStop}%
\bibitem [{\citenamefont {Hartmann}\ and\ \citenamefont
  {Strunz}(2017)}]{Hartmann1}%
  \BibitemOpen
  \bibfield  {author} {\bibinfo {author} {\bibfnamefont {R.}~\bibnamefont
  {Hartmann}}\ and\ \bibinfo {author} {\bibfnamefont {W.~T.}\ \bibnamefont
  {Strunz}},\ }\href {\doibase 10.1021/acs.jctc.7b00751} {\bibfield  {journal}
  {\bibinfo  {journal} {Journal of Chemical Theory and Computation}\ }\textbf
  {\bibinfo {volume} {13}},\ \bibinfo {pages} {5834} (\bibinfo {year}
  {2017})},\ \bibinfo {note} {pMID: 29016126},\ \Eprint
  {http://arxiv.org/abs/https://doi.org/10.1021/acs.jctc.7b00751}
  {https://doi.org/10.1021/acs.jctc.7b00751} \BibitemShut {NoStop}%
\bibitem [{\citenamefont {Orús}(2014)}]{ORUS2014117}%
  \BibitemOpen
  \bibfield  {author} {\bibinfo {author} {\bibfnamefont {R.}~\bibnamefont
  {Orús}},\ }\href {\doibase https://doi.org/10.1016/j.aop.2014.06.013}
  {\bibfield  {journal} {\bibinfo  {journal} {Annals of Physics}\ }\textbf
  {\bibinfo {volume} {349}},\ \bibinfo {pages} {117} (\bibinfo {year}
  {2014})}\BibitemShut {NoStop}%
\bibitem [{\citenamefont {Cirac}\ \emph {et~al.}(2020)\citenamefont {Cirac},
  \citenamefont {Perez-Garcia}, \citenamefont {Schuch},\ and\ \citenamefont
  {Verstraete}}]{cirac2020matrix}%
  \BibitemOpen
  \bibfield  {author} {\bibinfo {author} {\bibfnamefont {I.}~\bibnamefont
  {Cirac}}, \bibinfo {author} {\bibfnamefont {D.}~\bibnamefont {Perez-Garcia}},
  \bibinfo {author} {\bibfnamefont {N.}~\bibnamefont {Schuch}}, \ and\ \bibinfo
  {author} {\bibfnamefont {F.}~\bibnamefont {Verstraete}},\ }\href@noop {}
  {\enquote {\bibinfo {title} {Matrix product states and projected entangled
  pair states: Concepts, symmetries, and theorems},}\ } (\bibinfo {year}
  {2020}),\ \Eprint {http://arxiv.org/abs/2011.12127} {arXiv:2011.12127
  [quant-ph]} \BibitemShut {NoStop}%
\bibitem [{\citenamefont {J\o{}rgensen}\ and\ \citenamefont
  {Pollock}(2019)}]{Jorgensen2019}%
  \BibitemOpen
  \bibfield  {author} {\bibinfo {author} {\bibfnamefont {M.~R.}\ \bibnamefont
  {J\o{}rgensen}}\ and\ \bibinfo {author} {\bibfnamefont {F.~A.}\ \bibnamefont
  {Pollock}},\ }\href {\doibase 10.1103/PhysRevLett.123.240602} {\bibfield
  {journal} {\bibinfo  {journal} {Phys. Rev. Lett.}\ }\textbf {\bibinfo
  {volume} {123}},\ \bibinfo {pages} {240602} (\bibinfo {year}
  {2019})}\BibitemShut {NoStop}%
\bibitem [{\citenamefont {Fux}\ \emph {et~al.}(2021)\citenamefont {Fux},
  \citenamefont {Butler}, \citenamefont {Eastham}, \citenamefont {Lovett},\
  and\ \citenamefont {Keeling}}]{Fux2021}%
  \BibitemOpen
  \bibfield  {author} {\bibinfo {author} {\bibfnamefont {G.~E.}\ \bibnamefont
  {Fux}}, \bibinfo {author} {\bibfnamefont {E.}~\bibnamefont {Butler}},
  \bibinfo {author} {\bibfnamefont {P.~R.}\ \bibnamefont {Eastham}}, \bibinfo
  {author} {\bibfnamefont {B.~W.}\ \bibnamefont {Lovett}}, \ and\ \bibinfo
  {author} {\bibfnamefont {J.}~\bibnamefont {Keeling}},\ }\href
  {http://arxiv.org/abs/2101.03071} {\  (\bibinfo {year} {2021})},\ \Eprint
  {http://arxiv.org/abs/2101.03071} {arXiv:2101.03071} \BibitemShut {NoStop}%
\bibitem [{\citenamefont {Gribben}\ \emph {et~al.}()\citenamefont {Gribben},
  \citenamefont {Fux}, \citenamefont {Strathearn}, \citenamefont {Kirton},\
  and\ \citenamefont {Lovett}}]{Gribben2021}%
  \BibitemOpen
  \bibfield  {author} {\bibinfo {author} {\bibfnamefont {D.}~\bibnamefont
  {Gribben}}, \bibinfo {author} {\bibfnamefont {G.~E.}\ \bibnamefont {Fux}},
  \bibinfo {author} {\bibfnamefont {A.}~\bibnamefont {Strathearn}}, \bibinfo
  {author} {\bibfnamefont {P.}~\bibnamefont {Kirton}}, \ and\ \bibinfo {author}
  {\bibfnamefont {B.~W.}\ \bibnamefont {Lovett}},\ }\href@noop {} {\
  }\BibitemShut {NoStop}%
\bibitem [{\citenamefont {{The TEMPO collaboration}}(2020)}]{TimeEvolvingMPO}%
  \BibitemOpen
  \bibfield  {author} {\bibinfo {author} {\bibnamefont {{The TEMPO
  collaboration}}},\ }\href {\doibase 10.5281/zenodo.4428316} {\enquote
  {\bibinfo {title} {{TimeEvolvingMPO: A Python 3 package to efficiently
  compute non-Markovian open quantum systems.}}}\ } (\bibinfo {year}
  {2020})\BibitemShut {NoStop}%
\bibitem [{\citenamefont {Braun}(2002)}]{Braun}%
  \BibitemOpen
  \bibfield  {author} {\bibinfo {author} {\bibfnamefont {D.}~\bibnamefont
  {Braun}},\ }\href {\doibase 10.1103/PhysRevLett.89.277901} {\bibfield
  {journal} {\bibinfo  {journal} {Phys. Rev. Lett.}\ }\textbf {\bibinfo
  {volume} {89}},\ \bibinfo {pages} {277901} (\bibinfo {year}
  {2002})}\BibitemShut {NoStop}%
\bibitem [{\citenamefont {McCutcheon}\ \emph {et~al.}(2009)\citenamefont
  {McCutcheon}, \citenamefont {Nazir}, \citenamefont {Bose},\ and\
  \citenamefont {Fisher}}]{McCutcheon_2009}%
  \BibitemOpen
  \bibfield  {author} {\bibinfo {author} {\bibfnamefont {D.~P.~S.}\
  \bibnamefont {McCutcheon}}, \bibinfo {author} {\bibfnamefont
  {A.}~\bibnamefont {Nazir}}, \bibinfo {author} {\bibfnamefont
  {S.}~\bibnamefont {Bose}}, \ and\ \bibinfo {author} {\bibfnamefont {A.~J.}\
  \bibnamefont {Fisher}},\ }\href {\doibase 10.1103/physreva.80.022337}
  {\bibfield  {journal} {\bibinfo  {journal} {Physical Review A}\ }\textbf
  {\bibinfo {volume} {80}} (\bibinfo {year} {2009}),\
  10.1103/physreva.80.022337}\BibitemShut {NoStop}%
\bibitem [{\citenamefont {Benatti}\ \emph {et~al.}(2009)\citenamefont
  {Benatti}, \citenamefont {Floreanini},\ and\ \citenamefont
  {Marzolino}}]{Benatti_2009}%
  \BibitemOpen
  \bibfield  {author} {\bibinfo {author} {\bibfnamefont {F.}~\bibnamefont
  {Benatti}}, \bibinfo {author} {\bibfnamefont {R.}~\bibnamefont {Floreanini}},
  \ and\ \bibinfo {author} {\bibfnamefont {U.}~\bibnamefont {Marzolino}},\
  }\href {\doibase 10.1209/0295-5075/88/20011} {\bibfield  {journal} {\bibinfo
  {journal} {{EPL} (Europhysics Letters)}\ }\textbf {\bibinfo {volume} {88}},\
  \bibinfo {pages} {20011} (\bibinfo {year} {2009})}\BibitemShut {NoStop}%
\bibitem [{\citenamefont {Kast}\ and\ \citenamefont
  {Ankerhold}(2014)}]{Kast_2014}%
  \BibitemOpen
  \bibfield  {author} {\bibinfo {author} {\bibfnamefont {D.}~\bibnamefont
  {Kast}}\ and\ \bibinfo {author} {\bibfnamefont {J.}~\bibnamefont
  {Ankerhold}},\ }\href {\doibase 10.1103/PhysRevB.90.100301} {\bibfield
  {journal} {\bibinfo  {journal} {Phys. Rev. B}\ }\textbf {\bibinfo {volume}
  {90}},\ \bibinfo {pages} {100301} (\bibinfo {year} {2014})}\BibitemShut
  {NoStop}%
\bibitem [{\citenamefont {Sahrapour}\ and\ \citenamefont
  {Makri}(2013)}]{Makri2013}%
  \BibitemOpen
  \bibfield  {author} {\bibinfo {author} {\bibfnamefont {M.~M.}\ \bibnamefont
  {Sahrapour}}\ and\ \bibinfo {author} {\bibfnamefont {N.}~\bibnamefont
  {Makri}},\ }\href {\doibase 10.1063/1.4795159} {\bibfield  {journal}
  {\bibinfo  {journal} {The Journal of Chemical Physics}\ }\textbf {\bibinfo
  {volume} {138}},\ \bibinfo {pages} {114109} (\bibinfo {year} {2013})},\
  \Eprint {http://arxiv.org/abs/https://doi.org/10.1063/1.4795159}
  {https://doi.org/10.1063/1.4795159} \BibitemShut {NoStop}%
\bibitem [{\citenamefont {Benatti}\ \emph {et~al.}(2010)\citenamefont
  {Benatti}, \citenamefont {Floreanini},\ and\ \citenamefont
  {Marzolino}}]{Benatti}%
  \BibitemOpen
  \bibfield  {author} {\bibinfo {author} {\bibfnamefont {F.}~\bibnamefont
  {Benatti}}, \bibinfo {author} {\bibfnamefont {R.}~\bibnamefont {Floreanini}},
  \ and\ \bibinfo {author} {\bibfnamefont {U.}~\bibnamefont {Marzolino}},\
  }\href {\doibase 10.1103/PhysRevA.81.012105} {\bibfield  {journal} {\bibinfo
  {journal} {Phys. Rev. A}\ }\textbf {\bibinfo {volume} {81}},\ \bibinfo
  {pages} {012105} (\bibinfo {year} {2010})}\BibitemShut {NoStop}%
\bibitem [{\citenamefont {Rivas}(2017)}]{Rivas}%
  \BibitemOpen
  \bibfield  {author} {\bibinfo {author} {\bibfnamefont {A.}~\bibnamefont
  {Rivas}},\ }\href {\doibase 10.1103/PhysRevA.95.042104} {\bibfield  {journal}
  {\bibinfo  {journal} {Phys. Rev. A}\ }\textbf {\bibinfo {volume} {95}},\
  \bibinfo {pages} {042104} (\bibinfo {year} {2017})}\BibitemShut {NoStop}%
\bibitem [{\citenamefont {Hartmann}\ and\ \citenamefont
  {Strunz}(2020{\natexlab{b}})}]{Hartmann}%
  \BibitemOpen
  \bibfield  {author} {\bibinfo {author} {\bibfnamefont {R.}~\bibnamefont
  {Hartmann}}\ and\ \bibinfo {author} {\bibfnamefont {W.~T.}\ \bibnamefont
  {Strunz}},\ }\href {\doibase 10.1103/PhysRevA.101.012103} {\bibfield
  {journal} {\bibinfo  {journal} {Phys. Rev. A}\ }\textbf {\bibinfo {volume}
  {101}},\ \bibinfo {pages} {012103} (\bibinfo {year}
  {2020}{\natexlab{b}})}\BibitemShut {NoStop}%
\bibitem [{\citenamefont {Zanardi}\ and\ \citenamefont
  {Rasetti}(1997)}]{Zanardi_1997}%
  \BibitemOpen
  \bibfield  {author} {\bibinfo {author} {\bibfnamefont {P.}~\bibnamefont
  {Zanardi}}\ and\ \bibinfo {author} {\bibfnamefont {M.}~\bibnamefont
  {Rasetti}},\ }\href {\doibase 10.1103/physrevlett.79.3306} {\bibfield
  {journal} {\bibinfo  {journal} {Physical Review Letters}\ }\textbf {\bibinfo
  {volume} {79}},\ \bibinfo {pages} {3306–3309} (\bibinfo {year}
  {1997})}\BibitemShut {NoStop}%
\bibitem [{\citenamefont {Lidar}\ \emph {et~al.}(1998)\citenamefont {Lidar},
  \citenamefont {Chuang},\ and\ \citenamefont {Whaley}}]{Lidar_1998}%
  \BibitemOpen
  \bibfield  {author} {\bibinfo {author} {\bibfnamefont {D.~A.}\ \bibnamefont
  {Lidar}}, \bibinfo {author} {\bibfnamefont {I.~L.}\ \bibnamefont {Chuang}}, \
  and\ \bibinfo {author} {\bibfnamefont {K.~B.}\ \bibnamefont {Whaley}},\
  }\href {\doibase 10.1103/physrevlett.81.2594} {\bibfield  {journal} {\bibinfo
   {journal} {Physical Review Letters}\ }\textbf {\bibinfo {volume} {81}},\
  \bibinfo {pages} {2594–2597} (\bibinfo {year} {1998})}\BibitemShut
  {NoStop}%
\bibitem [{\citenamefont {Deutsch}(2018)}]{Deutsch_2018}%
  \BibitemOpen
  \bibfield  {author} {\bibinfo {author} {\bibfnamefont {J.~M.}\ \bibnamefont
  {Deutsch}},\ }\href {\doibase 10.1088/1361-6633/aac9f1} {\bibfield  {journal}
  {\bibinfo  {journal} {Reports on Progress in Physics}\ }\textbf {\bibinfo
  {volume} {81}},\ \bibinfo {pages} {082001} (\bibinfo {year}
  {2018})}\BibitemShut {NoStop}%
\bibitem [{\citenamefont {Garrison}\ and\ \citenamefont
  {Grover}(2018)}]{Tarun}%
  \BibitemOpen
  \bibfield  {author} {\bibinfo {author} {\bibfnamefont {J.~R.}\ \bibnamefont
  {Garrison}}\ and\ \bibinfo {author} {\bibfnamefont {T.}~\bibnamefont
  {Grover}},\ }\href {\doibase 10.1103/PhysRevX.8.021026} {\bibfield  {journal}
  {\bibinfo  {journal} {Phys. Rev. X}\ }\textbf {\bibinfo {volume} {8}},\
  \bibinfo {pages} {021026} (\bibinfo {year} {2018})}\BibitemShut {NoStop}%
\bibitem [{\citenamefont {Havel}(2003)}]{Havel_2003}%
  \BibitemOpen
  \bibfield  {author} {\bibinfo {author} {\bibfnamefont {T.~F.}\ \bibnamefont
  {Havel}},\ }\href {\doibase 10.1063/1.1518555} {\bibfield  {journal}
  {\bibinfo  {journal} {Journal of Mathematical Physics}\ }\textbf {\bibinfo
  {volume} {44}},\ \bibinfo {pages} {534} (\bibinfo {year} {2003})}\BibitemShut
  {NoStop}%
\bibitem [{\citenamefont {Breuer}\ \emph {et~al.}(1999)\citenamefont {Breuer},
  \citenamefont {Kappler},\ and\ \citenamefont {Petruccione}}]{Breuer_1999}%
  \BibitemOpen
  \bibfield  {author} {\bibinfo {author} {\bibfnamefont {H.-P.}\ \bibnamefont
  {Breuer}}, \bibinfo {author} {\bibfnamefont {B.}~\bibnamefont {Kappler}}, \
  and\ \bibinfo {author} {\bibfnamefont {F.}~\bibnamefont {Petruccione}},\
  }\href {\doibase 10.1103/physreva.59.1633} {\bibfield  {journal} {\bibinfo
  {journal} {Physical Review A}\ }\textbf {\bibinfo {volume} {59}},\ \bibinfo
  {pages} {1633–1643} (\bibinfo {year} {1999})}\BibitemShut {NoStop}%
\bibitem [{\citenamefont {Alicki}(1989)}]{Alicki1}%
  \BibitemOpen
  \bibfield  {author} {\bibinfo {author} {\bibfnamefont {R.}~\bibnamefont
  {Alicki}},\ }\href {\doibase 10.1103/PhysRevA.40.4077} {\bibfield  {journal}
  {\bibinfo  {journal} {Phys. Rev. A}\ }\textbf {\bibinfo {volume} {40}},\
  \bibinfo {pages} {4077} (\bibinfo {year} {1989})}\BibitemShut {NoStop}%
\end{thebibliography}%
\end{document}